\documentclass[sn-basic]{sn-jnl}

\usepackage{graphicx}%
\usepackage{multirow}%
\usepackage{amsmath,amssymb,amsfonts}%
\usepackage{amsthm}%
\usepackage{mathrsfs}%
\usepackage[title]{appendix}%
\usepackage{xcolor}%
\usepackage{textcomp}%
\usepackage{manyfoot}%
\usepackage{booktabs}%
\usepackage{algorithm}%
\usepackage{algorithmicx}%
\usepackage{algpseudocode}%
\usepackage{listings}%

\usepackage{bm}

\title{Spectral synthesis techniques for supernovae and kilonovae}

\author*[1]{\fnm{Anders} \sur{Jerkstrand}}\email{anders.jerkstrand@astro.su.se}

\affil*[1]{\orgname{The Oskar Klein Centre, Department of Astronomy, Stockholm University, AlbaNova}, \orgaddress{\city{Stockholm}, \postcode{10691}, \country{Sweden}}}

\newcommand{\blue}[1]{\textcolor{black}{{#1}}}
\newcommand{\bluetwo}[1]{\textcolor{black}{{#1}}}

\makeatletter
\input{aas_macros.sty}
\def\ref@jnl#1{{\jnl@style#1\ }}
\makeatother

\graphicspath{ {./figs/} }

\begin{document}

\maketitle

\begin{abstract}
\textbf{Supernovae (SNe) and kilonovae (KNe) are the most violent explosions in cosmos, signalling the destruction of a massive star (core-collapse SN), a white dwarf (thermonuclear SN) and a neutron star (KN), respectively. The ejected debris in these explosions is believed to be the main cosmic source of most elements in the periodic table. However, decoding the spectra of these transients is a challenging task requiring sophisticated spectral synthesis modelling. Here, the techniques for such modelling is reviewed, with particular focus on the computational aspects. We build from a historical review of how methodologies evolved from modelling of stellar winds, to supernovae, to kilonovae, studying various approximations in use for the central physical processes. Similarities and differences in the numeric schemes employed by current codes are discussed, and the path towards improved models is laid out.} 
\end{abstract}

\newpage
\setcounter{tocdepth}{3}
\tableofcontents

\newpage
\section{Introduction}

\subsection{Historical background and overview}

\blue{Explosive transients have always been one of the cornerstones of astronomy. Galactic supernovae have brought attention to the otherwise quiescent night sky for as long as human civilization has existed, with the oldest surviving records being those of Chinese astronomers observing SN 185 over 1800 years ago.}

\blue{It was the Swedish astronomer Knut Lundmark, who in his 1925 treatise \emph{``The motions and distances of spiral nebulae''} was the first to realize that there was a particular subgroup of novae that appeared fundamentally different to the usual ones, being much brighter. Lundmark refered to these a \emph{giant novae}. The name that would come to stick, however, was \emph{supernovae}, first used in \citet{Baade1934}. Baade and Zwicky speculated that supernovae represented the collapse of stellar cores to neutron stars, a remarkably apt inferrence coming just two years after the discovery of the neutron by James Chadwick in 1932. Ironically, all the supernovae that were known at that time were of the Type Ia variant, representing a completely different phenomenon; the thermonuclear destruction of a white dwarf. Today, we know that the collapse of the cores of massive stars to neutron stars (``core-collapse supernovae'') make up about 80\% of all supernovae by unit volume, with the Type Ia class the other 20\% \citep{Li2011}. But it was not until the mid-1980s that the various observational classes could be correctly associated with the right type of explosion \citep{Wheeler1985,Filippenko1986,Gaskell1986}.}
%
%

\begin{figure}[ht]
\centering
\includegraphics[width=0.75\linewidth]{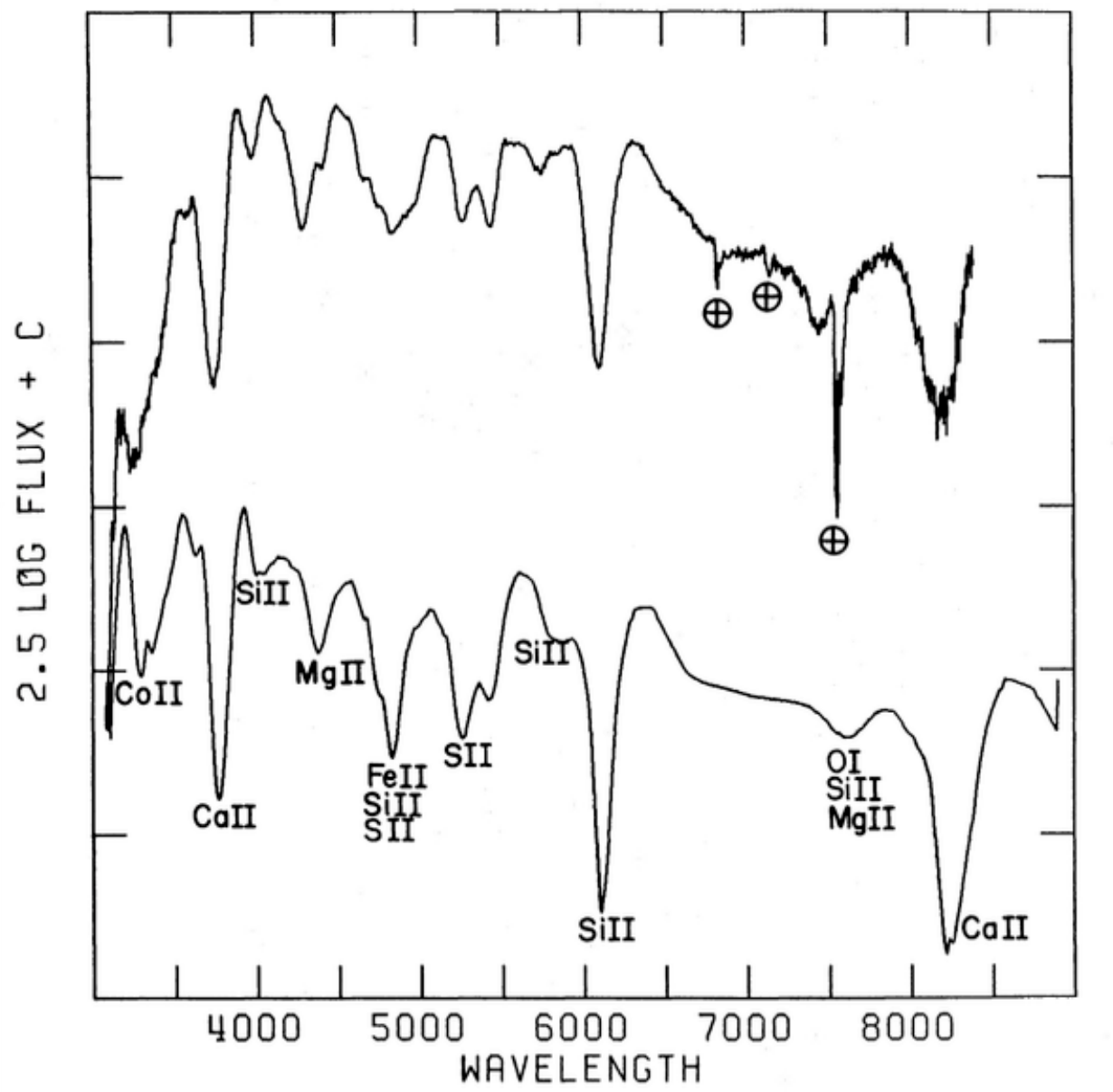}
\caption{Example of an early SN spectral model (bottom) for the photospheric spectrum of a white dwarf explosion model compared to an observed spectrum (top). From this modelling, the first identification of which lines are important for the different observed featured could be established. Image reproduced with permission from \citet{Branch1985}, copyright by AAS.}
\label{fig:branch}
\end{figure}

\blue{The key to that development, and much of supernova science since, was spectral observations and modelling of these spectra. One of the first papers (a conference proceeding) which presented supernova \emph{synthetic spectra} was \citet{Branch1980}. He was the first to develop and adapt the Schuster--Schwarzschild modelling approach (a frequency-independent photophere with a scattering atmosphere outside) to supernovae, identifying the use of important approximations such as the Sobolev formalism for line transfer. With this tool in place, quite detailed comparisons between model spectra of white dwarf explosion simulations and observations of Type I SNe (Fig.~\ref{fig:branch}) could be carried out starting with the works of \citet{Branch1983, Branch1985}. The tool built by David Branch and his group would eventually become the famous \texttt{SYNOW} code, which is available at \url{https://c3.lbl.gov/es/}. It is still frequently used today for line identifications and rapid model investigations.} 
%

\blue{As the ejecta expand the photosphere eventually disappears and the inner ejecta become visible, emitting nebular emission lines. The first step for computing spectral models in this phase was taken also in 1980, in the remarkable PhD thesis by Timothy Axelrod at Berkeley, supervised by Tom Weaver and Stan Woosley   \citep{Axelrod1980}. This work lays the foundation for the various pieces of physics going into such models, including non-thermal and NLTE (Non-Local Thermodynamic Equilibrium) physics. It still today sets the standard for many aspects of nebular-phase spectral modelling. The theoretical foundation by Axelrod has been used in several later nebular-phase codes, e.g. \citet{Mazzali2001,Maeda2006}.}

\blue{The explosion of SN 1987A brought more workers into the field. In Stockholm, Claes Fransson developed spectral models for its late emission \citep{Fransson1987}, and then extended this to Type Ib supernovae \citep{Fransson1989}. In London, Leon Lucy developed a Monte Carlo code for Schuster--Schwarschild modelling \citep{Lucy1987}. The Lucy code was further developed and applied in \citet{RuizLapuente1992, Mazzali1993,Mazzali2000}.}

\blue{Starting around 2005, significant developments took place for supernova spectral synthesis techniques, much of it inspired by the works of Leon Lucy over the preceding years to develop Monte Carlo techniques beyond the first simplified uses \citep{Lucy1999a,Lucy2003,Lucy2005}. The 3D LTE Monte Carlo codes \texttt{SEDONA} \citep{Kasen2006} and  \texttt{ARTIS} \citep{Kromer2009} appeared. The 1D NLTE codes (with radiative transfer) \texttt{SUMO} \citep{Jerkstrand2011,Jerkstrand2012}, \texttt{NERO} \citep{Maurer2011} and \texttt{CMFGEN} \citep[][describe its adaptations to SN applications]{Hillier2012} were developed. A first public Schuster--Schwarzschild code, \texttt{TARDIS}, was released  \citep{Kerzendorf2014}.}

\blue{Just as much of supernova modelling became an extension of methods, tools and concepts originally devised for stellar winds and H I regions, mostly during the 1970s \citep[e.g.][]{Lucy1970,Dalgarno1972}, it in turn became the spring-board for kilonova modelling. This era started in earnest with the paper of \citet{Metzger2010}, where the supernova code \texttt{SEDONA} was extended to be used for kilonovae. Other 3D LTE Monte Carlo codes capable of KN modelling were developed by \citet[][the code is not formally named, we will refer to it as \texttt{TH13} here]{Tanaka2013} and \citet[][\texttt{SuperNu}]{Wollaeger2013}. In this last paper, new technical steps were taken by the development and application of implicit Monte Carlo methods. The code \texttt{POSSIS} \citep{Bulla2019,Bulla2023} operates on similar principles as \texttt{ARTIS} and \texttt{TH13}, and the SN code \texttt{JEKYLL} \citep{Ergon2018} implements several computational efficiency improvements to the Lucy method. The \texttt{SUMO} code was recently adapted to KN modelling \citep{Pognan2022a}, which has opened up the path towards NLTE modelling of KNe. The first steps towards 3D NLTE modelling, so far for supernovae only, have also been taken \citep{Botyanszki2018,Shingles2020,vanBaal2023}.}

\blue{Looking back at the tapestry of developments, one can see how astrophysics grows by small extension steps into adjacent, related areas. That process takes time, as in decades. It is interesting to contextualize that ``application diffusion'' process against changes in computing power. Sometimes growth in computing power can drive such steps to be taken. But more often, they tend to be application-driven. For the latter case, it may happen that methods used eventually get outdated, with the original derivations occurring under the constraints of computing power limitations orders of magnitudes stricter than today. It is worth contemplating these aspects whenever a code or methodology is studied; sometimes simplifications and approximations done in the original model are no longer necessary, and modern computing power can be taken advantage of to improve upon the approach. This is one of several things we gain when studying the history and gradual development of methodologies.}

\subsection{Classification}

Supernovae come in two main types; the explosion of massive stars, called \emph{core-collapse supernovae}, and the explosion of white dwarfs, called \emph{thermonuclear supernovae}. 
The resulting SNe are classified in a system using both spectral and light curve properties \citep{Filippenko1997}. Presence of H lines in spectra leads to a ``Type II'' classification, and absence to a  ``Type I'' classification. For the Type I class, presence of He lines leads to ``Type Ib'' subclass, and absence to ``Type Ic''. A particular group of Type Ic SNe have unusually broad lines, and are referred to as ``Type Ic-BL'' SNe. The Type II's are divided into subclasses of Type IIP if the light curve has a plateau, and Type IIL if it instead has a linear decline. Finally, if narrow H lines are seen (thought to arise from the circumstellar medium rather than the SN), the classification is Type IIn.

The thermonuclear supernovae are classified as ``Type Ia''. The destruction of the white dwarf involves ignition of its large content  of carbon. Because conditions are degenerate, such nuclear burning ignition meets no damping by pressure expansion, and a runway occurs. The ignition can occur either when the white dwarf accretes matter from a companion star, or when it merges with another white dwarf. The smaller variation possible both for the exploding star (a white dwarf close to the Chandrasekhar mass), and the resulting nucleosynthesis (everything is burnt to the iron-group), means Type Ia SNe show more uniform properties and correlations than CCSNe. This allowed them to be the standard candle tool \citep{Phillips1993} that eventually led to the discovery of the accelerating expansion of the Universe \citep{Riess1998,Perlmutter1999}. However, rare subclasses also exist \citep{Taubenberger2017}.

While SNe have been observed and scientifically studied for over a century, \emph{kilonovae} were discovered only in 2017. These transients involve the merger of two neutron stars, or a neutron star and a black hole --- a much more rare phenomenon than the death of a massive star or a white dwarf. While most of the mass in the merging bodies will form a black hole, about 1\% or so ($\sim 0.05\, M_\odot$) will be ejected from the gravitation potential. This material has gone through the r-process --- rapid capture of neutrons forming heavier and heavier elements. The composition is therefore trans-iron elements. Many of the isotopes formed are radioactive, which provides an important power source to generate a bright transient from the ejected material. Because the mass is low, but the velocity high (about 10\% the speed of light), KNe rise and decline over just a few days, compared to weeks or months for SNe. With only one solid detection, and a handful of candidates, no classification system yet exists for them.

\subsection{Motivation and goals of supernova and kilonova spectral synthesis}

\begin{figure}[ht]
    \includegraphics[width=1\linewidth]{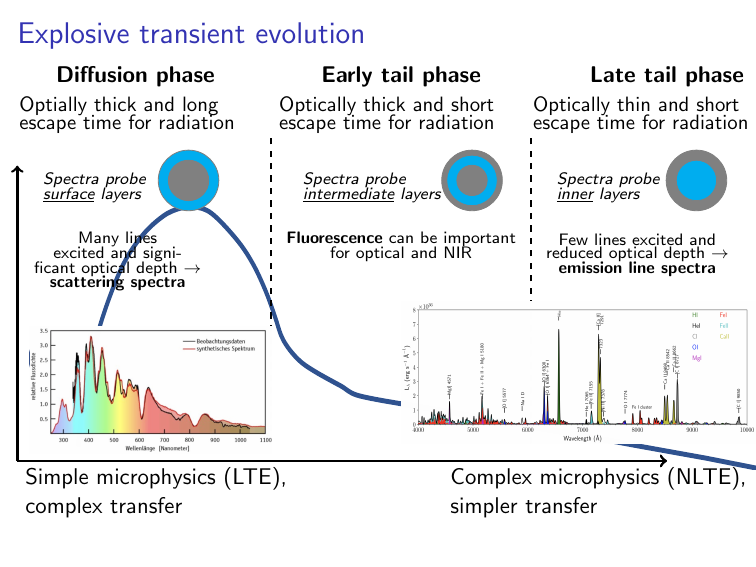}
    \caption{Characteristic evolution of SNe and KNe over the three phases of diffusion phase, early tail phase, and late tail phase.}
    \label{fig:scheme}
\end{figure}

Figure \ref{fig:scheme} shows a schematic illustration of the evolution of explosive transients, and what layers spectra probe at different phases. As optical depths decline with time, it becomes possible to see deeper and deeper into the nebula, and therefore is time-sequencing important to obtain full-ejecta information.

\begin{figure}[ht]
    \includegraphics[width=1\linewidth]{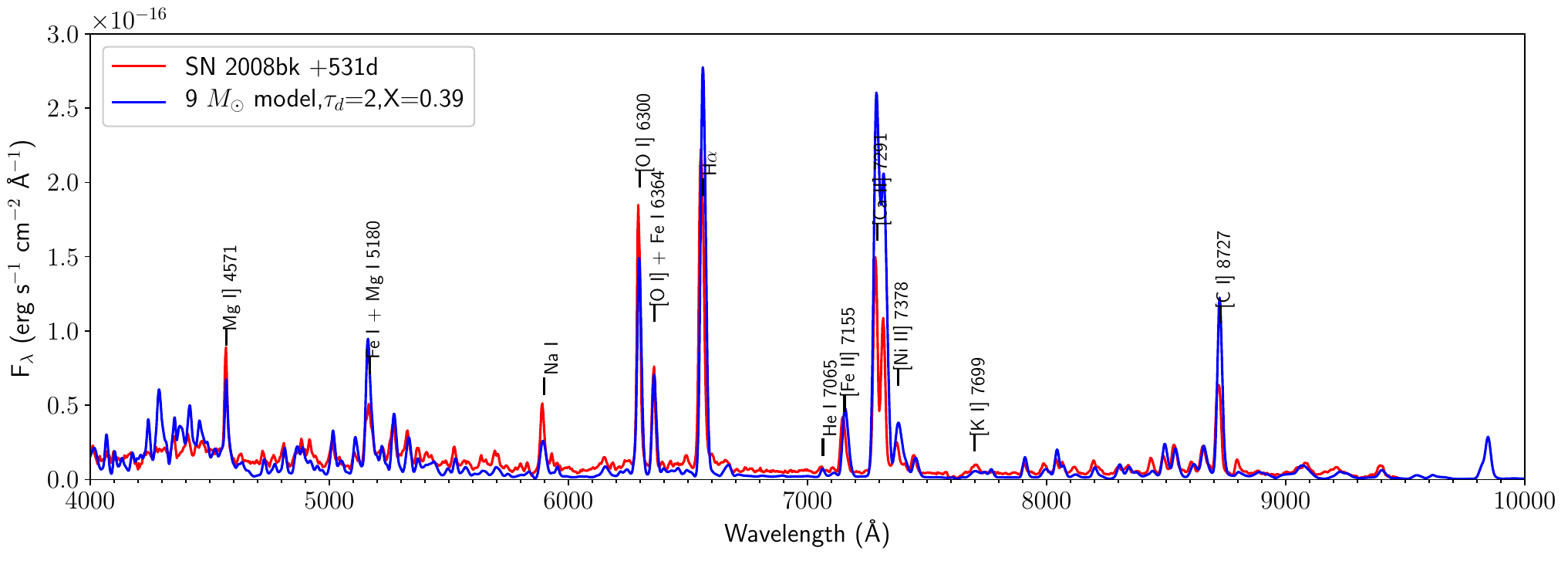}
    \caption{Observed spectrum of SN 2008bk (red), and a model spectrum (blue). Image reproduced with permission from \citet{Jerkstrand2018}, copyright by the author(s).}
    \label{fig:2008bk}
\end{figure}

Figure \ref{fig:2008bk} shows the spectrum of an observed Type II SN compared to a spectral synthesis model. The comparison demonstrates that current models can be quite successful in reproducing the overall spectral shapes of observations, and can thus be used to attempt detailed inferrances about composition and structure.

\blue{The motivation and goals for spectral synthesis modelling of supernovae and kilonovae are rich and multifaceted. Three of the main science drivers are discussed below.}\\ 

\noindent \textbf{\emph{1)  Identifying which elements produce which line features, and determining the elemental abundances in the ejecta}}. \blue{For many years, the challenges to directly infer elemental abundances directly from supernova spectra appeared almost unsurmountable. Supernovae provide formidable cosmic laboratories with a multitude of complex physics; rapid expansion that Doppler-blends lines, energy cascades over six orders of magnitudes, asymmetries, non-thermal and NLTE effects, molecule and dust formation. Traditional methods from nebular astrophysics run into difficulties and become difficult to meaningfully apply \citep[see][for a good discussion]{McCray1996}.}

\blue{This led to the curious situation of supernova explosion and nucleosynthesis models having been tested mainly through  
comparisons to the solar composition, stellar atmospheres abundances, and galactic chemical evolution modelling \citep[e.g.][]{Chiappini1997, Rauscher2002, Tominaga2007}, which depend on a long and complex injection and galactic mixing history of ejecta from a large number of SNe.} 

\blue{Over the last decade or so, however, supernova spectral synthesis models have reached, arguably, a degree of physical realism that this is no longer the case. Synthetic spectra of explosion models are now in quite good overall agreement with observations for several SN classes \citep[see][for reviews]{Jerkstrand2017,Sim2017}, and it is possible to attempt to infer abundances to the accuracy needed to directly test individual explosion models and nucleosynthesis theory. By looking into the modelling machinery, we can now also better understand \emph{how} lines are formed, and from this how to both understand uncertainties and to devise good analytic methods \citep[e.g.][]{Jerkstrand2012,Jerkstrand2015a,Jerkstrand2015b,Maguire2018}. Thus, one may expect that a paradigm shift is about to occur where this can be systematically done without being limited to more indirect comparisons to estimated stellar atmospheres abundances. There are today clear diagnostic methods established for H, He, C, N, O, Ne, Na, Mg, Si, S, Ar, K, Ca, Ti, Fe, Co, and Ni in supernovae, and direct full-ejecta spectral explosion modelling tests have been done for all major SN classes \citep[e.g.][]{Hoflich1998,Hoflich2002,Maeda2002,Maeda2006,Kasen2005,Kasen2007,Sim2010,Sim2012,Sim2013,Kromer2010,Kromer2013,Blondin2013,Blondin2023,Dessart2013,Dessart2016,Jerkstrand2012,Jerkstrand2015a,Jerkstrand2016,Jerkstrand2018,Shen2021,vanBaal2023}.}\\

\noindent \textbf{\emph{2)  Understanding the origin of the elements across the periodic table.}}
\blue{With the advent of AT2017gfo, the first kilonova, the upper two thirds of the period table have now also opened up for direct source analysis. While attempts for line identifications and abundance estimates have just begun, diagnostic potential has already been established for $_{34}$Se, $_{37}$Rb, $_{38}$Sr, $_{39}$Y, $_{52}$Te, $_{57}$La, $_{58}$Ce, $_{60}$Nd, and $_{74}$W \citep{Watson2019,Domoto2022,Sneppen2023,Hotokezaka2022,Gillanders2022, Hotokezaka2023,Pognan2023}. Kilonovae have both advantages and disadvantages when it comes to composition analysis compared to supernovae. The higher velocities ($\sim0.2$c compared to $\sim0.02$c) lead to more severe line blending, which complicates identifications. The atomic data for r-process elements is so far much less well known than for elements up to the iron-group, with wavelength uncertainties complicating line matching and A-value uncertainties complicating abundance determinations.  On the other hand, for KNe there are many more 3D hydrodynamic explosion models with realistic nucleosynthesis available to compute spectra for. The fact that all of the ejecta are radioactive in kilonovae also circumvents a long-standing difficulty of properly modelling the mixing between radioactive regions ($^{56}$Ni-rich) and the rest of the ejecta in supernova modelling.}

By systematic spectral modelling of SNe and KNe - event by event, class by class - a major objective is to synthesize a theory for the origin of elements based on direct nycleosynthesis inferrences.\\
\\
\noindent \textbf{\emph{3) Determining the progenitor stellar systems and the explosion mechanisms of supernovae and kilonovae.}} \blue{For collapsing massive stars, the inner parts of the star form a compact object (neutron star or a black hole), and the outer parts are violently ejected. That exotic region of a collapsing stellar core provides a cosmic laboratory where fundamental physics can be probed in regimes not accessible anywhere else in cosmos. The equation of state at the highest densities, strong-field general relativity, neutrino physics, including the enigmatic neutrino oscillations, accretion processes, magneto-hydrodynamics in the context of stellar rotation, and explosive nucleosynthesis are manifested over a few seconds to together produce the supernova phenomenon. Similarly, the explosion of white dwarfs probes a unique physical situation, either arising as two white dwarfs merge, or a single white dwarf accretes matter and initiates collapse as it exceeds the Chandrasekhar limit.  probes  By analysing supernova spectra, we can diagnose both the life of the progenitor star, through its hydrostatic nucleosynthesis yields, and its death, through its explosive nucleosynthesis yields and resulting morphology of the ejecta.}

\blue{In kilonovae, somewhat different high-energy physics is probed. There are here more distinct ejecta components \citep[see e.g.][for a review]{Shibata2019}, ranging from the ``contact-squeezed'' (probably relatively high electron fraction $Y_e=n_p/(n_p+n_n)$, where $n_p$ and $n_n$ are proton and neutron abundances, probes the light r-process), ``tidal tail'' (low $Y_e$, probes heavy r-process), ``disc shocks'' (probes hypermassive neutron star physics) and ``disc viscous'' components (probes accretion physics). The study of these outflows opens up a probe of  material having been as close to a black hole as matter can get, and can give us insights into the physics of accretion disks and their magnetohydrodynamic processes. The bulk of the ejecta are currently believed to originate from the accretion disk outflows (the last two channels mentioned above). The angular momentum-driven ``disc viscous'' outflow is particularly important, giving slower ejecta compared to the other channels, as the outflow occurs from the outer edge of the disk where the escape velocity is relatively lower. The lower velocities lead to formation of more narrow emission lines which blend less and therefore give more favorable conditions for identification and analysis.  Kilonova spectra give us a unique window on the r-process nucleosynthesis as it happens in real time, and a powerful diagnostic for compact object merger physics.} 

\subsection{Sketch of physical situation} 
Table \ref{table:properties} lists some of the basic physical parameters of CCSNe, TNSNe and KNe. For the phases we will mainly concern ourselves with here, some basic properties of the SN and KN nebulae are
\begin{itemize}
 \item \blue{\textbf{Homologous expansion}. Following an explosion, the faster fragments will soon be further away than the slower ones, and all of them will eventually travel on close to radial trajectories away from the explosion centre. When $\bm{r} \approx \bm{v}/t$, we say that the nebula is in homologous expansion. Every point then sees all other points receding away from it, like the galaxies in the Hubble flow. The characteristic velocity scale of the material is around 5000 km s$^{-1}$ for CCSNe and around 50,000 km s$^{-1}$ for KNe. KNe have about 100 times less ejecta mass than SNe ($\sim 0.05~M_\odot~\mbox{vs} \sim 5~M_\odot$), but about the same kinetic energy ($\sim 10^{51}$ erg), leading to a factor $\sim$10 higher velocities ($v \propto \sqrt{E/M}$). Homology is reached faster for more compact progenitors, taking a few hours or days for CCSNe (exploding system of size $10-10^3\ R_\odot$), but only minutes for KNe ($\sim 10^{-4}\ R_\odot$).}

\item \blue{\textbf{Radioactive power source}. The main power source for the electromagnetic display of many CSSNe and all TNSNe is believed to be the decay of $^{56}$Ni/$^{56}$Co. Most of the decay occurs in the form of gamma rays, which Compton scatter in the nebula, depositing their energy. Kilonovae are also powered by radioactivity, but here a large number of (r-process) nuclides contribute. The main mediators here are not gamma rays but leptons, and in some cases also $\alpha$-particles and fission fragments. Thus, the power source has here a different distribution with respect to the nebula as a whole, and a different time evolution and thermalization physics. Nevertheless --- modelling of both types of nebulae starts with modelling of how radioactivity powers a homologously expanding gas.}

\item \blue{\textbf{Semi-transparency, with localized line interactions.} Supernovae are often discussed and modelled in the theoretical limits of being optically thick (``photospheric'') and optically thin (``nebular''). The true situation is often somewhere in between, and very wavelength-dependent \citep{Jerkstrand2017handbook}. While the optical range clears of continuum opacity relatively quickly, line opacity can give lingering radiative transfer effects for years or decades \citep{Jerkstrand2011}. It has also been understood quite recently that the non-thermal nature of the powering makes UV opacity important, as much energy is reprocessed to UV emission also when temperatures are low \citep{Fransson2015}. At all times, much of the optical and IR spectra of both SNe and KNe are formed by fluorescence \citep{Shingles2023,Pognan2023}.}

\blue{The current picture is well described as that some lines and wavelength ranges can be understood and modelled with quite simple formation physics, e.g. pure scattering (P-Cygni lines) or optically thin emission, whereas others are more complex in formation. The situation is nuanced and much recent work has gone into trying to understand which regimes apply to which lines, and when.}

\blue{The large velocity scale covered by the ejecta (thousands of km/s) implies that line transfer (occurring over the thermal line width, of order 1--10 km/s) occurs over a region small compared to the overall size of the nebula. After a certain spatial distance the comoving (Lagrangian frame) rest wavelength of the photon will have redshifted out of resonance with the transition, and the ``Hubble flow'' dynamics means it can never come into resonance with that line again, anywhere else. Line interactions are therefore strictly local, and furthermore millions or sometimes billions of resonance scatterings in this local region can be simplified to a a single scattering description in the so called Sobolev formalism \citep{Sobolev1957,Castor1970}. This simplification has large consequences for convergence properties and algorithm choice, which we will get back to later.} 
\end{itemize}

\begin{table}[ht]
\centering
\caption{Characteristic properties of CCSNe, TNSNe, and KNe}
\label{table:properties}
\begin{tabular}{cccc}
\toprule
Property & CCSN & TNSN & KN \\
\midrule
$M_{ej} (M_\odot)$ & 2--15 & 1 & 0.05 \\
$v_{ej} (\mbox{km s}^{-1})$      & 2,000--8,000 & 8,000 & 30,000\\
Composition (Z) & 1--30 & 20--30 & 30--100\\
$\rho_{\rm peak} (\mbox{g cm}^{-3})$ & $10^{-13}$--$10^{-12}$ & $10^{-14}$--$10^{-13}$ & $10^{-14}$\\
$T_{\rm peak} (K)$ & 5,000--20,000 & 10,000 & 10,000\\
$\tau_{\rm trans} (d)$ & 50--200 & 50--100 & 10\\
\bottomrule
\end{tabular}
\end{table}

\subsection{Code classes}
\bluetwo{
Codes used for predicting explosive transient light curves and spectra can initially be divided into two groups depending on whether hydrodynamics is solved for or not. The codes including hydrodynamics are referred to as \emph{radiation hydrodynamics} codes. They often produce bolometric and photometric light curves, but not spectra. Examples include \texttt{KEPLER} \citep{Weaver1978}, the code of \citet{Hoflich1996}, \texttt{STELLA} \citep{Blinnikov1998}, \texttt{CRAB} \citep{Utrobin2004}, the Bersten code \citep{Bersten2011}, and \texttt{SNEC} \citep{Morozova2015}.}

\bluetwo{Codes not solving for hydrodynamics must assume a dynamic structure of the ejecta. This structure is, typically, homologous expansion. With the reduction in coding efforts, computing time, and convergence issues related to coupling in hydrodynamics, these codes can instead solve for the radiative transfer to higher accuracy, giving high-resolution spectra. These codes are referred to as \emph{radiative transfer} codes, and examples include \texttt{EDDINGTON} \citep{Eastman1993}, the Höflich code \citep{Hoflich1993}, \texttt{PHOENIX} \citep{Hauschildt1999}, \texttt{CMFGEN} \citep{Hillier2012}, \texttt{SEDONA} \citep{Kasen2006}, \texttt{ARTIS/TARDIS} \citep{Kromer2009,Kerzendorf2014}, \texttt{SUMO} \citep{Jerkstrand2011},  \texttt{TH13} \citep{Tanaka2013}, \texttt{SuperNu} \citep{Wollaeger2013}, \texttt{URILIGHT} \citep{Wygoda2019} and \texttt{POSSIS} \citep{Bulla2023}. 
Many of them do radiative transfer with Monte Carlo methods, especially useful for multidimensional modelling, but for 1D modelling also transfer equation solving can be done. A special subgroup of radiative transfer codes are those considering only local self-absorption in the transfer, through the Sobolev formalism,  ignoring the non-local transfer. This ansatz becomes reasonable only at late, nebular times, so these are all NLTE codes. Examples include the 1D codes of \citet{Mazzali2001} and \citet{Kozma1998-I} and the multi-D codes of \citet{Maeda2006}, \citet{Botyanszki2017}\footnote{In this work also the local self-absorption is ignored.}, and  \citet{vanBaal2023}.}

\bluetwo{The reason radiation hydrodynamics codes do not do detailed radiative transfer, and vice versa spectral synthesis codes do not do hydrodynamics, has not only to do with issues of coding complexity and convergence; it is the evolving physical situation that sets the needs.  Pressure forces, driving dynamics, are important when the density, temperature, and radiation field intensity are high. But those conditions also bring about LTE, broadly speaking, and the internal radiation field which governs the dynamics is well solved for by a simple diffusion approximation. When densities are lower, such that better transfer than diffusive is needed, the ejecta are typically also in their coasting stage (having reached close to the final velocities) and solving for hydrodynamics becomes unnecessary.}

\bluetwo{The natural division into certain discrete regimes for the physics of explosive transients has both allowed for progress, by having specific efforts tackle specific regimes, but also led to a literature of mostly piece-wise modelling efforts. Ideally, different modellers, specialising in different parts of the problem, would work together and pipeline their efforts. 
The recent large radiative transfer code comparison project \citep{Blondin2022} is an important step in this direction, comparing methods and outputs between codes, and identifying paths forward for the field. It continues in the spirit of \citet{Blinnikov1998}, who presented the first such efforts for supernova codes.}

\subsection{Review goals and structure}
\bluetwo{The purpose of this review is to give an overview of the basic physical ingredients for modelling spectra of supernovae and kilonovae, and the numeric implementation of these in modern spectral synthesis codes. Previous related reviews include those of SN spectral formation in the photospheric phases by \citet{Sim2017} and nebular phases by \citet{Jerkstrand2017handbook}. 
This topic is, of course, a vast one and the review only aims to cover the most central themes and aspects. In fact, one of the four central blocks of such modelling --- the radiative transfer --- is not treated in depth as a chapter in itself, but instead woven into the other three blocks (temperature, level populations, and radioactive powering). This is partly motivated by that that radiative transfer would truly need its own full review, but also that there exists already many extensive texts on the theoretical foundation and various radiative transfer solution methods \citep[e.g.][]{Baron1996,Pinto2000,Lucy2005,Hillier2012, Noebauer2019}.}

\bluetwo{We therefore instead go at some depth into the other three blocks, which have received less attention in previous reviews. In Sect.~\ref{sec:temperature} we study the law of energy conservation and its implementation, and how it serves as the most central equation governing the temperature evolution. In Sect.~\ref{sec:rateequations} we study how rate equations govern ion and level populations in NLTE modelling, which is now the frontier for both SN and KN modelling. In Sect.~\ref{sec:powering} we study the radioactivity that powers SNe and KNe, and how its treatment varies between the LTE limit and various types of NLTE.}

The target audience of the review are both beginning researchers who are entering the field of computational modelling of astrophysical transients, but also senior SN and KN researchers who have experience with transient observations and would like to develop a deeper understanding of how models are constructed and can help in interpretation of data. I hope also that the review is useful for the community of experienced modellers for an overview of where the field currently stands. For all groups, my hope is that a mixture of historical context, ingoing physics, numeric formulation, solution techniques, and example illustrations will give the right mixture of aspects to make the review a useful resource. 

\bluetwo{The following topics are not covered by this review in any significant depth:
\begin{itemize}
\item Light curve modelling.
\item Comparisons between models and observations.
\item Analytic and semi-analytic models.
\item Other powering situations than radioactivity. 
\item Atomic data.
\item Numerics of radiative transfer modelling of other explosive transients such as novae, Active Galactic Nuclei, or Tidal Disruption Events.
\end{itemize}}

\section{Temperature}
\label{sec:temperature}
\blue{The most fundamental quantity for SN and KN physical modelling is, arguably, the local gas temperature. Temperature governs the ionization and excitation, sets the regime for the spectral formation, and strongly influences the opacity, which in turn governs the radiative transfer. For a model to achieve good fidelity it needs an accurate machinery to compute the gas temperature. It is therefore suitable that we start our survey of the physical modelling of SNe and KNe by looking at how temperature is physically determined, and which equations, approximations, and numeric implementations are in use in current modelling.}

\blue{Due to the very large cross section for elastic particle collisions at low energies
, an equilibrium Maxwell--Boltzmann distribution is typically a good approximation for the bulk of the particles (electrons, atoms and ions), which are at $\lesssim$ eV energies in SNe/KNe. The quantity $1/2kT$ is the average kinetic energy per particle, per degree of freedom (three in total from three different spatial directions), in this distribution. This is the fundamental meaning of temperature in this context, and the equilibration is why it is a meaningful and central quantity.}

\subsection{The energy equation}
\blue{The fundamental physical law governing  temperature is energy conservation. The energy conservation equation, in the comoving frame, and ignoring conduction and dissipation which are unimportant in SNe and KNe, can be expressed as \citep[e.g.][their Eq.~(16.33)]{Hubeny2014}:}
\begin{equation}
\bluetwo{\frac{D e}{D t} + p\frac{D}{Dt}\left(\frac{1}{\rho}\right) =  \frac{4\pi}{\rho}\int_0^\infty   \left(\alpha_\nu J_\nu - \eta_\nu\right) d\nu + \epsilon},
\label{eq:firstlaw1}
\end{equation}
\blue{where $D/Dt$ is the Lagrangian derivative, $e$ is the internal energy per unit mass, $p$ is the gas pressure, $\rho$ is the density, $\alpha_\nu$ is the (total) absorption coefficient at frequency $\nu$, $\eta_\nu$ is the (total) emission coefficient, $J_\nu$ is the mean intensity of the radiation field, and $\epsilon$ is a source term (erg g$^{-1}$ s$^{-1}$) representing e.g. energy injection by radioactive decay.} 

\blue{The internal energy $e$ consists of translational kinetic energy by random motions $e_k$ plus the potential energy of excited/ionized states $e_p$\footnote{\blue{Note that radiation energy density is not included in $e$ --- this is because the RHS of Eq.~\eqref{eq:firstlaw1} only includes (net) terms that involve state change of material particles. One could add in also $e_{\rm rad}$, but then need to keep track of boundary fluxes.}} \citep[e.g.][]{Mihalas1984, Hillier2012}:} 
\begin{equation}
 \bluetwo{ e = e_k + e_p = \frac{3}{2}\frac{kT\left(1+x_e\right)}{\bar{A}m_p} + \frac{1}{\rho}\sum_{\rm ion,exc} n_i E_i} \,,
 \label{eq:eeq}
\end{equation}
\blue{where $x_e \equiv n_e/n_{nuclei}$ is the electron fraction, $\bar{A}$ is the mean atomic weight (number), $n_i$ is the number density of state $i$, and $E_i$ is the energy of that state. For atoms and ions ``excitation'' involves change of electronic orbitals, whereas for molecules the nuclei can also be excited into rotational and vibrational states.}

\blue{Absorption and emission can occur by a variety of processes, which can be divided into the categories of \emph{free-free}, \emph{free-bound/bound-free}, and \emph{bound-bound}, referring to the initial and final state of the electron involved in the process. Another division axis is along \emph{scattering} and \emph{absorption}, referring to whether photons merely change direction in the interaction process\footnote{\blue{In detail, there is always a non-zero energy exchange, but in scattering processes it is neglegibly small.}} or are destroyed. Because scattering does not change the states of the material particles, if $\alpha_\nu$ and $\eta_\nu$ have explicit  scattering components, these may be removed before use in Eq.~\eqref{eq:firstlaw1}. Temperature generally has a strong influence on $\alpha_\nu$ and $\eta_\nu$ which, together with the explicit $T$-dependency of $e$ (Eq.~\eqref{eq:eeq}), is why we can say that the energy equation is the key equation for setting the temperature. For example, the contribution to the emission integral by a line $ul$ in LTE from an ion $i$ is given by $4\pi/\rho \int j_\nu d\nu = (n_i/\rho) A_{ul} g_u \exp{\left(-E_u/kT\right)}/Z(T)$, where $Z(T)$ is the partition function.}


\blue{When NLTE is considered\footnote{\blue{In LTE it is assumed that energy exchange between the various pools is rapid, so it is not suitable to break out components of $e$.}}, one may physically equivalently,  
but computationally sometimes more expedient and numerically better conditioned, write an equation describing the evolution of the thermal kinetic energy stored in the pool of atoms, ions and thermal electrons \citep[e.g.][]{Kozma1998-I,Jerkstrand2011PhD}:}
%
%
%
%
%
%
%
\begin{equation}
\bluetwo{\frac{De_k}{Dt} + p\frac{D}{Dt}\left(\frac{1}{\rho}\right) = h - c \,,}
\label{eq:firstlaw2}
\end{equation}
\blue{where $h$ is the heating per unit mass (energy flow going towards increasing the kinetic energy of particles) and $c$ the radiative cooling per unit mass (energy flow going towards decreasing the kinetic energy of the particles). Adiabatic cooling is represented by the second term on the LHS. For example, if 50\% of the radioactive decay power (r.p) ends up as internal kinetic energy of the particles, $h_{r.p.}=0.5 \epsilon$. As another example, the cooling through a line transition is $c_{lu} = \rho^{-1} E_{ul}\left(q_{lu}(T)n_e n_l - q_{ul}(T) n_e n_u E\right)$, where $q_{lu}$ and $q_{ul}$ are upward and downward collision rates.}  

\blue{Note that charge exchange reactions and chemical reactions (molecule and dust formation and destruction) are also associated with heating and cooling which needs to be accounted for if these processes are included. The reader is referred to \citet{Hillier2003} for further discussion of the choice of energy equation.}

\blue{The \emph{thermal equilibrium} approximation 
corresponds to ignoring the two time-derivative terms in the energy equation (Eq.~\eqref{eq:firstlaw1} or, equivalently, Eq.~\eqref{eq:firstlaw2}), so the whole LHS becomes zero.  This changes the equation from a first-order initial value problem into an algebraic equation. If any source term entries ($\epsilon$) are considered as ``radiation'' (e.g. the $\alpha$, $\beta$ and $\gamma$ particles from radioactive decay), or if the source term is zero, thermal equilibrium is equivalently sometimes called \emph{radiative equilibrium}, as (from Eq.~\eqref{eq:firstlaw1}) the absorption of radiation $(\epsilon + 4\pi/\mathbf{\rho} \int \alpha_\nu J_\nu d\nu)$ balances the emission of radiation $(4\pi/\mathbf{\rho} \int \eta_\nu d\nu)$.}

\blue{For the case of Eq.~\eqref{eq:firstlaw2}, the thermal equilibrium condition is simply}
\begin{equation}
   \bluetwo{h = c.}
\end{equation}
\blue{The expression ``algebraic equation'' is here to be taken as a heuristic description. Both heating and cooling terms are, in practice, implicitly obtained by solving associated large numeric problems. For example, heating may be determined by a full Monte Carlo simulation of the gamma-ray Compton scattering process, and cooling by solving thousands of linked rate equations.}

\blue{When is the thermal equilibrium approximation suitable? It holds when the time-scales of both heating and cooling are fast compared to the time-scales over which the fundamental physical situation --- density and powering --- changes. The latter are characterised by the expansion and power source time-scales, $\rho/\dot{\rho}$ and $\epsilon/\dot{\epsilon}$, respectively. For the case of homologous expansion and single-isotope radioactive power these expressions equal $\rho/\dot{\rho}=t/3$ and $\epsilon/\dot{\epsilon}=\tau_{\rm decay}$. Section~4.4 in \citet{Jerkstrand2011PhD} contains a further discussion and some illustration of the accuracy of the approximation for SNe, and \citet{Pognan2022a} makes a study of it for KN applications. For the KN case one can show that, for $t^{-1.3}$ r-process raw radioactive decay power and a $\left(1+t/t_b\right)^{-1.5}$ thermalization factor \citep[][$t_b$ is a thermalization timescale]{Kasen2019,Waxman2019}, $\epsilon/\dot{\epsilon}$ becomes $t/\left(1.3 + 1.5\left(1+t/t_b\right)^{-1}\times t/t_b\right)$. This expression limits to $t/1.3\ (=0.77t)$ for small $t$ and $t/2.8\ (=0.36t)$ for large $t$. Thus, the homologous time-scale $t/3$ always puts more stringent constraints for KNe.}

\blue{Several radiative transfer codes, e.g. \texttt{STELLA}, \texttt{SuperNu}, \texttt{CMFGEN}, and (since 2022) \texttt{SUMO} have the capacity to retain the time-derivative terms in the energy equation (\texttt{STELLA} and \texttt{SuperNu} use Eq.~\eqref{eq:firstlaw1}, \texttt{SUMO} uses Eq.~\eqref{eq:firstlaw2}, whereas \texttt{CMFGEN} checks that both Eq.~\eqref{eq:firstlaw1} and Eq.~\eqref{eq:firstlaw2} are fulfilled) --- and thus can avoid the issue of the accuracy of the thermal equilibrium approximation. Numerically, implicit time-differencing gives equations of the same form as when doing thermal equilibrium solutions; the implicit time derivative just adds in more source terms, known from the previous time step. Thus, doing time-dependent solutions has from the implementation point-of-view no real additional complexity or convergence issues compared to equilibrium modelling. The difference is only the need to evolve the problem as an IVP through many time steps. If just a single epoch is of interest, having to do a large number of previous epochs is then the ``cost'' of dropping the approximation. But if one desires a full time series of spectra anyway, there is no real extra cost.}

\blue{Solving the time-dependent energy equation is done by a finite difference scheme. \texttt{STELLA} and \texttt{CMFGEN} use a \emph{fully implicit} time discretization \citep[all terms defining the derivative evaluated at the new time step,][]{Blinnikov1998,Hillier2012}, whereas \texttt{SuperNu} applies a \emph{semi-implicit} one \citep[some terms defined at the new time step, some at the previous,][]{Wollaeger2013}. The performance of fully implicit versus semi-implicit schemes can often vary significantly with application \citep[e.g,][]{Stone1992}.  
When applied to a single ODE with constant coefficients, textbooks show us that a fully implicit differencing scheme is unconditionally stable and first order accurate (meaning that accuracy is linearly proportional to the numeric step size), whereas a semi-implicit scheme is unconditionally stable but second order accurate (meaning that accuracy is quadratically dependent on numeric step size, which is the highest order achievable for unconditional stability). However, these statements do not hold universally for all problems. Here we are dealing with an ODE with non-constant coefficients, and in addition this equation is typically co-solved with other non-linear equations. As discussed by \citet{Press1992}, implicit methods are then usually, but not always, stable, and accuracy cannot be so precisely pre-stated in general. 
\citet{Hoflich1993} parametrizes the degree of implicitness, but states that in practice the fully implicit limit is used for all model runs. \citet{Lucy2005} analyses accuracy improvements one may obtain by using information from two previous timesteps instead of one, for the case of solving the energy equation for the radiation field energy, and the first-order moment equation for radiative transfer.}

\blue{When a time-dependent energy equation is used, there are normally also other equations in the system with time-derivative terms retained (giving a set of coupled ODEs). The choice of time-step then comes down to which of these equations requires the shortest step for stability and accuracy. In addition, the task of identifying an optimal discretization method (among implicit, semi-implicit, Runge--Kutta, Richardson extrapolation, predictor-corrector, etc.) becomes very complex. In the SN/KN literature so far, the most common approach is therefore to keep it simple with an implicit method using a fixed, moderate-sized time-step, typically around 10\%. This time step ($\Delta t=0.1t$) is always shorter than the homologous expansion time-scale ($0.33t$) meaning that time resolution with respect to density changes is controlled. The other time-scale to keep track of is the power source time-scale, which is the decay time $\tau_{\rm decay}$ in the case of single-isotope radioactivity (111\,d for $^{56}$Co). Desiring to stay in the regime $\Delta t \leq$ $\tau_{\rm decay}/\left(\mbox{factor\ few}\right)$ then means that 10\% stepping is sufficient for  epochs up to a few years. Conveniently, this is also when other more long-lived isotopes like $^{57}$Co ($\tau_{\rm decay}=392$\,d) and $^{44}$Ti ($\tau_{\rm decay}=87$\,y) take over in SNe. For KNe, we can estimate $\tau_{\rm decay} = (0.36-0.77)t$ as discussed above. Thus, $\sim$10\% time-stepping has a quite solid physical anchoring for both SN and KN applications, for any conceivable range of epochs.} 

\blue{A final note is, however, that having $\Delta t \ll \left[0.33t,\tau_{\rm decay}\right]$ is a first, rough requisite; what really matters is to well resolve the evolution of the physical state variables. Should these vary with time more rapidly (e.g., a sudden episode of dust formation that significantly changes the cooling function), a yet better time-resolution may be needed. Ideally is an adaptive time-step control implemented which can monitor this and redo the time-step if a violation occurs.}

%
%
%
\subsection{Heating}
\blue{When working with $e$ (energy equation \ref{eq:firstlaw1}), the term ``heating'' refers to processes that increase the internal energy, whether it is kinetic energy of the material particles, or excitation/ionization energy. When working with $e_k$ (energy equation \ref{eq:firstlaw2}), it refers to processes in the first category only.}

\blue{In layers exposed to radioactive decay particles (see Sect.~\ref{sec:powering} for details), heating  by impact and cascading of these will be important. Heating by absorption of the diffuse radiation (photoelectric absorption, free-free absorption, and bound-bound absorption followed by collisional deexcitation) can, however, also play a role. Thus, heating should be computed as the sum of a radioactive heating estimate and a radiation field heating estimate.}

\subsubsection{NLTE}
\underline{Radioactive heating.}
\blue{In NLTE one needs to compute the fractions of radioactive decay energy that goes to random thermal motions, and which go to excitations and ionizations. However, for the purposes of the energy equation, as long as the gas is at least singly ionized or so \citep{Kozma1992} no big error is made if one takes the full deposition power to go to heating. In SNe, the ionization decreases with time though, so this approximation gets progressively worse and beyond some point in time the fraction needs to be computed --- more details on this in Sect.~\ref{sec:powering}.} 
\blue{For KNe, it has recently been understood that ionization turns around and increases again after the diffusion phase ends \citep{Hotokezaka2021,Pognan2022a}. The ionization state is always high enough that no big error is incurred if the 100\% heating approximation is retained at all times for the energy equation.}\\
\\
\noindent \underline{Radiation field heating.}
\blue{By whichever radiative transfer method is used, radiation field heating by bound-free and free-free processes are straightforwardly computed.  For bound-bound, line absorption generate photoexcitation which lead to increased upper level population in the next iteration. This in turn modifies the net bound-bound cooling rate for the electrons on transitions involving that level (Eq.~\eqref{eq:firstlaw2}). In an average sense, the level push-ups induced by photoexcitations will lead to a smaller net cooling as more downward (heating) paths become available and fewer upward (cooling) ones. Thus, here line absorptions do not enter explicitly as a heating term in the energy equation, but instead they will degrade the cooling term. This is what happens in NLTE codes like \texttt{SUMO} and \texttt{CMFGEN}.}

\subsubsection{LTE}
\underline{Radioactive heating.}
\bluetwo{In LTE modelling one works with Eq.~\eqref{eq:firstlaw1}, and the radioactive heating is just the total locally deposited energy, with no need to specify its components.}\\
\\
\noindent \underline{Radiation field heating.}
\bluetwo{For the radiation field heating estimate, Monte Carlo codes  typically use the volume accumulator method of \citet{Lucy1999a} to estimate $J_\nu$:
\begin{equation}
J_\nu = \frac{1}{4\pi \Delta \nu V}\sum_{\Delta \nu} \epsilon \times \Delta l,
\end{equation}
where $\Delta \nu$ is the frequency bin width, $V$ is the cell volume, $\epsilon$ is the packet energy, and $\Delta l$ the travel segment.
To compute a radiative heating rate from $J_\nu$, LTE codes require specification of what fraction of absorbed radiation thermalizes and what fraction scatters, i.e. how the total absorption coefficient  breaks down into absorption (thermalization) ($\alpha_{\rm abs,\nu}$) and scattering ($\alpha_{\rm scatt,\nu}$). 
The radiative heating depends only on the absorption part:}
\begin{equation}
\bluetwo{h_{\rm rad-field}^{\rm LTE} = \rho^{-1} \int_0^\infty 4\pi J_\nu \alpha_{\rm abs,\nu} d\nu}.
\end{equation}

\bluetwo{While this division is straightforward for continuum processes, SNe and KNe tend to become dominated by lines for the opacity. Determining what happens following a line absorption is difficult and computationally demanding - and it turns out that the importance of fluorescence clashes with the omission or approximative description of this process in many LTE codes - more on this in Sect. \ref{sec:cooling}. 
}




\subsection{Cooling}
\label{sec:cooling}

\blue{When working with $e$ (energy equation \ref{eq:firstlaw1}), the term ``cooling'' refers to processes that decrease the internal energy, whether it is kinetic energy of the material particles or excitation/ionization energy. When working with $e_k$ (energy equation \ref{eq:firstlaw2}), it refers to processes in the first category only.}

\blue{Cooling will occur by various radiative processes, of which \emph{line cooling} 
is typically the most important in SNe and KNe. Other cooling processes include recombination and free-free emission, which however typically contribute by of order a few percent only.}

\subsubsection{NLTE}
\blue{The line cooling by electrons (used for Eq.~\eqref{eq:firstlaw2}) is given by:}
\begin{equation}
c = \rho^{-1} h\nu_0 n_e \left(q_{up}(T) n_l - q_{\rm down}(T)n_u\right).
\label{eq:cooling}
\end{equation}
The first thing to note is that if one inserts the exact LTE values for $n_l$ and $n_u$ ($n_u/n_l = g_u/g_l \exp(-E_{ul}/kT)$)), $c$ becomes identical zero, because $q_{up} = q_{\rm down} g_u/g_l \exp{\left(-E_{ul}/kT\right)}$. But the upper level still emits photons at a rate of $A$ (or $A \times \beta_s$, more below), which must represent (at least partially) cooling. In fact, it is this radiative leakage that gives the upper level a population slightly below the LTE limit that makes all the difference. That deviation from LTE is proportional to $1/n_e$ in magnitude if collisions dominate. Thus, as $n_e$ becomes large, Eq.~\eqref{eq:cooling} shows that the cooling is the small difference between between two large terms. Clearly care both in formulation and numerics is needed here.

Calculating cooling by the direct summing of level populations times $A$ or $A\times \beta_s$ (instead of using Eq.~\eqref{eq:cooling}) can be done only in thermal excitation-only NLTE making a consistent choice in the rate equations, as in e.g. \citet{Hotokezaka2021}. However, as soon as other processes are allowed to affect the NLTE solutions (e.g. recombination), one instead has to work with the electron net collision rates.

\blue{In general, how much a given atom or ion can cool (by being a target of collisional excitation by the thermal electrons) depends on its NLTE state, which in turn depends on both the number density of electrons, the number density of the parent ion (affects recombination inflows), and the radiation field (affects photoexcitation and photoionization rates). However, in the low-density limit most atoms/ions are in the ground state, and one can assume that each collisional excitation leads to radiative decay (cooling). There then exists a unique cooling function $\Lambda$ that depends on $T$ only, so that cooling by ion $i$ is 
}
\begin{equation}
c_i = \rho^{-1} \Lambda_i(T) n_i n_e \,.
\end{equation}

\begin{figure}[ht]
\centering
\includegraphics[width=0.49\linewidth]{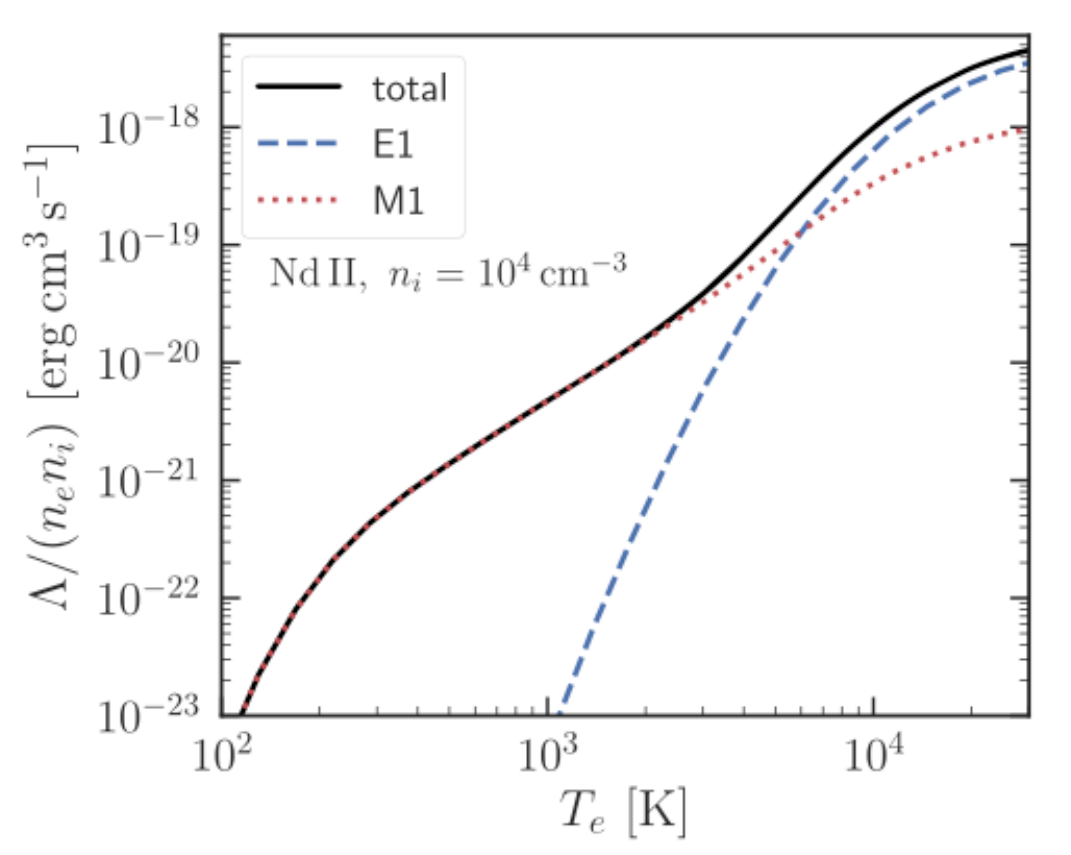}
\includegraphics[width=0.49\linewidth]{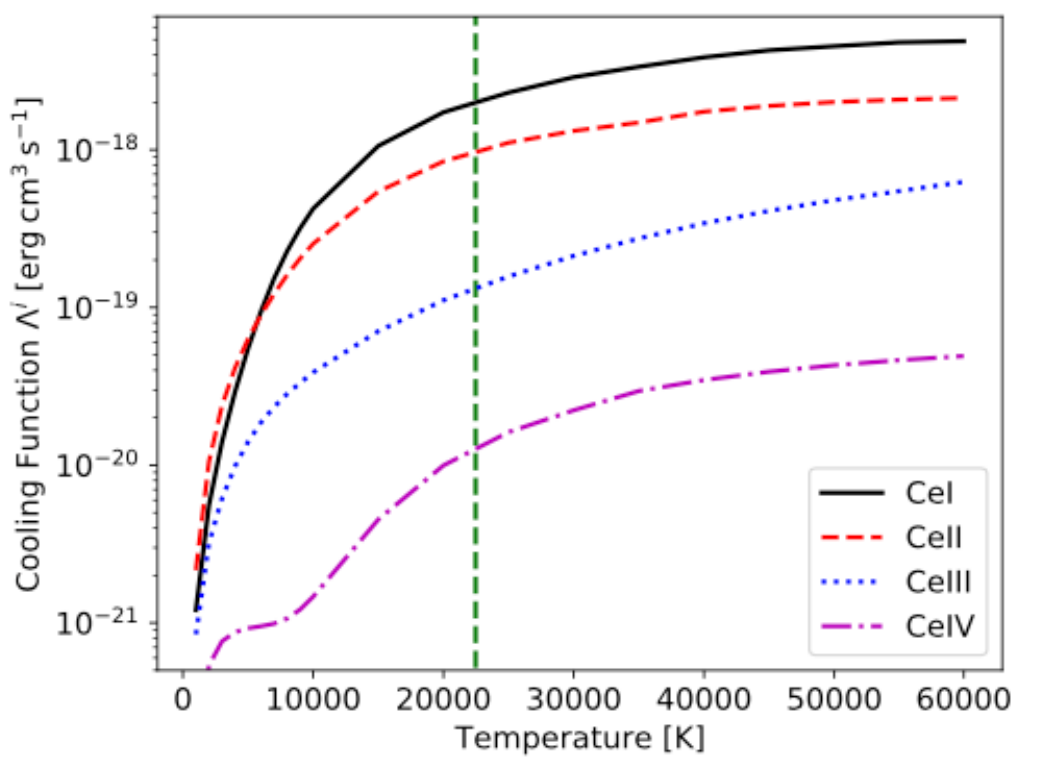}
\caption{\bluetwo{\emph{Left:} Cooling function of Nd II in the low-density limit, from \citet{Hotokezaka2021}. \emph{Right:} Cooling functions of different ions of Ce, in the low density limit, from \citet{Pognan2022a}.}}
\label{fig:Hoto2021}
\end{figure}

\blue{Figure~\ref{fig:Hoto2021} shows the low-density cooling functions of Nd II, as well as of different ions of Ce. The rapid growth with temperature is a characteristic feature of cooling functions due to the exponential terms that enter expressions for the collisional excitation rates. It is what keeps the temperature range obtained in astrophysical nebulae quite limited, with H II regions, planetary nebulae, SN and KN ejecta all being in the range $\sim$ 3000--30,000\,K for the most part. The Nd II curve illustrates that both forbidden (red, dotted) and allowed (blue, dashed) transitions can be important to include in the cooling modelling. The Ce curves illustrate that cooling can be sensitive to the ionization state of the gas, so there is an intrinsic strong coupling betweeen temperature and ionization.}

\subsubsection{LTE} 
In LTE, a subtlety affects the formulation of the line cooling. One plausible approach would to take the line cooling as the Boltzmann level populations times $A\beta_s$ (and $h\nu_0)$, which is the effective decay rate. Another plausible approach is to set the source function equal to the Planck function. One may show that this corresponds to taking the Boltzmann level populations times $A$. The source if this discrepancy is that the LTE approach at its very basic level lacks self-consistency in its formation \citep[see e.g. discussion in Chapt.~2 of][]{Mihalas1978}. Here, self-trappings change the radiation field from a Planck function, and the level populations from pure LTE, but that cannot be accommodated for in the framework. It is also significantly easier/faster to use $S_\nu = B_\nu$, and one may argue that engaging in other more expensive treatments has little meaning as yet further self-inconsistencies are added. However, the Kirchoff--Planck treatment may overestimate the cooling (and thus underestimate temperatures) when optically thick lines are important.

\blue{With the $S_\nu = B_\nu$ approach, cooling in LTE is computed as} 
\begin{equation}
\bluetwo{
    c^{\rm LTE} = \frac{4\pi}{\rho} \int_0^{\infty} \alpha_{abs,\nu} B(\nu,T) d\nu},
\end{equation}
\blue{where}
\begin{equation}
\bluetwo{\alpha_{\rm abs,\nu} =\alpha^{abs}_{bb,\nu}  + \alpha_{bf,\nu} + \alpha_{ff,\nu}}.
\end{equation}

\blue{The second and third terms on the RHS are due to continuum heating (bound-free and free-free, respectively), which have well-known expressions. The first term on the RHS represents absorption in lines, meaning photoexcitation followed by collisional de-excitation.
The ``true'' absorption (``thermalization'') probability $p^{abs,true}_{bb}$ is} 
\begin{equation}
\bluetwo{
    p^{\rm abs,true}_{bb} =  \frac{\sum_{\rm down} q_{\rm down}n_e}{\sum_{\rm down} \left[ q_{\rm down}n_e + A\beta_s\right]} \,,    \label{eq:ptherm_direct}
    }
\end{equation}
\blue{which is typically small, in the range $10^{-6}-10^{-4}$ as long as allowed de-excitation transitions ($A\gtrsim 10^6\ \mbox{s}^{-1}$) exist  \citep{Kasen2006}.}

\blue{However, it is expensive to compute $p_{bb}^{abs,true}$ for all lines. Many LTE codes therefore use a parameterized constant value $p_{bb}^{abs}=\epsilon_0$, letting the photon thermalize with probability $\epsilon_0$ and resonance scatter with probability $1-\epsilon_0$. An additional advantage of this is that lines can be binned and added up to an ``expansion opacity'', as what follows after a line absorption becomes the same for all of them, and therefore the order between the lines in a bin does not matter}. Such an expansion opacity is computed as, for wavelength bin $i$,
\begin{equation}
\alpha_i = \frac{1}{ct} \frac{1}{\Delta \lambda_i} \sum_{j} \lambda_j \epsilon_0  \left(1-\exp{\left(-\tau_{s,j}\right)}\right).
\end{equation}

\blue{Surprisingly, the value of $\epsilon_0$ must be high ($\approx 1 $) for realistic light curves and spectra to be produced \citep{Blinnikov1998,Kasen2006,Kozyreva2020}. This is interpreted to be because the lack of fluorescence in this treatment gives an artificial lack of wavelength transformations for the photons. A high $\epsilon_0$ value is a way to heuristically mimic the fluorescence process, but this comes at the cost of introducing errors in the energy equation, as more radiative energy is thermalized than it should be.}

\blue{Improvement on the situation can be done by treating at least some of the lines in detail, including the fluorescence. This is done by \texttt{SEDONA} \citep{Kasen2006}, which splits
\begin{equation}
\alpha_{bb,\nu}^{\rm abs} = \alpha_{bb,\nu}^{\rm direct} + \alpha_{bb,\nu}^{\epsilon_0} = \alpha_{bb,\nu}\times \left(f_{\rm direct}p_{bb}^{\rm abs,true} + (1-f_{\rm direct})\epsilon_0\right).
\end{equation}
Using a thermalization probability $p_{bb}^{\rm abs,true} \ll 1$ for some lines and $\sim 1$ ($\epsilon_0$) for others means that the $\epsilon_0$-treated ones will be most influential for determining the temperature. In LTE with thermal equilibrium, and assuming that lines are dominant for the opacity, we have a condition}
\begin{equation}
\bluetwo{
h_{\rm radioactivity} + \int_0^\infty \frac{4 \pi}{\rho} J_\nu \alpha_{bb,\nu}^{\rm abs} d\nu = \int_0^\infty \frac{4\pi}{\rho} B_\nu(T) \alpha_{bb,\nu}^{\rm abs} d\nu.
}
\label{eq:scaleexample}
\end{equation}
\blue{If heating is radiatively dominated (with the microphysically correct thermalization probability), we see that the scale of $\alpha_{bb}^{abs}$ does not actually matter --- we can multiply this function with any number we want and the LHS and RHS of Eq.~\eqref{eq:scaleexample} would change by the same factor in the limit that $h_{\rm radioactivity}$ is unimportant. LTE codes with expansion opacities therefore obtain temperatures quite robust to this uncertainty in the photospheric phase, especially in the layers outside of the radioactive material, which are also the most important ones for the spectral formation.}

\blue{When radioactivity dominates the heating,  however, we have (for a roughly gray $\alpha_{bb}^{abs}$):}
\begin{equation}
\bluetwo{
aT^4  \approx \frac{h_{\rm radioactivity}}{\alpha_{bb}^{abs}}.
}
\end{equation}
\blue{If $\alpha_{bb}^{abs}$ is too large by a factor $10^4$ (using $\epsilon_0 \sim\ 1$ instead of the microphysically accurate $p_{bb}^{abs,true} \lesssim 10^{-4}$), the temperature estimate will therefore be a factor 10 too low. This is likely one driving factor (combined with deviation from LTE for level populations) for why LTE codes give significantly lower SN and KN temperatures than NLTE ones from a quite early phases \citep[e.g.][]{Pognan2023,Blondin2023}. One can also note that KNe are quite different with respect to SNe in that here all the ejecta layers are radioactive. This raises also the possibility that LTE KN temperatures in the photospheric phases may be less accurate than in SN modelling.}



\subsection{Temperature from radiation field}
\blue{A gray model for a (source-free) plane-parallel stellar atmosphere gives a temperature stratification going from $T_{\rm eff}$ at the photosphere to $0.81 T_{\rm eff}$ at the surface \citep{Mihalas1970}. From this, the choice of temperature in an isothermal model is often $T_e=0.9 T_{\rm eff}$, an average value for the layer. Non-gray models tend to give somewhat lower atmospheric temperatures so the coefficient can also be somewhat lower \citep [e.g. 0.7 in][]{Lucy1970}.}

\blue{In diffusion-phase SNe and KNe a photosphere can analogously be defined from where the optical depth to the surface is 2/3, although certain issues become more severe here compared to the stellar atmospheres case \citep[stronger frequency-dependency of the photospheric depth, and a bigger differences between thermalization depth and photosphere, see][for a good discussion]{Sim2017}. For the analogous Schuster--Schwarzschild modelling (inner boundary with overlying source-free atmosphere) \citet{Mazzali1993} proposed that for SNe a more suitable approximation is $T_e = 0.9 T_R$, where}
\begin{equation}
    T_R \equiv \frac{\langle h \nu \rangle}{3.82k}.
    \label{eq:T_R}
\end{equation}
\blue{Here $\langle h \nu \rangle$ is the mean energy of photons in the radiation field, obtained from either a radiative transfer solution or a Monte Carlo simulation. $T_R$ can be said to be the temperature of a blackbody with the same ``color'' as the radiation field.}

\blue{Another variant was proposed by \citet{Kromer2009}, who in their ``simple temperature/ionization'' methodology use $T_e = T_J$, with}
\begin{equation}
    T_J \equiv \left(\frac{\pi J}{\sigma}\right)^{1/4},
    \label{eq:T_simple}
\end{equation}
\blue{which is the temperature of a blackbody field with the same intensity as the radiation field. Here $J$ is the bolometric radiation field strength and $\sigma$ is the Stefan-Boltzmann constant. As the radiation field gets more dilute with time, Eq.~\eqref{eq:T_simple} will inevitably start to underestimate the temperature even if gas-photon coupling is still strong.}  

\begin{figure}[ht]
    \centering 
    \includegraphics[width=0.49\linewidth]{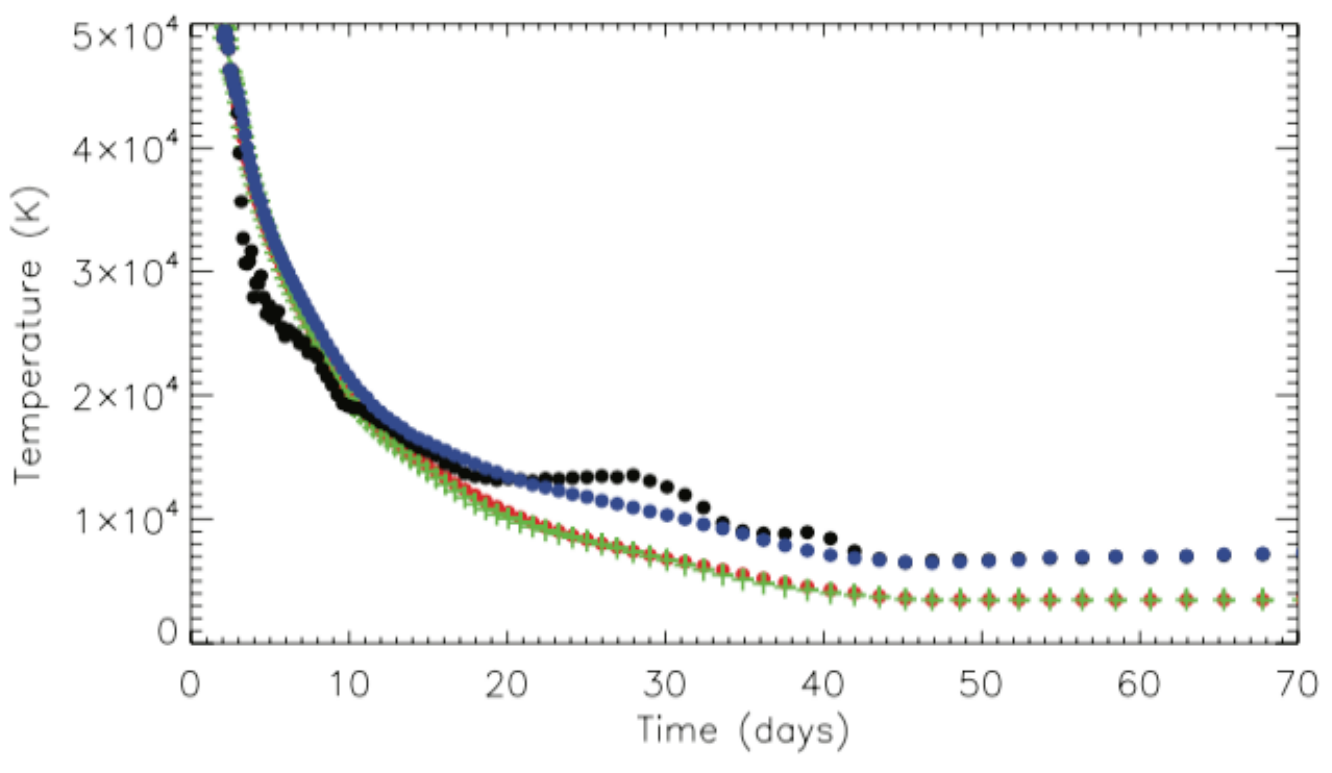}
    \includegraphics[width=0.49\linewidth]{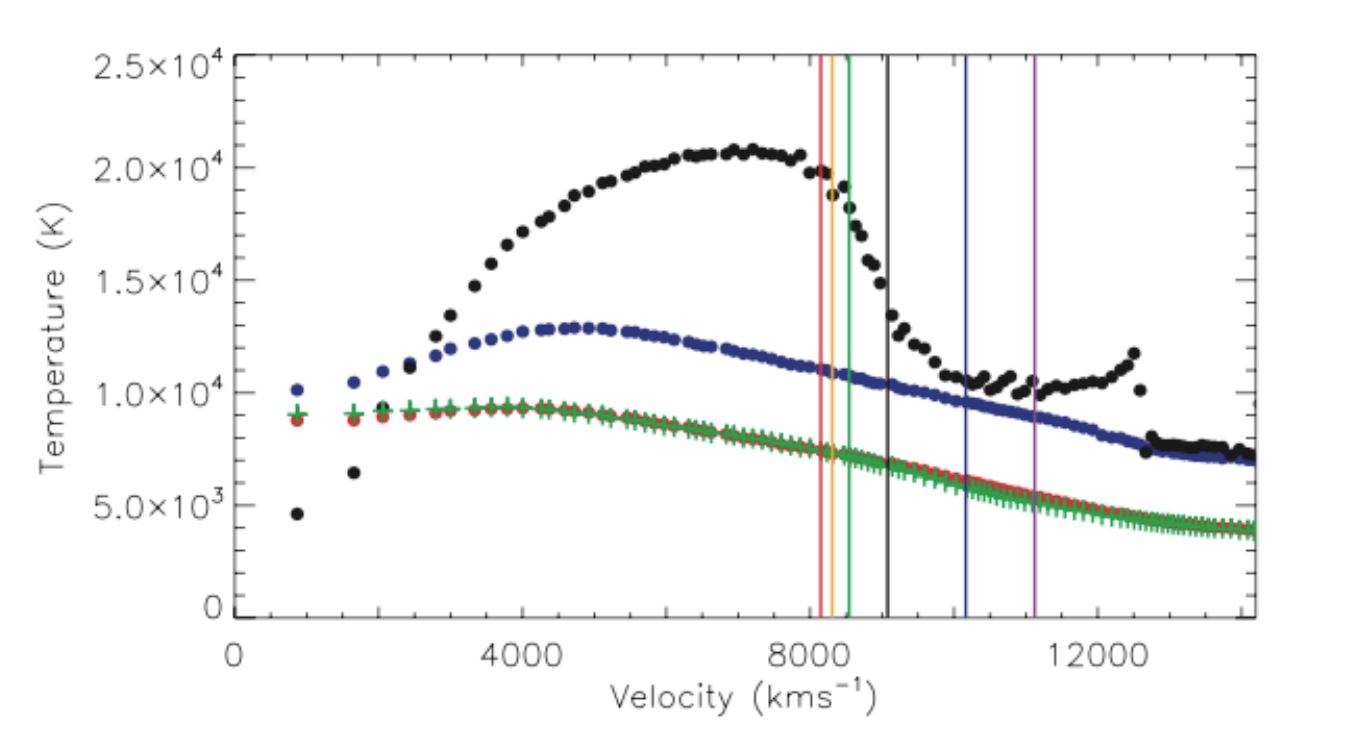}
    \caption{\emph{Left:} \bluetwo{Comparison between temperature time-evolution at velocity coordinate 9600 km s$^{-1}$ in Type Ia LTE SN model using different treatments. Electron temperatures $T_e$ using Eq.~\eqref{eq:T_simple} are plotted as green pluses, while those from thermal equilibrium are plotted as black points. The blue and red points are $T_R$ and $T_J$ in the thermal equilibrium model. \emph{Right:} Same as the left figure, but showing profiles (temperature versus velocity) at 31 days. The black/violet/blue/green/red/orange vertical lines show the mean radii of last scattering in the $U,B,V,R,I$ bands. From \citet{Kromer2009}.}}
    \label{fig:KS09}
\end{figure}

\blue{\citet{Kromer2009} compare results of this method with a more accurate (named ``detailed temperature/ionization'') method in which $T_e$ is instead solved from balancing heating with cooling as described above (Fig. \ref{fig:KS09}). 
For the radiative heating estimate, $4\pi \int \chi_\nu J_\nu d\nu$, in this detailed approach, the authors implement the method of \citet{Mazzali1993} for a 2-parameter model of the radiation field (there used to estimate ionization rates):
\begin{equation}
    J_\nu = W B_\nu(T_R) \,,
\end{equation}
where 
$W \equiv J/B(T_R)$ is called the dilution factor. 
Thus, here the requirement of Eq.~\eqref{eq:T_simple} that the radiation field should be as strong as a blackbody is relaxed, while the requirement of a similar SED shape is retained. The strong coupling assumption is also relaxed as with the thermal equilibrum the radiation field ``temperature'' can be different than the gas temperature.} 
%

\blue{Due to their simplicity and particular ease of implementation in Monte Carlo codes, Eqs.~\ref{eq:T_R} and \ref{eq:T_simple} have been quite widely used in the modelling literature, not least for extensive parameter space modelling of photospheric SN spectra using the codes of \citet{Mazzali1993,Mazzali2000,Kerzendorf2014}. Moreover, they have also carried over to be used for many currently available 3D, full-ejecta models of kilonovae \citep[e.g.][]{Tanaka2013,Bulla2023}.}

\blue{Figure \ref{fig:KS09} shows the \citet{Kromer2009} comparison of temperatures obtained through  \ref{eq:T_simple},  and from thermal equilibrium, for a Type Ia model. The agreement is quite good for 0--15\,d, whereas after this Eq.~\eqref{eq:T_simple} gives an underestimate by a factor $\sim$2, due to its neglect of the dilution as discussed above. We can also see that using $T_R$ instead of $T_J$ to estimate $T_e$ tends to give a better approximation at all phases. While this gives some information about what errors the simplifications of Eqs.~\eqref{eq:T_R} and \ref{eq:T_simple} incur, for one particular SN type (Type Ia), it is currently not known what the level of accuracy is for CCSNe or KNe.} 

\subsection{Numeric aspects and some comparisons}
%
%
\bluetwo{Energy conservation is defined by a single equation (Eq.~\eqref{eq:firstlaw1} or equivalently Eq.~\eqref{eq:firstlaw2}) but this contains many terms, most of which are highly non-linear in temperature. In NLTE, the constituent terms are not explicit functions, but evaluated at different temperatures by solving the associated rate equations (described in the next chapter). Newton--Raphson is a standard algorithm choice, but for single non-linear equations there are also other options \citep[see e.g. Ch. 9 in][]{Press1992}. Newton--Raphson is in particular to prefer if there is a good starting guess, and when it comes to temperature it is relatively straight-forward to roughly pre-estimate its value. The implicit nature of the problem in NLTE means the derivative is computed numerically (i.e., the heating and cooling terms typically do not have explicit, manageable analytic expressions, and therefore cannot yield analytic temperature derivatives).} 

\bluetwo{Stability is enhanced by both damping and capping steps. In the thermal equilibrium iterations described by \citet{Lucy1999a,Lucy2003}, a step damping factor of 0.5--0.8 and a step cap of $|\Delta T|<200$ K per iteration are reported to be beneficial for robust convergence. In addition, an adaptive scheme is stated to be needed that allows a further reduction of both these in (rare) cases where corrections do not evolve healthily. 
\citet{Lucy2003} uses a temperature convergence criterium of $\Delta T <$1 K, and also checks that $(h-c)/c$ is small.}

\bluetwo{It is beneficial for convergence to solve the (single) energy equilibrium equation together with the level population equations \citep[e.g.][]{Lucy2003}, or at least some of them if the total number is too high. The thermal equation increases the system size by only one, while allowing for sometimes significantly fewer iterations. The coupling becomes in particular powerful if the main cooling species are identified and co-solved for.}

\begin{figure}[ht]
\centering 
    \includegraphics[width=0.49\linewidth]{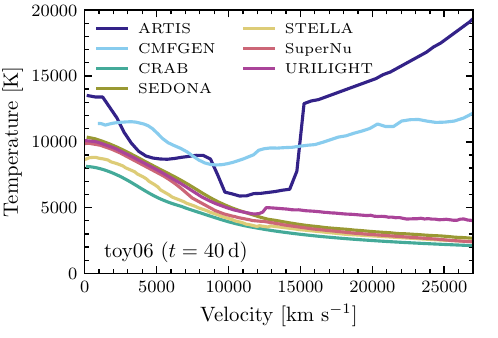}
     \includegraphics[width=0.49\linewidth]{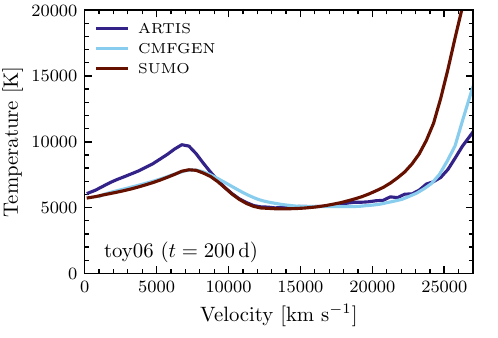}
     \caption{\bluetwo{Temperature profiles obtained by various codes for a Type Ia test model at 40\,d (left) and 200\,d (right). Image reproduced with permission from \citet{Blondin2022}, copyright by the author(s).}}
     \label{fig:blondin_T}
\end{figure}

\bluetwo{We close this section by  showing a comparison of temperature profiles computed with a variety of codes for a Type Ia SN test model \citep[Fig.~\ref{fig:blondin_T}, from][]{Blondin2023}. As can be seen, temperature predictions for even a simple test model (this one contains only Fe, Co, Ni) vary quite significantly between codes. All models shown here compute temperature from the first law of thermodynamics, but clearly estimates of heating and cooling terms vary quite significantly. The comparison illustrates the challenges for temperature modelling even for state-of-the-art radiative transfer tools, and much work remains to be done to arrive at robust, accurate determination of this fundamental quantity in SNe and KNe.}

\newpage
\section{Rate equations}
\label{sec:rateequations}
We now move on to study numeric techniques for solving rate equations, needed for NLTE modelling. LTE codes make use of the Boltzmann equations for excitation which for given values of $T$ and $\rho$ (and composition) provide any level population by a formula depending only on the energy and statistical weight of the level and the partition function, well known quantities. LTE ionization is computed from the the Saha--Boltzmann equations, where also the electron density enters. This makes the equations non-linear, but their solution is a straightforward procedure which will not be discussed further here. 

\begin{figure}[htbp]
\centering
\includegraphics[width=0.9\linewidth]{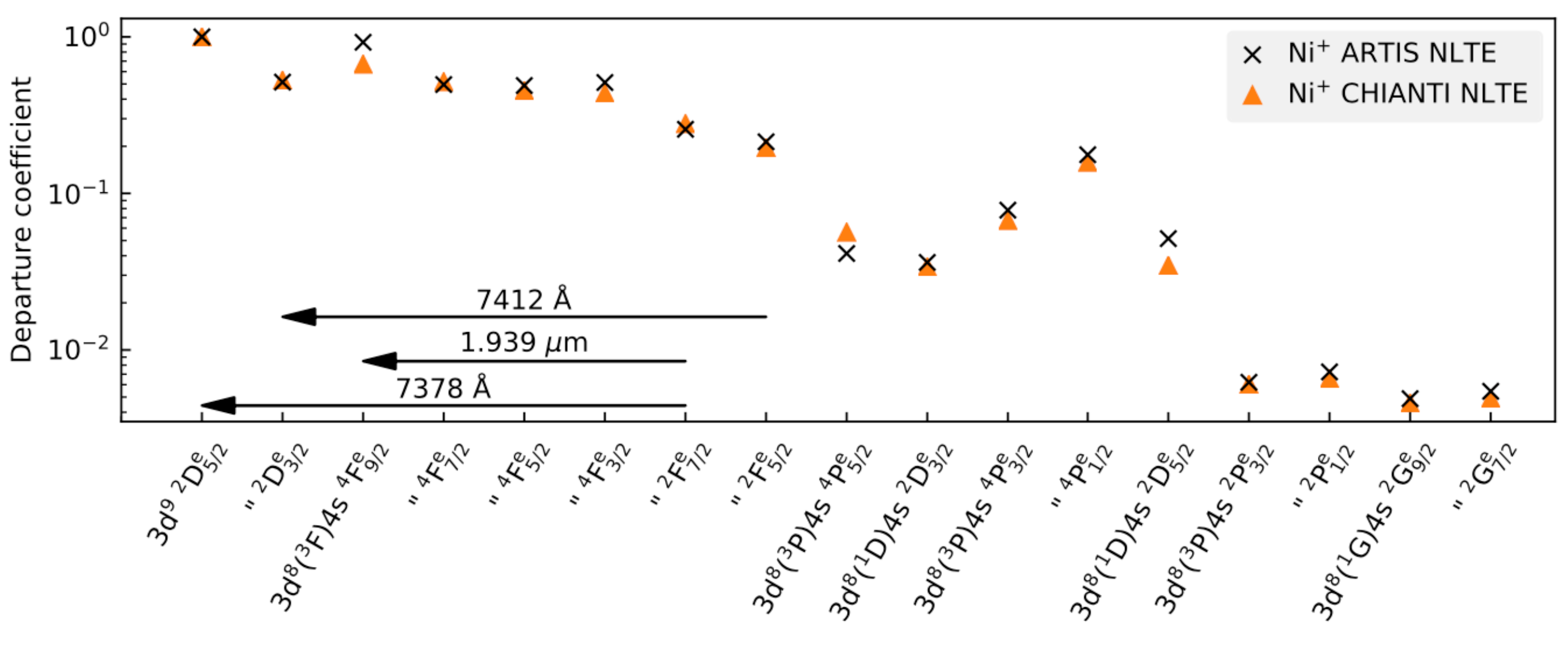}
\includegraphics[width=0.9\linewidth]{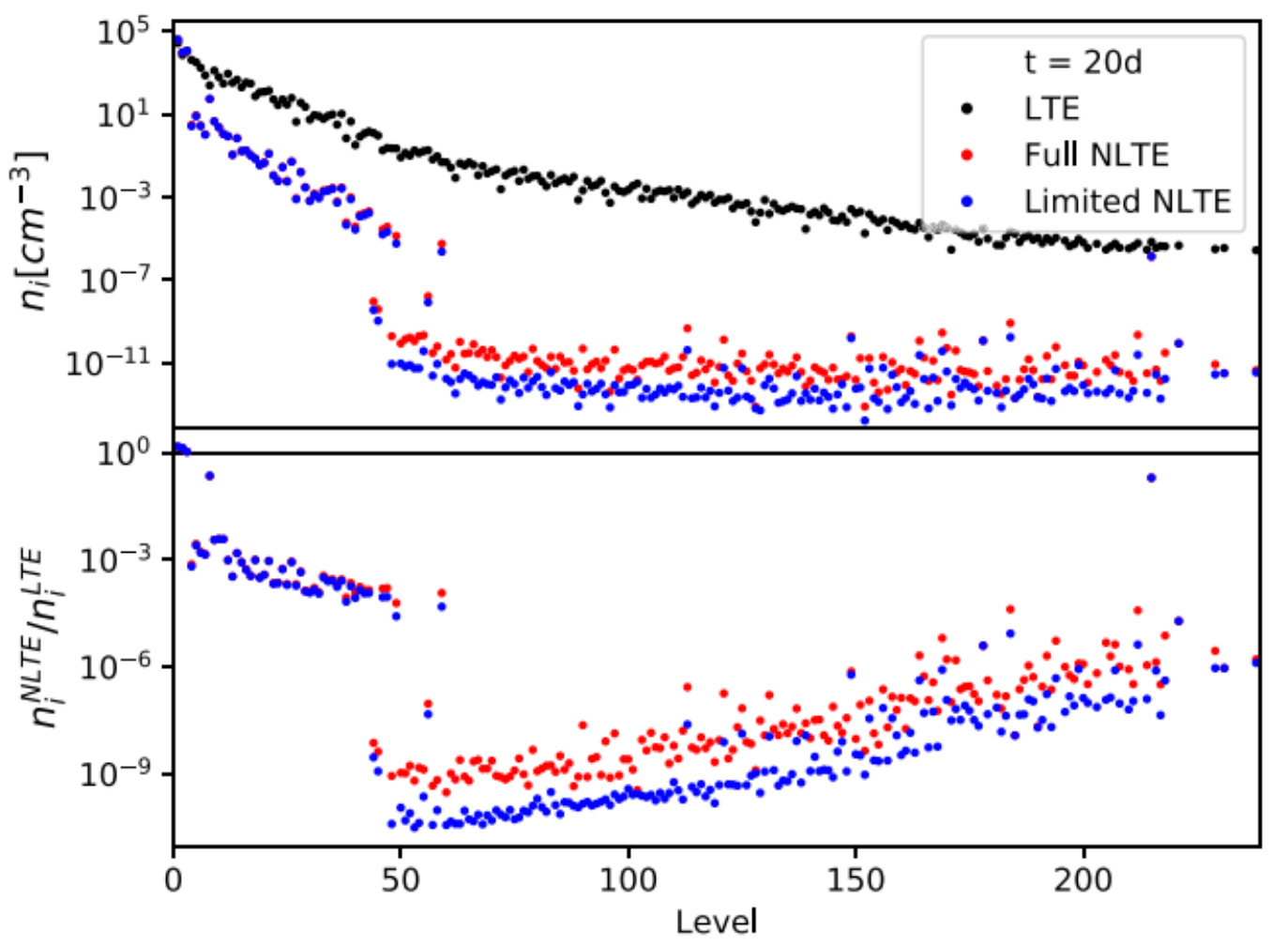}
\caption{\emph{Top:} Departure coefficients for the first 17 levels of Ni II in a Type Ia model at 330\,d (the two types of symbols referes to using two different sources of atomic data).  \blue{\emph{Bottom:} LTE level population solutions for Ce\,II, compared to NLTE values, in a KN model at 20d. ``Limited NLTE'' means NLTE excluding photoexcitation and photodeexcitation rates. The upper panel shows the number densities, and the lower panel the departure coefficients. Image reproduced with permisision from [top] \citet{Shingles2020} and [bottom] \citet{Pognan2022b}, copyright by the author(s).}}
\label{fig:pognan2022_fig2}
\end{figure}

\blue{After some time populations start to deviate significantly from LTE. Figure \ref{fig:pognan2022_fig2} shows computed departure coefficients (NLTE level populations relative to LTE ones) for Ni\,II in a Type Ia model at 330\,d (top), and for Ce\,II in a KN model at 20\,d (bottom). We can see how NLTE effects are strong in both cases, and increase with excitation energy. A few take-away points are:}

\begin{enumerate}

\item \blue{The NLTE populations are in general very different from their LTE ones. This tells us that LTE modelling beyond some point in time is not expected to give even approximately correct results; NLTE gives a completely different state of the material.}

\item Departure coefficients initially decline with increasing level number, driven by the fact that there are more and more possible radiative de-excitation paths opening. However, a trend reversal can arise due to recombination.

\item \blue{The Ce plot (bottom panel) shows that with enough levels included, the levels fall into two distinct groups; those with relatively high level populations and departure coefficients of order $10^{-3}$ (level numbers $\lesssim 50$ in this example) and those with much lower populations and departure coefficients $10^{-10}$--$10^{-6}$ (level numbers $\gtrsim 50$ in this example). This split arises as levels over $\sim$50 in Ce\,III start to have allowed-transition de-excitation paths, whereas the lower lying ones have only forbidden-transition ones. This situation is not to be confused with that the lower group is necessarily more important for emission; it is $n\times A$ that governs emission, and the upper group is depopulated  because of their large $A$ values.}

\end{enumerate}

The ionization structure will in general also deviate strongly from LTE, in addition to the excitation structure devitations shown here. This is because ionization occurs mainly by non-thermal electrons, and recombination by radiative and di-electronic processes, both different to the thermal processes assumed to dominate in LTE.

\subsection{Formulation}
We will focus on NLTE rate equation systems for the internal states of atoms and ions in \emph{steady state}, which means that balance holds between inflow and outflow. For non-steady state effects, the reader is referred to discussions in \citet[][late evolution of SN 1987A]{Fransson1993}, \citet [][photospheric phase of Type II SNe]{Utrobin2005,Dessart2008}, \citet[][nebular phase of Type Ia SNe]{Fransson2015}, and \citet[][nebular phase of kilonovae]{Pognan2022a}. Discussion and examples are taken for atomic modelling; works on solving rate equations for molecules can be found in \citet{Liu1992,Liu1995,Liljegren2020,Liljegren2023,Rho2021}.\\
\\
\blue{There are many rate equations for each grid cell, one for each included level. As there may be dozens of elements, maybe half-a-dozen ionization stages, and hundreds or even thousands or levels per ion, the count quickly gets large, often to $N \gtrsim 10^5$ per cell. Normally one splits the full equation system up into blocks, for example may each ion (e.g. Fe II) define a separate block which then has $N=10^2-10^3$ equations. These blocks are then solved sequentially and iteratively. This split and iteration is done because matrix storage scales with $N^2$ and inversion time with $N^3$. Thus, 100 equations would cost $\propto 10^4$ in storage if coupled, but $\propto 10 \times 10^2 = 10^3$ if split into blocks of ten, similarly the number of operations would be $\propto 10^6$ if coupled but $\propto N_{iter} 10^3$ if split into ten blocks; clearly if $N_{iter}$ (number of iterations needed) is not too large a saving is made}. When solving for each ion, all other quantities (populations of levels in other elements and ions, temperature, electron density) are held fixed. 
\\
\\
\noindent \blue{Writing down a balance between inflow (LHS) and outflow (RHS) for level $i$ in a $N$-level block gives an equation of form}
\begin{equation}
\sum_{j=1, \neq i}^N n_j R_{j,i} + s_{i,\rm ext} = n_i \times \left[ \sum_{j=1, \neq i}^N R_{i,j} + R_{i,\rm ext}\right],
\label{eq:inflowoutflow}
\end{equation}
\blue{Here $R_{i,j}$ are reaction rates (unit s$^{-1}$) between levels $i$ and $j$, $s_{i,\rm ext}$ (cm$^{-3}$s$^{-1}$) is a block-external inflow rate (a source term) and $R_{i,\rm ext}$ is a reaction rate (s$^{-1}$) to block-external states (a sink term). For single-ion blocks, $s_{i,\rm ext}$ could for example contain constributions by recombination inflows, and $R_{i,\rm ext}$ could contain contributions by photoionization outflows.}

\blue{The components of the different terms arise from a variety of collisional and radiative processes, see e.g. Chapt.~4 in \citet{Jerkstrand2011PhD}, for an overview. In the optically thin limit, no $R_{i,j}$ terms depend on any of the level populations in the block so the system is linear, 
$\mathbf{R}\mathbf{n} = \mathbf{s_{\rm ext}}$, where the $\mathbf{R}$ matrix contains the $R$ terms and the $\mathbf{s_{\rm ext}}$ array the $s_{i,\rm ext}$ terms.
If line optical depths are included and treated in the Sobolev formalism \citep{Sobolev1957,Castor1970}, the effective spontaneous radiative decay rate from level $i$ to $j$ is $A\times \beta^s_{i,j}$, where $\beta^s_{i,j}$ is the \emph{Sobolev escape probability}, which for homologous expansion is
\begin{equation}
\beta^s_{i,j} = \frac{1- \exp{\left(-\tau^s_{i,j}\right)}}{\tau^s_{i,j}} \,,
\label{eq:betas}
\end{equation}
where the \emph{Sobolev optical depth} $\tau^s_{i,j}$ is
\begin{equation}
\tau^s_{i,j} = \frac{1}{8\pi}\frac{g_i}{g_j} A_{i,j} \lambda_{i,j}^3 n_j t \left(1 - \frac{g_j}{g_i}\frac{n_i}{n_j}\right),
\end{equation}
which introduces a non-linearity as the $R$ component for the spontaneous decay ($R_{i,j}^{\rm spont}=A_{i,j}\times \beta_{i,j}^s$) now depends on the level populations $n_i$ and $n_j$ themselves}. Another source of non-linearity can arise if one considers also continuum destruction probabilities. If photons are allowed to be destroyed by photoionization on e.g. a level $k$ in the block before escaping the Sobolev resonance, another term $\beta_c(n_k)$ would enter \citep{Hummer1985,Chugai1987,Jerkstrand2011PhD}.

\subsection{Solution with Newton--Raphson method}
\blue{The non-linearity introduced by self-absorption discussed above is dealt with quite effectively by \emph{fixed-point iteration} \citep[compute $\beta$ values for given populations, hold fixed, solve linear system, repeat,][]{Lucy1991}, but one may also choose to 
solve for the next corrective step in a multi-dimensional Newton--Raphson scheme,}
\begin{equation}
\mathbf{J} \cdot \Delta \mathbf{n} = -\mathbf{e} \,,
\end{equation}
\blue{where $\mathbf{J}$ is the Jacobian of $\mathbf{R}$ ($N\times N$ array of all partial derivatives) and $\mathbf{e}$ is the length-$N$ array of current errors of the rate equations (residuals of total inflow minus total outflow for each level, Eq.~\eqref{eq:inflowoutflow}). To minimize numeric errors, and save computation time, it is beneficial to derive analytic Jacobian entries when possible (while this is typically not doable for the energy equation (Sect.~2), it is manageble for each rate equation, but on the other hand there is a very large number of them). 
If doing numeric ones, an optimal derivative step size can be derived \citep{Heath2005}; the rule of thumb is a step size $dx/x \sim \sqrt{\epsilon}$, where $\epsilon$ is the relative accuracy of the function evalution. If this is similar to the machine precision, then in single precision $dx/x \sim  10^{-4}-10^{-3}$}.

In many algorithms the rate equations are coupled to other non-linear equations (e.g. continuum radiative transfer equations), necessitating Newton-Raphson also for the case of optically thin lines. 

\subsubsection{Convergence criteria and issues}
\blue{A universal convergence criterium is that the relative corrections should have become smaller than some fraction $\alpha_n$; $|\Delta n_{i}/n_i| < \alpha_n$. The \texttt{SUMO} code for example uses $\alpha_n=0.01$ (the largest relative population change between iterations must have fallen below 1\%). In Monte Carlo simulations, statistical noise can sometimes make high-lying rarely interacting states never reach these kind of convergence levels. A simple handling of this is to limit the $\alpha_n$ condition to a group of low-lying states exclusively. For example does \citet{Lucy2003} apply the condition to the 5 lowest-lying states only, albeit with a more stringent value $\alpha_n=10^{-5}$.} 

\blue{One may also pose a condition on the maximum allowed error,  $max{\left(\bm{e}\right)}$,  relative to the inflow and outflow rates (LHS and RHS of Eq.~\eqref{eq:inflowoutflow})}. 
It sometimes happens that the errors have reached a small regime, but the NR stepping is jumping between two solutions bracing the true one. One may then consider a convergence criterium based on this second condition alone, possibly flagging the solution as having a larger error than desired. One can also attempt to change the NR step size to break the jumping loop.

\blue{The matrix equation $\mathbf{R} \cdot \mathbf{n} = \mathbf{s}$ (or $\mathbf{J} \cdot \mathbf{\Delta n} = -\mathbf{e}$)  has either a single solution, no solution, or an infinite number of solutions. The latter two cases occur if the matrix is singular, i.e. if at least one equation corresponds to another equation scaled with a (non-zero) constant.
If that constant is non-unity, there is no solution. If it is unity, there is an infinity of solutions.} 

\blue{True matrix singularity, for real physical processes, is not expected to occur unless some term is forgotten or something is formulated
incorrectly. 
However, limitations to the numeric precision of the computer can lead to two types of problems \citep[e.g.][]{Press1992}:
    1) ``False" matrix singularity due to machine precision limitation.
    2) Accumulated roundoff errors, giving a solution with unacceptably large errors.
Both types of problems are more likely to occur the closer to singular the true matrix is. The second problem also gets more severe for larger equation systems, and large dynamic range (difference between largest and smallest values of solution).}

\blue{In addition to this, even if the system is neither singular nor close to singular, issues with how blocks of the full equation system are split out and iterated with each other can lead to a (single) non-physical solution. The most obvious case of this is when the level population vector $\mathbf{n}$ obtains one or more negative entries. }
\blue{Obtaining negative solutions is fundamentally a symptom of the same underlying problem as obtaining no solution --- the equation system describes constraints that have no physical solution. 
Such a situation can arise if some quantities, external to the block, are unable to make the necessary adaptations to keep up with other changes in the system. It is therefore clearly important \emph{how} one decouples the full problem into an iteration between blocks. Another way negative solutions can arise is when the linearization overshoots the solution, e.g. if the starting condition is too poor, or the derivative not accurate enough.}

\blue{This situation is one motivation for co-solving the level populations of all the ions of a given element, instead of one by one. Such an approach is implemented by both \citet{Shingles2020} and \citet{vanBaal2023} --- here negative level populations are rare, whereas in \texttt{SUMO}, which solves ion by ion, they can happen quite often and corrective action become necessary.
There is, however, always \emph{something} that is held fixed whichever equation block one is solving. ``No solution'', based on real issues with the equation system rather than numeric accuracy-related ones, can therefore always arise and needs to be dealt with in the codes. Examples of strategies are (1) stay with the previous block solution (and hope that the issues resolved in subsequent iterations), (2) reset to some specified solution (``reset initial guess''), and (3) reduce the step size for some quantities. One may also (4) attempt to backtrack on updates to other blocks. For example --- instead of using the new photoionization and photoexcitation rates as estimated in the previous radiation field solution, one may try a step between the previous values, that gave a valid solution, and the new ones.}

\blue{Because classes of algorithms come in a large number of flavours --- and small changes can make a big difference, depending on application, there is basically an infinity of choices. Not only that --- the performance of any given algorithm (stability, run time) varies over a practically infinite space (position-dependent density and composition, as well as starting guesses for all the parameters). Therefore, what is perhaps more meaningful than trying to compare and rank algorithms, is to devise a \emph{layered approach} in the algorithm design, where a variety of fall-back strategies are resorted to in case the currently attempted one fails. At low level, this could for example be to retry with a different step size. At high level, it could be to redefine the whole equation block currently being attempted to solve for.}

\subsubsection{Coupling to the radiation field} 
\blue{The rate equations involve radiation field quantities; photoexcitation rates which depend on the line-averaged mean intensities $\bar{J}_{ij}$, $i$ and $j$ denoting the two levels involved, and photoionization rates which depend on $J_\nu$ convolved with the cross sections $\sigma_\nu$ for the given bound-free transition. In regular Lambda iteration one completely decouples solutions to level populations and to radiative transfer, i.e. for each level population solution the radiation field terms are held constant. While this is a method with poor convergence properties for high-density situations and with resolved line transfer, the situation is much more benign in Sobolev modelling, especially for the low-density tail phase. This is because the Sobolev approximation treats multiple line scatterings as single ones, making the problem equation-wise simpler and numerically much easier to solve by iteration. It gives mostly robust and fast convergence for both SNe and KNe. The \texttt{SUMO} code implements a flexible way to select the degree of coupling by allowing the user to specify which (low-lying) levels are to be solved for in NLTE, and which (high-lying) are present for on-the-fly scattering and fluorescence but do not participate in collisionally induced emission. By reducing the NLTE coupling in this way can the Lambda iteration be used up to quite high densities, at least for Sobolev modelling.}

\blue{For diffusion-phase modelling of SNe and KNe, better coupling than in classic Lambda iteration is preferable and sometimes necessary.
If the radiation field is discretized into $M$ frequencies, and the $J_k$ terms ($k=1,M$) are treated as variables to be solved for jointly with the level populations, the NR equation system ($N$ rate equations plus $M$ radiative transfer equations) takes the form}
\begin{equation}
    \begin{bmatrix}
    j_{1,n_1} &...& j_{1,n_N}& j_{1,J_1}& .. & j_{1,J_M}\\
    .. \\
    j_{N,n_1} &...& j_{N,n_N} &j_{N,J_1} &..& j_{N,J_M} \\
    t_{1,n_1} &...& t_{1,n_N}& t_{1,J_1}& .. & t_{1,J_M}\\
    .. \\
    t_{N,n_1} &...& t_{N,n_N} & t_{N,J_1} &..& t_{N,J_M} \\
    \end{bmatrix}
 \cdot
 \begin{bmatrix}
  \Delta n_1 \\
  .. \\
  \Delta n_N \\
  \Delta J_1 \\
  .. \\
   \Delta J_M \\
 \end{bmatrix}
 =
 \begin{bmatrix}
 -e_{n_1} \\
 .. \\
 -e_{n_N} \\
 -e_{J_1} \\
 .. \\
 -e_{J_M} \\
 \end{bmatrix}
\end{equation}
\blue{where $j$ denotes Jacobian entries from the rate equations and $t$ correspondingly from the radiative transfer equations.
One may also add $n_e$ and $T$ as variables by the two equations of charge and energy conservation \citep[e.g.][] {Hillier1990}. The issue is that the transfer equations depend on the $\Delta n$ population changes of \emph{all} elements and levels, so for this coupling to give real convergence improvement it is preferable that the $N$ rate equations include all levels for all species. This puts severe constraints on the number of levels per species one can treat.}

\blue{In one of the pioneering works to address the matter-radiation coupling, \citet{Hillier1990} used the transfer equation to eliminate the radiation field $\Delta J$ terms in the expressions involving the level population $\Delta n$ terms. In principle does the radiation field depend on population $\Delta n$ terms at all depths. Retaining such a full dependency corresponds to ``complete linearization'' where one solves for population corrections at all depths simultaneously. It quickly becomes a prohibitive approach for any realistic SN/KN modelling problem, as one has tens of thousands of levels or more, and $\gtrsim$100 zones even in 1D, giving an equation system exceeding size of a million and having $>10^{12}$ entries. It can however be used for small, specific problems \citep[e.g.][]{Hillier1983, Hillier1989}.} 

\blue{A ``local'' operator limit corresponds to retaining dependencies of $J$ in a given cell only on the level populations in that cell. Then the level populations can be solved cell-by-cell. This is attractive from the perspective of ease of parallelization, and also for a straightforward extension to multi-D. It is the operator used by \texttt{CMFGEN}
for supernova applications.  
For non-local coupling, the simplest scheme is nearest-neighbor, which in gives a block-diagonal equation system. It is used by \texttt{CMFGEN} for stars (Hillier, priv. comm.).}

\blue{Monte Carlo codes are by construction doing the radiative transfer in a separate computational step, and cannot direcly make use of the equation-based coupling schemes discussed above. However, they can achieve good convergence by other means. One key technique is the use of \emph{indestructible energy packets}  \citep{Lucy1999a,Lucy2005}. The indestructability corresponds to enforcing radiative equilibrium which must hold at convergence, and enforcing this from the beginning improves the convergence situation. Another method to improve coupling in NLTE is the use of so called \emph{macro-atoms} \citep{Lucy2002}. An extensive review of these and other techniques in MC transfer is given by \citet{Noebauer2019}.}

\paragraph{Photoexcitation rates.}
\blue{Photoionization involves a continuum state (the ionized electron) and the rates therefore depend on integration of the radiation field over a quite broad range of frequencies. As such, it is relatively straightfoward to produce good estimators for photoionization rates with low or moderate resolution transfer equation solving, and in Monte Carlo simulations with small or moderate packet counts.}

\blue{Photoexcitation, however, depends on the radiation flux over a very small frequency range; that of the line profile whose thermal width is of order 1 km/s (so $\Delta \nu/\nu_0 \sim 1/3\times 10^5$). Here, things get trickier. It is, for example, by no means obvious that a Monte Carlo simulation will have produced at least one packet propagating through each line in each cell. By not relying on counts but instead using information from all packets passing through lines (whether they interact or not) the statistics is improved \citep{Lucy1999a} --- but nevertheless one cannot be sure packets make such pass-throughs in all lines. Here, it is suitable to make checks whether rates have been generated at all, and if not perhaps choose to use the previous (non-zero) rate estimate instead. Or, if there are too many such instances, increase the packet count in the simulations. This illustrates a general aspect of Monte Carlo modelling that it comes with need for some special considerations compared with solving the transfer equation. On the other hand, solving the transfer equation without use of the Sobolev approximation requires resolving each line which puts much higher requirements on the number of frequency points.}

\blue{A related issue is that storage of the photoexcitation rates for all transitions in each cell can become prohibititely expensive in 3D. Therefore, 3D modelling solving rate equations is so far limited to either ignoring these rates \citep{Botyanszki2017,Botyanszki2018,vanBaal2023} or estimating them from a low-resolution model of the radiation field \citep{Shingles2020}. Accurate treatment of photoexcitation rates in 3D NLTE modelling is a yet unsolved problem.}

\subsubsection{Further aspects}

\paragraph{Starting guesses.} \blue{In root finding with multi-dimensional Newton--Raphson it is important to have a good starting guess. Borrowing from Numerical Recipes: \emph{``Carefully crafted initial estimates reward you not only with reduced computational effort, but also with increased understanding and self-esteem''}. This last point is valuable to regularly remind ourselves us, as the steadily increasing power of computers and algorithms may lead to loss of understanding of the basic physical behaviour of the system unless we stay involved.}

\blue{For problems that are somewhat close to an LTE situation, LTE populations are the natural choice for the starting conditions. At low enough densities, however, LTE fails so severely that it is not useful for this purpose. Instead, expectations for the rough behaviour of the system, coming from a combination of theory and experience, provide the best starting guesses. For example, we know from previous solutions computed in the literature that nebular-phase CCSNe have ionization degrees of a few percent. We can combine this with the basic property that levels with forbidden deexcitation paths only will have much higher abundances than levels with at least one allowed deeexcitation path. Figure~\ref{fig:pognan2022_fig2} illustrates this behaviour. A rough starting guess in the low density limit for a level $i$ is then $n_i \approx \left(\sum_{j=1}^{i-1} A_{i,j}\right)^{-1} \times max\left[n_g C_{g,i}(T),\alpha_i n_e n_+\right]$, where $\alpha_i$ is the recombination rates, $C_{g,i}$ is the collisional excitation rate from ground (``g'') to level $i$, which scales with $\exp{\left(-kT/E_i\right)}$ \citep[see][for more details]{Jerkstrand2017handbook}.}  

\begin{figure}[ht]
    \centering   \includegraphics[width=0.8\linewidth]{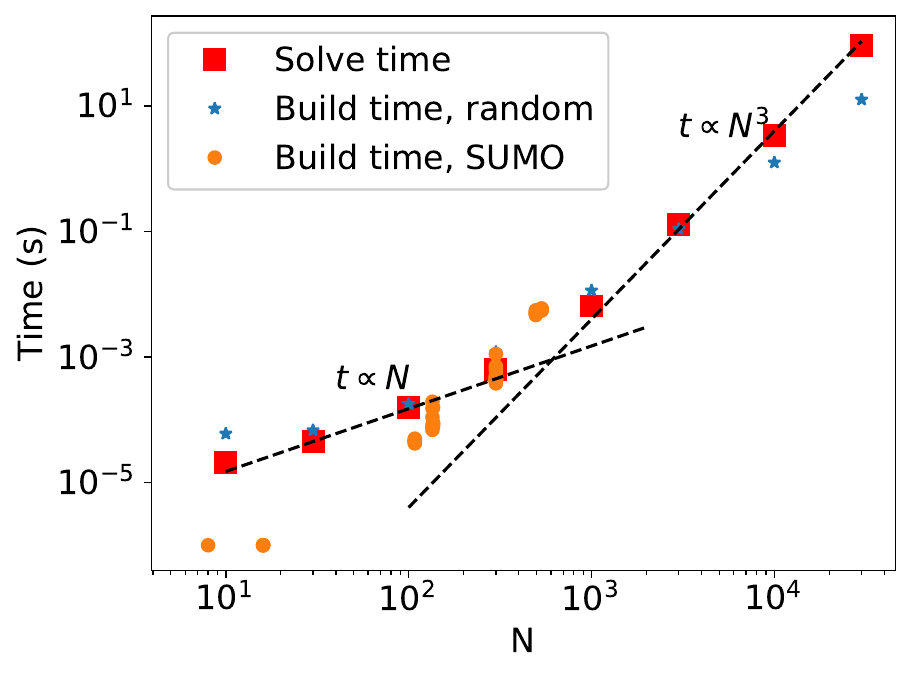}
    \caption{\blue{Solution time for \texttt{DGESV} (red squares) and build times (blue stars) with randomly generated values in $\mathbf{R}$ and $\mathbf{s}$, versus equation system size $N$. Also shown are some build times for different elements in a \texttt{SUMO} model (orange circles). Computed on a MacBook Pro Intel core i9 with \texttt{gfortran} compilation.}} 
    \label{fig:timingtests_DGESV}
\end{figure}

\paragraph{Specific software for linear equation system solving.} 
\blue{The \texttt{BLAS/LAPACK} libraries provide robust and fast algorithms for basic linear algebra operations like solving a linear equation system. \texttt{DGESV} is the standard routine when the matrix has no particular simplifying properties to its structure (e.g., tridiagonal, symmetric). It employs LU-decomposition, which is a $N^3$ operation, and therefore becomes slow for large equation systems. However, as Fig. \ref{fig:timingtests_DGESV} shows, in practice the $N^3$ scaling will not become problematic in until $N$ gets quite large, $N \gtrsim$ few hundred. For $N\lesssim$ 500, the LU decomposition is subdominant to other data operations and the scaling is close to linear. For higher $N$, the $N^3$ scaling is closely reproduced. $N\sim 500$ is close to a sweet-spot for balance between accuracy and compute time. For most elements up to iron one is also close to the ionization edge at $N \sim 500$, in full LS-coupling, so increasing this further is not expected to increase accuracy by much. At the same time, reducing $N$ to lower values will not give significant total run-time improvements, especially when surrounding operations (loading the matrices) are considered as well. For  trans-iron elements, especially open $f$-shell ones, $N\sim 500$ may not be sufficient and avoiding the $N^3$ regime requires application of techniques such as superlevels (more below).}

\blue{There exists an advanced version of \texttt{DGESV}, called \texttt{DGESVX}, that may be a suitable second ansatz in a layered strategy. \texttt{DGESVX} allows for equation rescaling, iterative refinement, and other features that may enable solutions when \texttt{DGESV} fails for close-to-singular matrices.} 

\blue{Various operations and pre-conditions can improve the numeric situation. For example, \texttt{CMFGEN} performs two operations prior to the LU-decomposition solver step (Hiller, priv. comm.); 1) Recasting of the equations to solve for relative steps rather than absolute ones. 2) Pre-conditioning of the matrix with \texttt{DGEEQ} of \texttt{LAPACK}. Another example is that \texttt{ARTIS} solves for departure coefficients rather than absolute abundances. The value of such operations can be understood from inspection of the level populations example in Fig. \ref{fig:pognan2022_fig2}: solutions can span 15 orders of magnitude in NLTE which can pose severe numeric challenges if the equations are solved for in ``raw'' form.}


\paragraph{Iterative methods.} \blue{An alternative to directly finding the exact solution to a linear equation system is to iterate approximate solutions until sufficient accuracy is reached. There is a vast array of methods and the most suitable one depends on the properties of the particular matrix. For rate equations, we have no general simplifying properties of the structure. In such cases, among the non-stationary methods, Conjugate Gradient on the Normal Equations (CGNE) or Generalized Minimal Residual (GMRES) are the most relevant to use.} 

\blue{So far, in no numeric approach to SN/KN spectral synthesis, that the author is aware of, have iterative methods been implemented. This is because $N$ has
been kept below $\sim 10^3$, and as Fig. \ref{fig:timingtests_DGESV} shows, in that regime 1) The solve time is still in the $\propto N$ regime not quite having transitioned to $N^3$ yet, and 2) Matrix build times are anyway comparable to the solve time. However, as computational capabilities improve, solutions for $N\gtrsim 10^4$ systems will become relevant, in particular for KN modelling, and for these LU decomposition may become replaced by iterative methods as the standard choice.} 

\paragraph{Superlevels.} \blue{The number of levels, summed up over all elements and ion stages, quickly becomes large. This can give rise to challenges, the nature of which depends on the algorithm. For example, \texttt{CMFGEN} solves for all level populations in a cell simultaneously to allow for coupling between them (through the radiation field) (Hiller, priv. comm.). At full (fine-structure) level resolution, the matrix therefore becomes much too large to even store. \texttt{SUMO}, on the other hand, solves ion by ion. Storing is then not a problem, but nevertheless the $N^3$ scaling of LU-decomposition makes large $N$ models costly when $N$ gets large. The situation becomes especially problematic for lanthanides which can have tens of thousands of levels all well below the ionization threshold.}

\blue{One approach to address this is to use \emph{superlevels} \citep{Anderson1989}. This means that one bundles certain groups of levels into a single effective level. Typically one assumes an LTE distribution within the superlevel. 
Looking at the transitions between the fine-structure levels in a multiplet, typically the radiative rates are low and the collisional rates large. 
These conditions favor relative LTE distributions.}

\blue{As an example of the scale reduction that can be achieved by superlevel use, the full-composition Type Ia models in \citet{Blondin2023} have 10,605 levels grouped into 2,338 superlevels --- saving a factor $\sim$16 in storage and up to a factor $\sim$64 in matrix inversion time.} 

\blue{One should note that if one wants to retain individual, wavelength-separated lines connected to the superlevel, the energy discrepancies arising in the thermal balance (using the single specified energy of the superlevel) and radiative transfer (using several energies all slightly different from the single superlevel one) needs addressing. \citet{Hillier2012} discuss techniques for this for both hot wind and supernova applications.}


\section{Radioactive powering}
\label{sec:powering}
\blue{The radioactive decays in SNe and KNe inject high-energy particles - gamma rays, electrons, positrons, and, for the heaviest r-process elements, also $\alpha$-nuclei and fission fragments. These particles define the de-facto power source for the nebula already from an early point in time --- the energy deposited by the explosion itself is quickly adiabatically degraded and is not able to create much radiation (the only exception to this are the Type IIP SNe in which the large radii of the red supergiant progenitors make explosion-deposited energy important for the first 2--3 months).}

These high-energy particles transfer their kinetic energy to the ambient gas by the three processes of  ionization, excitation, and heating. They may also lose energy by radiative losses such as Bremsstrahlung (this radiation may or may not be reabsorbed by the ambient gas), and by adiabatic losses. Ionizations by gamma-rays will eject bound electrons with quite high energies, called primaries.  
These can in turn cause further ionizations, these electrons are called secondaries. This cascading means a complete theoretical and numerical treatment of the physical situation is a significant challenge.

\noindent LTE codes do not solve rate equations and therefore information about what fractions of the energy loss that go into the various channels (ionization, excitation, and heating, and their specific subchannels) cannot be used. The only option in this case is to allocate all energy deposition to heating. If the degradation (in this context often called \emph{thermalization} because all the energy is assumed to become heat) is fast and local, nothing needs to be solved for at all --- the total radioactive decay power in each cell is simply used as a heating contribution term going into the temperature equation (Sect.~\ref{sec:temperature}). 

Typically non-locality (transport) of gamma-rays are allowed for, however, as they can travel quite some distance in the ejecta, and in some cases partially escape already in the diffusion phase. Their main energy deposition mode is Compton scattering, in which typically a large fraction of the gamma-ray energy is transferred to an electron, creating a distribution of 0.1--1\,MeV primary non-thermal electrons. The Compton scattering process is straightforward to solve the transfer equation for or to simulate with Monte Carlo methods.  

Baryons and leptons, on the other hand, have much shorter mean-free-paths and are, in addition, quite easily trapped even by weak magnetic fields \citep{Axelrod1980}. They are therefore often assumed locally absorbed. 

\blue{As densities decrease with time as the nebula expands, at some point does the degradation start to be slow and/or non-local also for the leptons and baryons. Non-locality for leptons occurs only after about 5--10 years for Type II CCSNe, when NLTE is anyway necessary. An example is the situation in SN 1987A at an age of 8y, when positrons from $^{44}$Ti-decay power the metal-rich SN core and the spectrum probes their degree of spatial transport \citep{Jerkstrand2011}. For KNe, it was established in \citet{Barnes2016} that time-dependent effects (slow degradation) are important already after a few weeks or months.} 

\subsection{LTE modelling}
We will now study the different processes by which radioactive decay particles, and the particles generated by their interactions, deposit their energy to the ejecta. A sandbox to explore the various loss mechanisms have been coded up and is available at \url{https://github.com/eliotayache/elec_degrad}. It is useful to build intuition for how the different loss channels operate depending on energy and density, and which ones need consideration for a given problem at hand.

\subsubsection{Ionization and excitation}
\blue{The energy loss of a fast proton, or other nuclear projectile, due to collisional ionization and excitation of the ambient medium was worked out by Hans Bethe in the early 1930s, soon after the necessary quantum physics was in place. His work built upon the semi-classical foundation for the process established by Niels Bohr in 1913. The formula Bethe arrived at \citep[see][for a detailed rederivation and discussion]{Fano1963} is a good approximation for projectile energies much larger than the ionization potentials \citep[$\gtrsim$ 500\,eV,][]{Axelrod1980}, and holds also for relativistic motion. This means it is useful for determining the distance (or equivalently, time) over which radioactive decay particles deposit the bulk of their energy (over 99.9\% of the energy has been lost by the time a 1\,MeV primary goes below 500\,eV). 
Another property of the Bethe regime is that the energy loss occurs in a continuous manner, with small fractional losses for the primary in each ionization or excitation event. This property has consequences for analytic and numeric solution approaches.} 

\blue{With adaptation of the SI formula of \citet{Longair2011} to cgs form\footnote{This requires a $(4\pi)^2$ correction in addition to setting $\epsilon_0=1$, when mapping from $r_e$.}, the Bethe formula is, for single species gas (the "np" superscript refers to nuclear projectile):}
\begin{equation}
\frac{dE}{dx}^{\rm Bethe,np}(E) = \frac{4 \pi  q^4}{m_e} z_p^2 n_{eb}\frac{1}{v(E)^2} \left[\ln{\left(\frac{2\gamma(E)^2 m_e v(E)^2}{\bar{I}}\right)}-\frac{v(E)^2}{c^2}\right]\ \mbox{erg}\ \mbox{cm}^{-1},
\label{eq:bethe}
\end{equation}
\blue{where $n_{eb}$ is the number density of targets (all bound electrons per unit volume), $q$ is the unit charge (which is 1 in cgs units, with dimension cm$^{3/2}$g$^{1/2}$s$^{-1}$), $z_p$ is the (integer) charge of the projectile, and $\bar{I}$ is a mean potential, considered as a parameter that typically needs experimental input for setting\footnote{\blue{A compilation of values is available at \href{https://journals.sagepub.com/toc/crub/os-19/2}{https://journals.sagepub.com/toc/crub/os-19/2}. The values are determined from condensed phase experiments, but when occasional comparisons to gas and liquid states are done no big differences are seen.}}. 
In 1933 Felix Bloch showed that a reasonable approximation is $\bar{I} \approx 10\ \mbox{eV} \times Z$, where $Z$ is the atomic number. 
If that is inserted in the equation the formula is referred to as the Bethe--Bloch formula. The quantity $dE/dx$ is sometimes called the \emph{stopping power}. It is sometimes presented in units of erg\, g$^{-1}$\, cm$^2$, obtained by division with $\rho$.}
\blue{The value of the constant factor is $4\pi q^4/m_e=7.34\times 10^{-10}\, \mathrm{erg\, cm^{4}\, s^{-2}}$. The corresponding energy loss per unit time is $L(E)\equiv dE/dt =dE/dx\times v$.}

\blue{For electron projectiles, the formula is (again converting the SI form of \citet{Longair2011} to cgs, and going to energy per time form):}
\begin{equation}
L^{\rm Bethe,e}(E) = \frac{2\pi q^4}{m_e} n_{eb} \frac{1}{v(E)}\left[\ln{\left(\frac{\gamma(E) m_e v(E)^2 E_{\max}}{2\bar{I}^2}\right)} + f(\gamma(E))\right]
\end{equation}
\blue{where $f(\gamma(E))$ is a factor varying between 0.3 and -0.2 at the energies relevant here ($\gamma=1-3$), and
$E_{\max}$ is the maximum transferable energy per collision, which for electron-electron collisions is given by
$E_{max} = \frac{\gamma^2 m_e v^2}{1 + \gamma}$,}
\blue{so the loss can also be written}

\begin{equation}
L^{\rm Bethe,e}(E) = \frac{4\pi q^4}{m_e} n_{eb} \frac{1}{v(E)}\left[\ln{\left(\frac{E}{\bar{I}}\right)} -0.19 + g(\gamma(E)) \right]
\label{eq:Bethe-e}
\end{equation}
\blue{where $g(\gamma)$ is a relativistic correction term only important for high energies.}

\blue{Figure \ref{fig:bethebloch} shows the loss rate $L^{\rm Bethe,e}(E)$ for a target number density of $n=10^7$ cm$^{-3}$ (the scaling is inverse with $n$). The loss rate has a minimum close to the characteristic injection energies of primaries, at or a bit below 1\,MeV. The loss rate increases by a factor of a few going to lower energies, with maxima at about $10\bar{I}$. We can see that for the $\lesssim$ 1\,MeV regime the relativistic correction term $g(\gamma)$ may be ignored. Finally, note the moderate impact of $\bar{I}$; even a factor 10 change leads to a factor $<2$ change in the loss rate, due to the logarithmic dependency.}


\blue{For $\alpha$ particles and fission fragments, we may use Eq.~\eqref{eq:bethe} instead of Eq.~\eqref{eq:Bethe-e}. These heavy particles are always in the non-relativistic regime.}

\subsubsection{Heating}
\blue{Energy transfer to the free, thermal electrons occurs in a continuous manner as no bound states are involved - the long-range Coulomb potential leads to a smooth energy transfer process. Due to the repulsion and their small mass, the free thermal electrons are able to move away from the path of the projectile to some extent, which also contributes to this. As such we can assume their energy gain is not sufficient to make them leave the thermal pool and no cascading is obtained as in the ionization case.}

\blue{The modelling of heating losses by fast particles has a long history in astrophysics, with the original motivation being the study of how cosmic rays impact nebulae. \citet{Spitzer1969} derived a heating loss formula (for the secondary electrons created in the cosmic ray ionizations, see also \citealt{Dalgarno1999}):}
\begin{equation}
L_{e,\rm heat}(E)= \frac{4 \pi q^4}{m_e} n_{e,th} \frac{1}{v(E)}  \Lambda(E)~\mbox{erg\ s}^{-1},
\end{equation}
\blue{where $\Lambda(E)$ is a dimensionless quantity tabulated in \citet[][a book, unavailable online]{Spitzer1962}. Note the similarity of the functional form to Eq.~\eqref{eq:bethe}.}
%
A widely used formula for $\Lambda(E)$ for SN/KN modelling is the
result of \citet{Schunk1971},
\begin{equation}
\Lambda(E)^{\rm SH} = \ln{\left(\frac{m_e v(E)^2}{\bar{h}\sqrt{4\pi q^2 n_{e,th}/m_e}}\right)}.  
\end{equation}
\blue{Note here the additional appearance of $n_{e,th}$. The \citet{Schunk1971} formula corresponds to the non-relativistic limit of a more general expression derived by \citet{Inokuti1978},} 
\begin{equation}
\Lambda(E)=
\Lambda(E)^{\rm SH}
+ \frac{1}{2} \ln{\left(\frac{1}{1-\beta^2}\right)} - \frac{1}{2}\beta^2,
\end{equation}
where $\beta=v/c$. The Inokuti formula in turn builds from \citet{Fano1963}, but adds in collective excitations of the plasma. The additional terms in the \citet{Inokuti1978} formula gives only small corrections below a few MeV (Fig. \ref{fig:bethebloch}). Thus, for the purposes of SN/KN radioactivity modelling the \citet{Schunk1971} formula should be sufficient. 
The $\Lambda(E)$ factor varies over the range $15-30$ for relevant combinations of $E$ ($\sim 10^2-10^6$\,eV) and $n_{e,th}$ ($\sim 10^3-10^{10}$ cm$^{-3}$) for SN/KN applications. Thus, it is a reasonable approach to fix it to a constant, this is done e.g. in the KN modelling of 
\citet{Barnes2016}\footnote{The $\lambda_{ee}$ factor there corresponds to $\Lambda$ here. Note that \citet{Barnes2016} use a formula where the $1/v$ factor has been expressed as $\sqrt{m_e/2E}$; that relation has however poor accuracy for $E \gtrsim$0.1\,MeV and the $1/v$ expression is preferable.}.
%

\subsubsection{Discussion}
\begin{figure}[ht]
\centering
\includegraphics[width=0.8\linewidth]{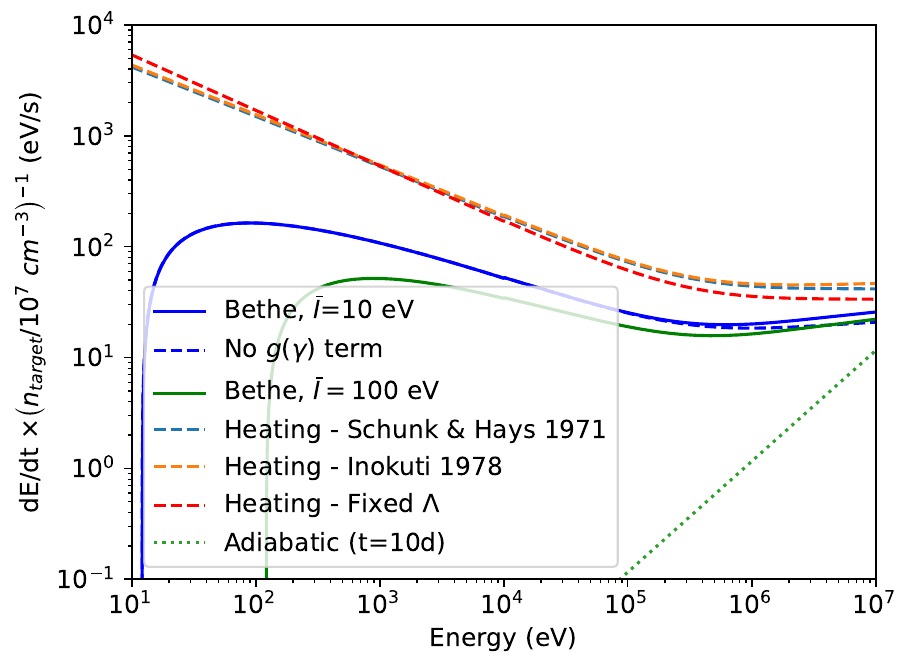}
\caption{Electron energy loss rates as function of projectile energy. The collisional loss rates scale with $n_{\rm target}^{-1}$, with $n_{\rm target}=n_{e,th}$ for heating and $n_{\rm target}=n_{eb}$ for ionization/excitation, which is therefore embedded in the abscissa values for these. The adiabatic loss rate is just $E/t$ (so does not depend on $n_{targets}$).} 
\label{fig:bethebloch}
\end{figure}

Figure \ref{fig:bethebloch} plots the loss rates per target electron. For significantly ionized H and He-rich gases, $n_{e,th}$ would be larger than $n_{eb}$, and heating is then the dominant loss mechanism for all energies. However, for heavy composition layers, models and observations indicate that the gas becomes at most a few times ionized (i.e., Fe would exist as Fe I-IV ions). In that case many more electrons remain bound than free (for example 23 bound and 3 free if all Fe is in the Fe IV state). The effective ratio of ionization/excitation to heating scales with $n_{eb}/n_{e,th}$ (23/3 in the example) meaning most loss occurs by ionization/excitation in such a situation.

Another point to take note of that the distribution of energy loss over the channels is not the same as the final distribution of energy going into the different channels. For example, if a 1\,MeV electron transfers 100\,eV to a bound electron in a $\bar{I}=10$\,eV atom, it has lost 100\,eV by an ionization process, but the ionization energy of the gas has only increased by 10\,eV. The other 90\,eV resides now in another high-energy electron, and only a fraction of that will eventually be added on as further ionization energy as this particle cascades down.

\subsubsection{Radiative losses}
Radiative losses are for the most part small or neglegible compared to the collisional ones.  
The main radiative loss mechanism is Bremsstrahlung \citep{Barnes2016}. It can be described by \citep{Seltzer1986}
\begin{equation}
L_{\rm rad,Brems}(E) = r_e^2 \alpha \sum_i n_i Z_i^2 v(E)\left(E + m_e c^2\right) \phi(E,Z)~~\mbox{erg}~\mbox{s}^{-1},
\end{equation}
\blue{where summation is done over all ions $i$ with charge $Z_i$. The $\phi(E,Z)$ function is tabulated in \citet{Seltzer1986} and has a typical value of $\sim$10.}\\

\subsubsection{Formulation and numerics}
\blue{Radioactive decay leads to injection of high-energy particles with both a total power and energy distribution being  functions of time. For a given type of decay particle, this may be represented as a sum of decays per unit mass $\dot{n}_i(t)$ at specific energies $E_i$, so total decay power $\dot{\epsilon}(t)=\sum_i E_i \dot{n}_i(t)$ erg s$^{-1}$ g$^{-1}$.}

\blue{The energy loss rate per particle is given by}
\begin{equation}
\frac{dE}{dt} = -L_{\rm atom}(E,t) - L_{\rm elec}(E,t) - L_{\rm rad}(E,t) -x(E)\frac{E}{t},
\label{eq:Eloss}
\end{equation}
\blue{where the last term is due to adiabatic losses \citep{Waxman2018}. Its pre-factor $x$ has limits of 1 and 2 in the ultra-relativistic and non-relativistic limits, respectively. It is not significant initially, but because it evolves as $t^{-1}$ compared to (for fixed gas state) $t^{-3}$ for the collisional loss terms, it will eventually come to dominate.} 

\blue{For a given decay particle of energy $E_0$ at creation, does a unique mapping $\hat{E}(E_0,t_0,t)$ --- the energy of the particle at any later time $t$  have --- exist? Such a mapping would come from solving the Eq.~\eqref{eq:Eloss} IVP
with initial condition $E(t_0)=E_0$.
However, the ionization, excitation, and heating losses depend not only on composition but also on the time-dependent gas ionization state, which in turn depends on the solution to the degradation evolution for all decay particles. Therefore, a full solution to this problem requires simultaneous time-evolving solution of the gas state (corresponds to determining the temperature in LTE) and the degradation equations. No such work yet exists in the literature. And clearly, the answer to the posed question is no --- the answer is density and composition dependent, and in addition is the evolution dependent on all decay particles.} 

Anything of generic nature must therefore assume a fixed gas state, or a parameterized time evolution of the gas state. In the approaches in the KN literature so far those assumptions are 

\begin{enumerate}
    \item  Simplified loss functions taken as constants independent of energy and time \citep{Waxman2018}.
    \item All elements are in the singly ionized state \citep{Barnes2016,Hotokezaka2020}.
    \item All elements are in the neutral state \citep{Kasen2019}.

\end{enumerate}
The $\hat{E}_i(E_0,t_0,t)$ functions can then be uniquely determined and
the power at time $t$ becomes, ignoring secondary electrons,
\begin{equation}
P(t) = \sum_i \int_0^t \dot{n}_i(t')\left[L_{i,\rm atom}(t',t) + L_{i, \rm elec}(t',t)\right]
dt'.
\label{eq:Pt}
\end{equation}
\blue{The solution of Eq.~\eqref{eq:Eloss}, needed to compute Eq.~\eqref{eq:Pt}, involves solving a first-order IVP with non-constant coefficients. Analytic limits based on the simplifications outlined above are extensively discussed in \citet{Barnes2016,Waxman2018,Waxman2019,Kasen2019,Hotokezaka2020}. A public code to solve the equations for KN ejecta \citep{Hotokezaka2020} is available \citep{kenta_hotokezaka_2020_3601589}.} 

\subsubsection{Examples}
\begin{figure}[ht]
\centering
\includegraphics[width=0.49\linewidth]{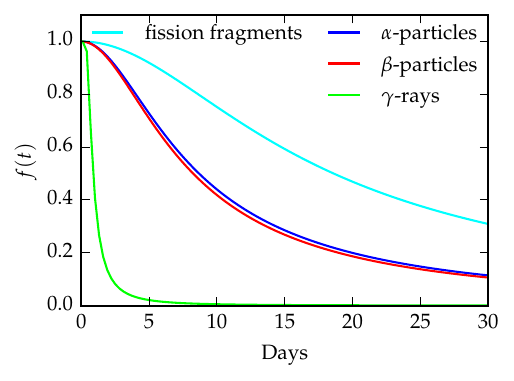}
\includegraphics[width=0.49\linewidth]{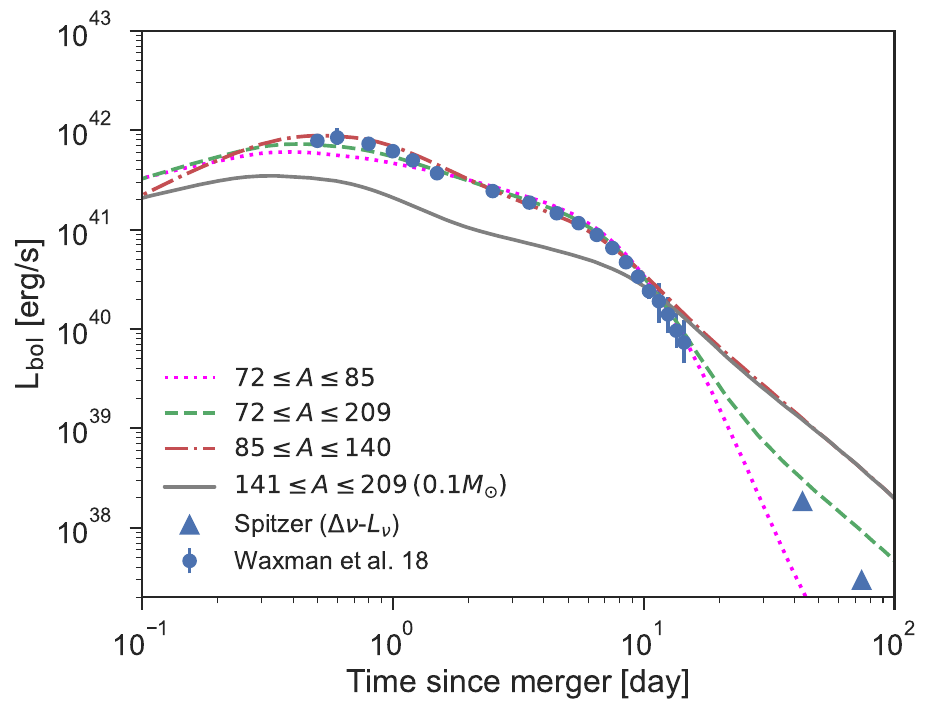}
\caption{\blue{\emph{Left:} Thermalization efficiency versus time for a homologous expansion trajectory corresponding to typical KN densities ($\rho(1d)=8\times 10^{-15}$g cm$^{-3}$). \emph{Right:} Observations of AT2017gfo compared to models with different nuclear compositions, and with time-dependent thermalization. The figure demonstrates that the late-time KN data is able to constrain the composition. Images reproduced with permission from [left] \citet{Barnes2016} and [right] \citet{Hotokezaka2020}, copyright by AAS.}}
\label{fig:barnes2016_fig9}
\end{figure}

\blue{Figure \ref{fig:barnes2016_fig9} (left) shows a calculation from \citet{Barnes2016} of the so called \emph{thermalization efficiency} in a KN, defined as the power at time $t$ (Eq.~\eqref{eq:Pt}) divided by the radioactive decay power at the same time. Observations of AT2017gfo covered up to 20\,d in optical/NIR and yet later (up to 74\,d) with the Spitzer mid-infrared telescope. Over observable epochs, thermalization is clearly a time-dependent phenomenon requiring solutions of Eqs.~\eqref{eq:Eloss} and \eqref{eq:Pt}.}

\blue{The RHS of Fig.~\ref{fig:barnes2016_fig9} shows different models, with time-dependent thermalization, compared to observations of AT2017gfo, from \citet{Hotokezaka2020}.}

\subsection{NLTE modelling: Boltzmann solutions}
\blue{For NLTE modelling the specific power going into the different channels of ionization, heating and excitation must be solved for. The ionization channel breaks down into subchannels for each ion-electron shell pairs (e.g. O I $K$, O I $L$, O I $M$, O II $K$...), whereas the excitation channel breaks down into the set of considered bound-bound transitions for each atom/ion. Electrons at each energy in the degradation cascade will contribute differently to these different channels --- the degradation energy spectrum therefore needs to be solved for. Free electrons in the ejecta span energies from the thermal population ($\sim kT \sim$eV) to the primary electrons released in radioactive decays or other high-energy process ($\sim$\,MeV). Given the six orders of magnitude in the energy cascade, this is a challenging task.} 

Current codes solve the Boltzmann equation under the two assumptions 

\begin{enumerate}
    \item \emph{Steady state} (no time-dependent degradation). 
    \item \emph{Locality} (no spatial transport).
\end{enumerate}
This ansatz is valid if the particles lose their energy over temporal and spatial scales much smaller than those of changing physical conditions. The first such solutions in the SN context were by \citet{Meyerott1978} and \citet{Axelrod1980}. Later works include those of \citet{Kozma1992, Xu1991, Lucy1991}. The first full solutions for KNe were done with the \texttt{SUMO} code.

\blue{Whereas LTE modelling can bundle non-thermal energy transfer to excitation and ionization into a single treatment, in NLTE it needs separation. The use of the Bethe formalism with an empirical $\bar{I}$ is therefore less useful, because those empirically determined values of $\bar{I}$ involve the combined effects of excitation and ionization. In addition, even though the low-energy regime where the Bethe formalism starts to fail ($\lesssim$ 500\,eV) is unimportant for thermalization time/distance, it is important for the ionization and excitation rates needed in NLTE. For these reasons, one now needs cross sections for individual processes, also extending down to low energies ($\sim$ 1--500\,eV).}

\subsubsection{Ionization terms}
\blue{Let us begin with the total cross sections for ionization (integrated over all energy transfer values). A commonly used semi-empirical formalism is that of \citet{Lotz1967}. In the initial work of \citet{Axelrod1980}, the following flavour of this formalism was used:}
\\
\begin{equation}
\sigma^{\rm tot}(E) = \frac{\rm const}{\beta^2 m_e c^2}\sum_{i=1}^{N \rm shells} \frac{n_i}{I_i}\left[\ln{\left(\frac{\beta^2m_e c^2}{2I_i}\right)} - \ln{\left(1-\beta^2\right)}-\beta^2\right],
\end{equation}
\blue{where 
$\rm const=6.9 \times 10^{-38}$ erg$^2$ cm$^2$
, $n_i$ is the number of electrons in shell $i$ per atom, $\beta=v/c$ and $I_i$ is the ionization potential of shell $i$.} 
\blue{Later works for SN modelling have have used dedicated theoretical or experimental determinations for the total cross sections when available, but the \citet{Lotz1967} formalism still finds use for KN modelling \citep{Pognan2022a}.} 

\blue{A complication arises for inner shell ionizations, because  refilling of the created "hole" leads to Auger electron ejection and/or X-ray fluorescence. Different codes take different approaches to dealing with this. 
The \texttt{SUMO} code, for example, lets the inner shells contribute to the ionization cross section, and the excess energy is used to raise the ionization rate (which corresponds to treating all electrons like valence ones). In \citet{Shingles2020}, an improved treatment is devised, implemented into the \texttt{ARTIS} code, by coupling in a feedback loop for the Auger electrons. No large impact on Type Ia model spectra were seen, likely because for iron-group elements the contribution by the inner $K$ and $L$ shells is not so large, with 10 electrons in total compared to the 16 $M$-shell electrons. The impact may be larger for both lighter and heavier elements, when the valence count is not so dominant as in the iron-group. 
}\\
\\
\textbf{Differential cross sections.} \blue{The ionizations lead to the creation of secondary high-energy electrons. To solve for the Boltzmann distribution, we need to know the energy distribution of these, i.e. not just the total ionization cross sections  $\sigma^{\rm tot}(E)$ discussed above, but also the differential ones $\sigma(E,\epsilon)$, where $\epsilon$ is the energy transfer. Note the unit for these is cm$^2$erg$^{-1}$.}

\blue{Stopping power depends on $\int \sigma(E,\epsilon) \epsilon d\epsilon$, i.e. the first moment of the differential cross section. The Bethe formula for stopping power, Eq.~\eqref{eq:Bethe-e} (valid for energy transfers $I \ll \epsilon \ll E_0$) derives from the first moment of a differential cross section following a "Rutherford" form $\sigma \propto 1/\epsilon^2$.
This gives the $\ln(E)$ factor for the stopping power, because we need to integrate the loss over $1/\epsilon$ ($1/\epsilon^2 \times \epsilon$).}

\blue{More detailed information can be obtained from experiment.
\citet{Opal1971} measured the secondary electron energies (up to 200\,eV) resulting from the impact of high-energy electrons (up to 2 keV) on several molecules and inert atoms. The differential cross section was found to be well  described by the formula}

\begin{equation}
\sigma(E,\epsilon) = \rm const(E) \times \frac{1}{1+\left(\epsilon/\bar{E}\right)^2}, 
\label{eq:opal}
\end{equation}
\blue{where $\bar{E}$ was found from the experiments to lie in the range $\left(0.5-1\right)I$ depending on element and projectile energy. This equation takes the Rutherford form in the high energy limit, but levels out in the low-energy limit. One should note that also inner shells may contribute electrons to this experimental ejection process, and also that in the experiment no distinction can be made between the two post-collisions free electrons, necessitating a definition of secondary electrons as the less energetic post-collision one.} 

\blue{The primaries in the Opal experiment had much lower energies ($\lesssim$ 200\,eV) than the primaries in radioactive decays ($\sim$1\,MeV). For high-energy primaries, \citet{Manson1975} reported differential cross sections for proton impact on He, up to the MeV range. The differential cross section is found to have a distinct steepening for high secondary energies compared to Eq.~\eqref{eq:opal}; for $E_0=100$ keV impact that steepening begins at  $\epsilon \approx 100$\,eV, for $E_0=300$ keV at $\epsilon \approx 300$\,eV, and for $E_0=1$\,MeV at  $\epsilon \approx 1$ keV; thus the steepening occurs uniformly at $\epsilon/E_0 \approx 10^{-3}$. A power law fit to the steeper part has an exponent of between -4 to -5. Assuming the behaviour is similar for electron projectiles, one has a prescription available for the differential cross section over the whole energy interval for the projectile.}

\subsubsection{Excitation terms}
\blue{In excitations the energy transfer is fixed to the transition energy $\Delta E$. What is needed is just the cross section for the process as function of projectile energy. For large projectile energies this is, in Bethe theory for electron collisions, given by \citep{Mott1949,vanRegemorter1962}}
\begin{equation}
\sigma^{tot}(E) =
3 \pi q^4 f_{osc} \frac{1}{E \Delta E} \ln{\left(\frac{4E}{\Delta E}\right)}~~~~,E \gg \Delta E
\label{eq:mott}
\end{equation}
\blue{This expression is used for all energies in the Boltzmann solver developed by \citet{Kozma1992} and used by the \texttt{SUMO} code, for those transitions lacking specific cross section data. Note that the Bethe limit is more readily obtained for line transitions than ionization continua as the line transitions have lower energy.} 

\blue{For light elements (up to iron) the energy going into excitations is typically comparable to, or smaller by a factor few, than the energy going to ionizations \citep{Kozma1992}. At the relatively neutral conditions in nebular CCSNe, which leads to the thermal electrons taking less than $\sim$50\% of the power, the emissivity from non-thermal excitation can therefore be significant.} 

\blue{For r-process elements, no calculations yet exist. However, one may comment that the ionization state of KN ejecta appears significantly higher than in CCSNe \citep{Hotokezaka2021,Pognan2022a}. This means that direct emission from non-thermal ionizations and excitations carries less power compared to cooling emission following the non-thermal heating (Chapter \ref{sec:temperature}}). While the non-thermal ionizations are crucial for setting the ionization balance and determining which ions emit the cooling, non-thermal excitations have no such indirect impact. On the other hand, some of the r-process elements have orders of magnitude more line transitions than any light elements - which would boost the importance of non-thermal excitations. A clear picture of the overall effect awaits detailed calculations.


\subsubsection{Solution approach}
\blue{\paragraph{(Lack of) thermal electron coupling.} If the thermal electrons ($E \lesssim $1\,eV) are to be included in the Boltzmann solution, one must allow transfers that both decrease and increase the primary electron's energy - and also account for the creation and removal of electrons by ionization and recombination terms. There will be coupling between all energies in general, i.e. a ``full'' matrix. There are no published solutions for such an approach in the literature.}

\blue{Instead, all published solutions treat the thermal electrons separately. This is possible because quite few of the secondary electrons have such low energies that they are in the thermal regime at creation. For the Opal prescription of secondary electron distributions (Eq.~\eqref{eq:opal}) and a standard choice $\bar{E} = 0.6 I$, the low-end 10th 
percentile of the distribution is at $\sim 0.2I$.  Then even the lowest ionization potentials ($I \sim 6$\,eV) give less than 10\% of secondaries below 1\,eV. We should therefore be able to put a clear distinction, around 1\,eV for typical conditions, between a thermal electron group below this threshold and a non-thermal one above.}\\
\\
\paragraph{Formulation and numerics.}
\blue{Let $z$ be the flux of non-thermal electrons at energy $E$ per unit area and energy (cm$^{-2}$\, s$^{-1}$\, erg$^{-1}$). It is related to the spectral number density $n_E$ (cm$^{-3}$\, erg$^{-1}$) by $z = n_E v(E)$. If the source term $S_E$ (containing injection by radioactive decay) is given in units of cm$^{-3}$\, erg$^{-1}$\, s$^{-1}$, the steady-state distribution for $z$ is determined by}

\begin{align}
    S_E(E) + 
     \sum_i n_i \left[
     \int_{I_i}^{E_{\max}} z(E') \frac{ d \sigma_i^{\rm ion+exc}}{d \epsilon}(E',E'-E) dE' \right.
     \nonumber \\ 
    \left. + \int_{E+I_i}^{E_{\max}} z(E') \frac{ d \sigma_i^{\rm ion}}{d \epsilon}(E',I_i+E) dE'\right] 
    \nonumber \\ 
     = z(E) \sum_i n_i \int_0^E \frac{ d\sigma_i^{\rm ion+exc}}{d \epsilon}(E,\epsilon) d \epsilon - \frac{d}{dE}\left[z(E)\frac{dE}{dx}_{\rm heat}(E)\right] 
    \label{eq:SF}
\end{align}
\blue{Here, the first term on the LHS represents injection by radioactive decays, the second term injection into the bin due to downscattered primaries, and the third term creation of secondaries following ionization events. On the RHS, the first term is loss from the bin due to ionizations and excitations, followed by a  term representing heating losses \citep[see][for derivation]{Xu1991}.}
\blue{This last term, not treated by \citet{Spencer1954},  
changes Eq.~\eqref{eq:SF} from an integral equation into an integro-differential one. Note that for excitations, the differential cross section can be treated as a delta function in the integral, in practice creating a separate sum term for these events.} 

\paragraph{Radiation terms.}
\blue{Including radiative losses, such as Bremsstrahlung or synchrotron, can be done by adding further (negative) source terms, which will be of the same general form as the heating loss term as long as the radiation process can be described as continuous. Self-absorption of this radiation may conceivably be added in if it creates non-thermal electrons. Neither radiation process is expected to be important for SNe, whereas for KNe Bremsstrahlung may play some role \citep{Barnes2016}. To date, radiation terms have not been implemented in any solutions in the literature.}

\paragraph{Solution approach.} \blue{As long as the thermal electron pool is uncoupled, and energy gain by the non-thermal electrons (by collisions with higher-energy ones) is ignored, existing electrons can only move downward in energy. Eq.~\eqref{eq:SF} can therefore be solved from injection energy down to the lowest treated energies in single steps, not requiring matrix storage and inversion. It is a first-order integro-differential IVP, solvable by e.g. implicit Euler.}

\blue{The discontinuous nature of bound-bound and bound-free transitions (e.g. the excitation cross section in Eq.~\eqref{eq:mott} discontinuously drops to zero at the threshold energy as the peak lies below threshold at $0.68 \Delta E$) calls for some special care in the numerics. One should also pay attention to the fact that the differential ionization cross sections are only large at transfer energies of order of the ionization potential. Therefore, the integrations in Eq.~\eqref{eq:SF} can involve integration from e.g. 10\,eV to 1\,MeV, using a kernel function that is only large up to say 100\,eV for the energy transfer. Resolving this integral over the energy transfer argument means that the primary energy must be resolved to the same precision. In short: it will be problematic to use anything but a linear energy grid, and the resolution of this must be better than the ionization potentials.} 

\blue{It turns out that if the output needed are the various fractions of energy going into different channels, that can be calculated by an alternative formalism where one steps up in energy instead of downwards \citep{Xu1991,Lucy1991}. In this formalism the degradation spectrum $z(E)$ is never explicitly solved for. One advantage of this approach is that the fractions will converge towards final values quasi-independent of the injection energy. Thus, there is a possibility to stop the integration with some criterion on the change rates. It is also more efficient if one needs to compute the solution many times, for different injection energies. However, different types of numeric issues show up in this approach compared to the degradation solution approach --- see Sect.~4 in \citet{Xu1991} for an in-depth discussion.}

\subsubsection{Some results}
\begin{figure}[ht]
\includegraphics[width=0.5\linewidth]{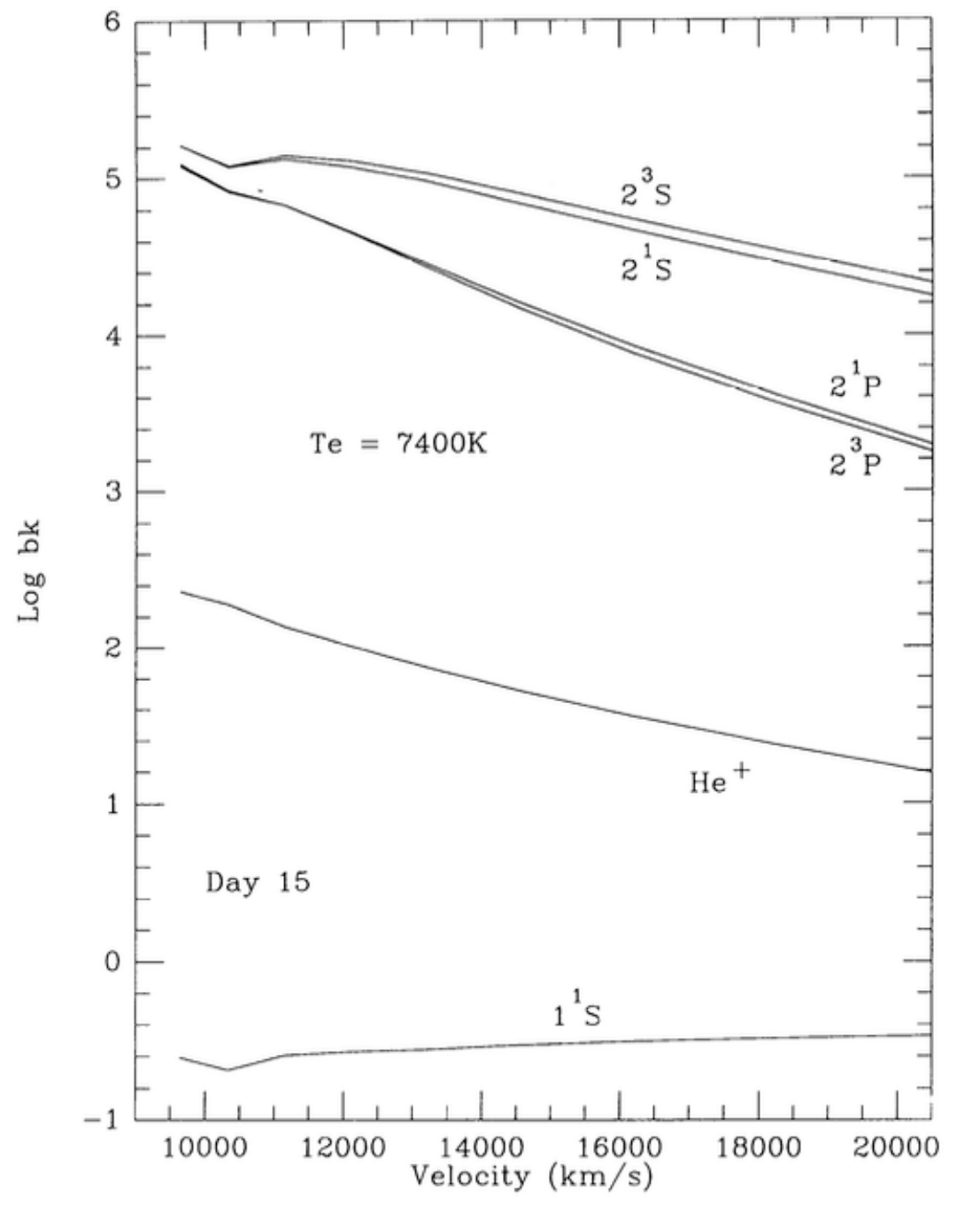}
\includegraphics[width=0.49\linewidth]{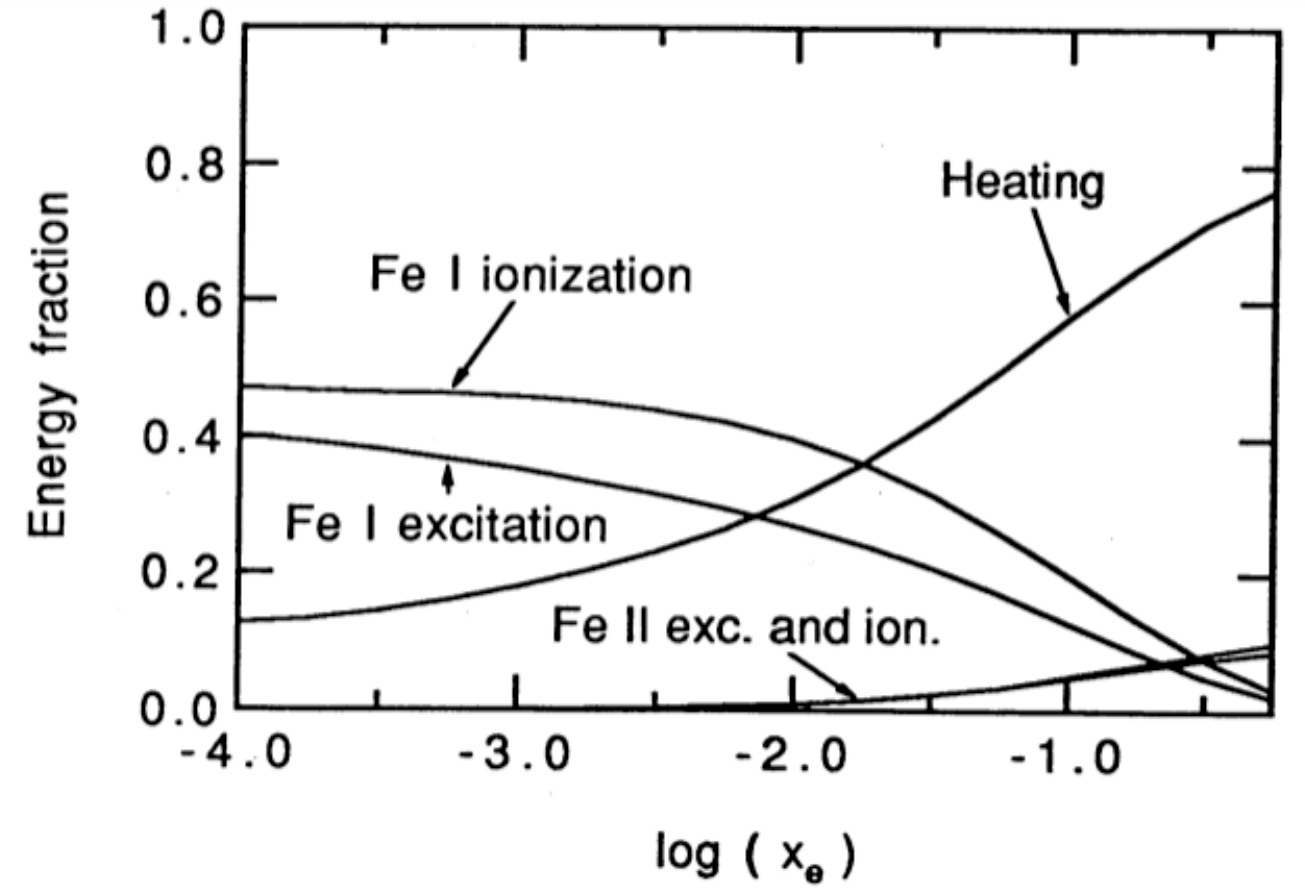}
\caption{\emph{Left:} Departure coefficients for He in a Type Ib CCSN model at 15d (photospheric phase) when non-thermal effects are considered. \emph{Right:} Fraction of non-thermal energy going into channels of heating, ionization and excitation, versus electron fraction in a pure Fe\,I +  Fe\,II gas (so $x_{\rm FeII}=x_e$). Images reproduced with permission from [left] \citet{Lucy1991} and [right] \citet{Kozma1992}, copyright by AAS.}
\label{fig:examples_Boltzmann}
\end{figure}

\blue{The LHS of Fig.~\ref{fig:examples_Boltzmann} shows the level population boosts in the metastable states of He\,I (emitting the important 1.08 and 2.056 $\mu$m lines) when non-thermal excitations and ionizations are considered \citep{Lucy1991}. The departure coefficients reach $\log~b = 4-5$ which illustrates how powerful these mechanisms can be for certain elements and levels.}

\blue{The RHS shows the fraction of radioactive power going into different channels as function of ionization state of the gas. We see how heating goes over 50\% already at electron fractions $x_e \equiv n_e/n_{\rm atoms}\gtrsim$3\%. This is why temperature calculations are not very sensitive to the details of non-thermal degradation in the diffusion phase ($x_e \gg$ 3\%) or even in the nebular phase of CCSNe ($x_e \sim$ 10\%). We also see that the atomic stopping contribution is about equally distributed between ionization and excitation for iron.}

\newpage
\section{Summary and outlook}
We have reviewed some of the key aspects for computing spectral models of explosive transients; supernovae and kilonovae. The emphasis has been on the computational techniques and considerations, and to sort out which approximations are in use and what they mean.

\begin{figure}[ht]
    \includegraphics[width=1\linewidth]{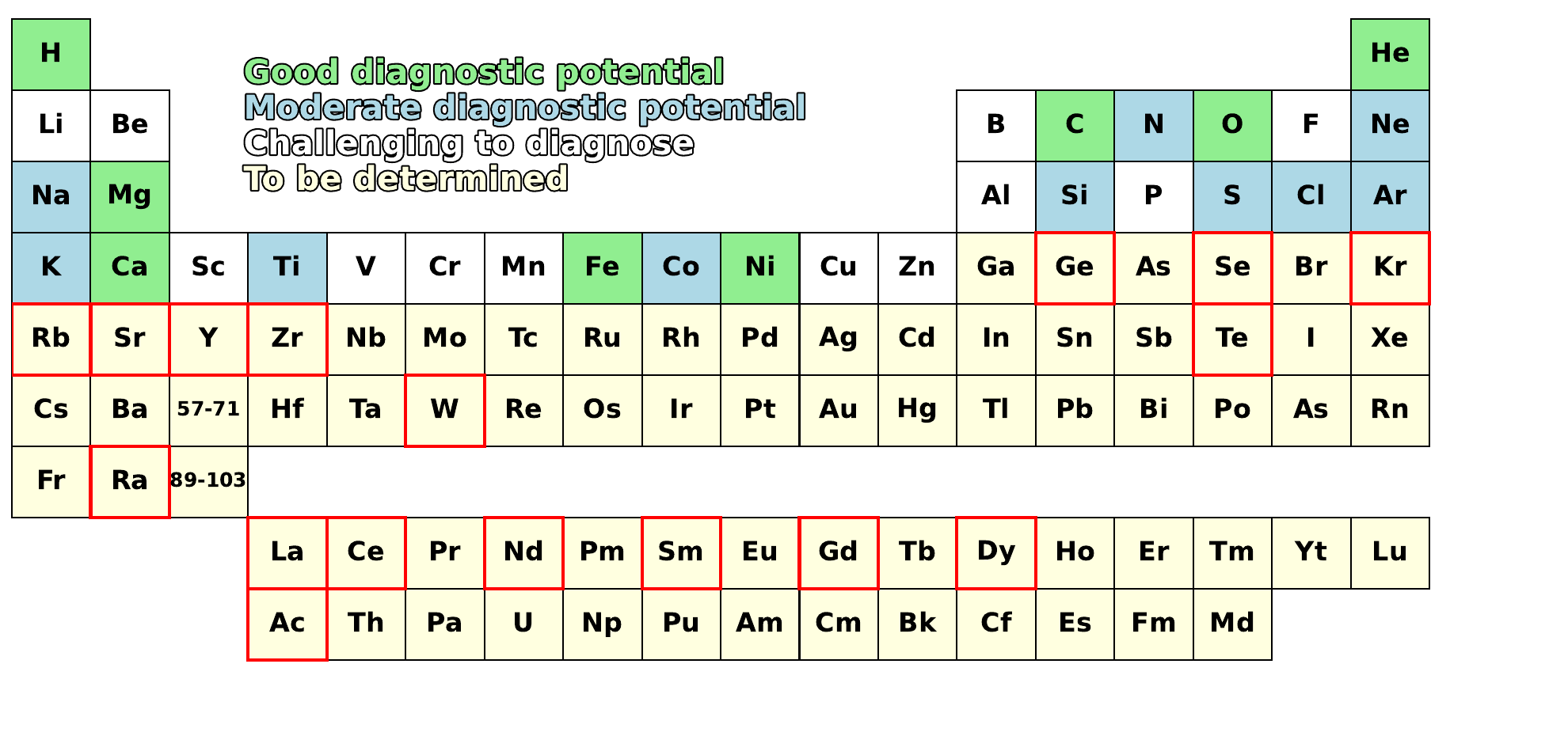}
    \caption{Diagnostic situation for elements in the periodic table from SN and KN spectral modelling as applied to observational data.}
    \label{fig:periodic}
\end{figure}

The main goal of spectral modelling is to, eventually, be able to infer the composition of SN and KN ejecta from observations. Figure \ref{fig:periodic} shows an overview of where the field stands in its current ability to do this.

For the $Z=1-30$ elements, SN spectral studies since the early 1980s have by now brought about a relatively mature picture of which signatures we observe and do not observe in spectra, and what we can infer from them. The elements are divided into the three categories of ``good diagnostic potential'', ``moderate diagnostic potential'' and ``challenging to diagnose''. The first category typically has at least two clear lines whose formation is relatively well understood and is abundance constraining. The last has either no clear lines, or perhaps a single line that is however sensitive to model uncertainties. The middle category is somewhere in between. It may be seen as a success of the field over its first 45 years that 19 out of these 30 elements are now blue or green \citep[for recent efforts see e.g.][]{Dessart2021,Fang2022,Liljegren2023,Barmentloo2024,vanBaal2024}. This accomplishment has been made possible by the combination of high-quality spectral data and the development and application of synthesis codes addressing the various challenges outlined in this review.

For the trans-iron elements, spectral searches and analysis began only with the first KN in 2017. It is at this point not possible to meaningfully assess the diagnostic potential of many of these elements --- we have neither sufficient observations nor spectral models with r-process elements either in SNe or KNe. However, for KNe this is now rapidly coming about. I have marked in red the r-process elements that in the current literature have been identified as plausible candidates for spectral features in AT2017gfo, or appearing to have potential for identification in KN spectra \citep{Watson2019,Domoto2021,Domoto2022,Hotokezaka2022,Hotokezaka2023,Sneppen2023,Vieira2023,Pognan2023,Pognan2025,Gillanders2024,Banerjee2025}.

To increase the blue and green in this diagram, on the radiative transfer modelling side we need to steadily improve on the physical treatments of the various foundational processes - hopefully this review contributes to visualizing the roadmap for this endeavor. Equally important is the development of accurate atomic data that goes as input to these model (in particular for the trans-iron elements), development of 3D hydrodynamic models for various types of explosions and mechanisms, and high-quality observations over the whole UV to mid-infrared range and over the different epochs probing different layers of the ejecta. By the combined and linked efforts in these categories we can be optimistic to eventually be able to determine the origin of elements as directly observed at their production sites in supernovae and kilonovae.

\appendix

\bmhead{Acknowledgements}

The author acknowledges useful discussion with D. Kasen, D.J. Hillier, and S.E. Woosley, and funding from the European Research Council
(ERC) under the European Union's Horizon 2020 Research and
Innovation Program (ERC Starting Grant 803189 -- SUPERSPEC), the Swedish Research Council (grant 2018-03799) and Knut and Alice Wallenberg Foundation grant ``Gravity Meets Light''. 

\phantomsection
\addcontentsline{toc}{section}{References}
\bibliography{references}

@ARTICLE{Banerjee2025,
       author = {{Banerjee}, Smaranika and {Jerkstrand}, Anders and {Badnell}, Nigel and {Pognan}, Quentin and {Ferguson}, Niamh and {Grumer}, Jon},
        title = "{Nebular Spectra of Kilonovae with Detailed Recombination Rates. I. Light r-process Composition}",
      journal = {\apj},
         year = 2025,
        month = oct,
       volume = {992},
       number = {1},
          eid = {19},
        pages = {19},
          doi = {10.3847/1538-4357/adf6ba},
archivePrefix = {arXiv},
       eprint = {2501.18345},
 primaryClass = {astro-ph.HE},
       adsurl = {https://ui.adsabs.harvard.edu/abs/2025ApJ...992...19B}
}

@ARTICLE{Vieira2023,
       author = {{Vieira}, Nicholas and {Ruan}, John J. and {Haggard}, Daryl and {Ford}, Nicole and {Drout}, Maria R. and {Fern{\'a}ndez}, Rodrigo and {Badnell}, N.~R.},
        title = "{Spectroscopic r-Process Abundance Retrieval for Kilonovae. I. The Inferred Abundance Pattern of Early Emission from GW170817}",
      journal = {\apj},
         year = 2023,
        month = feb,
       volume = {944},
       number = {2},
          eid = {123},
        pages = {123},
          doi = {10.3847/1538-4357/acae72},
archivePrefix = {arXiv},
       eprint = {2209.06951},
 primaryClass = {astro-ph.HE},
       adsurl = {https://ui.adsabs.harvard.edu/abs/2023ApJ...944..123V}
}

@ARTICLE{Pognan2025,
       author = {{Pognan}, Quentin and {Wu}, Meng-Ru and {Mart{\'\i}nez-Pinedo}, Gabriel and {da Silva}, Ricardo Ferreira and {Jerkstrand}, Anders and {Grumer}, Jon and {Fl{\"o}rs}, Andreas},
        title = "{Actinide signatures in low electron fraction kilonova ejecta}",
      journal = {\mnras},
         year = 2025,
        month = jan,
       volume = {536},
       number = {3},
        pages = {2973-2992},
          doi = {10.1093/mnras/stae2778},
archivePrefix = {arXiv},
       eprint = {2409.16210},
 primaryClass = {astro-ph.HE},
       adsurl = {https://ui.adsabs.harvard.edu/abs/2025MNRAS.536.2973P}
}

@ARTICLE{vanBaal2024,
       author = {{van Baal}, Bart F.~A. and {Jerkstrand}, Anders and {Wongwathanarat}, Annop and {Janka}, Hans-Thomas},
        title = "{Diagnostics of 3D explosion asymmetries of stripped-envelope supernovae by nebular line profiles}",
      journal = {\mnras},
         year = 2024,
        month = aug,
       volume = {532},
       number = {4},
        pages = {4106-4131},
          doi = {10.1093/mnras/stae1603},
archivePrefix = {arXiv},
       eprint = {2404.01763},
 primaryClass = {astro-ph.HE},
       adsurl = {https://ui.adsabs.harvard.edu/abs/2024MNRAS.532.4106V}
}

@ARTICLE{Dessart2021,
       author = {{Dessart}, L. and {Hillier}, D.~J. and {Sukhbold}, T. and {Woosley}, S.~E. and {Janka}, H. -T.},
        title = "{Nebular phase properties of supernova Ibc from He-star explosions}",
      journal = {\aap},
         year = 2021,
        month = nov,
       volume = {656},
          eid = {A61},
        pages = {A61},
          doi = {10.1051/0004-6361/202141927},
archivePrefix = {arXiv},
       eprint = {2109.12350},
 primaryClass = {astro-ph.SR},
       adsurl = {https://ui.adsabs.harvard.edu/abs/2021A&A...656A..61D}
}

@ARTICLE{Fang2022,
       author = {{Fang}, Qiliang and {Maeda}, Keiichi and {Kuncarayakti}, Hanindyo and {Tanaka}, Masaomi and {Kawabata}, Koji S. and {Hattori}, Takashi and {Aoki}, Kentaro and {Moriya}, Takashi J. and {Yamanaka}, Masayuki},
        title = "{Statistical Properties of the Nebular Spectra of 103 Stripped-envelope Core-collapse Supernovae}",
      journal = {\apj},
         year = 2022,
        month = apr,
       volume = {928},
       number = {2},
          eid = {151},
        pages = {151},
          doi = {10.3847/1538-4357/ac4f60},
archivePrefix = {arXiv},
       eprint = {2201.11467},
 primaryClass = {astro-ph.HE},
       adsurl = {https://ui.adsabs.harvard.edu/abs/2022ApJ...928..151F}
}

@ARTICLE{Barmentloo2024,
       author = {{Barmentloo}, Stan and {Jerkstrand}, Anders and {Iwamoto}, Koichi and {Hachisu}, Izumi and {Nomoto}, Ken'ichi and {Sollerman}, Jesper and {Woosley}, Stan},
        title = "{Nebular nitrogen line emission in stripped-envelope supernovae - a new progenitor mass diagnostic}",
      journal = {\mnras},
         year = 2024,
        month = sep,
       volume = {533},
       number = {2},
        pages = {1251-1280},
          doi = {10.1093/mnras/stae1811},
archivePrefix = {arXiv},
       eprint = {2403.08911},
 primaryClass = {astro-ph.HE},
       adsurl = {https://ui.adsabs.harvard.edu/abs/2024MNRAS.533.1251B}
}

@ARTICLE{Gillanders2024,
       author = {{Gillanders}, J.~H. and {Sim}, S.~A. and {Smartt}, S.~J. and {Goriely}, S. and {Bauswein}, A.},
        title = "{Modelling the spectra of the kilonova AT2017gfo - II. Beyond the photospheric epochs}",
      journal = {\mnras},
         year = 2024,
        month = apr,
       volume = {529},
       number = {3},
        pages = {2918-2945},
          doi = {10.1093/mnras/stad3688},
archivePrefix = {arXiv},
       eprint = {2306.15055},
 primaryClass = {astro-ph.HE}
}

@BOOK{Mihalas1978,
       author = {{Mihalas}, Dimitri},
        title = "{Stellar atmospheres}",
         year = 1978,
        edition = {2nd},
      address = {San Francisco},
    publisher = {W.H. Freeman},
       adsurl = {https://ui.adsabs.harvard.edu/abs/1978stat.book.....M}
}

@ARTICLE{Riess1998,
       author = {{Riess}, Adam G. and {Filippenko}, Alexei V. and {Challis}, Peter and {Clocchiatti}, Alejandro and {Diercks}, Alan and {Garnavich}, Peter M. and {Gilliland}, Ron L. and {Hogan}, Craig J. and {Jha}, Saurabh and {Kirshner}, Robert P. and {Leibundgut}, B. and {Phillips}, M.~M. and {Reiss}, David and {Schmidt}, Brian P. and {Schommer}, Robert A. and {Smith}, R. Chris and {Spyromilio}, J. and {Stubbs}, Christopher and {Suntzeff}, Nicholas B. and {Tonry}, John},
        title = "{Observational Evidence from Supernovae for an Accelerating Universe and a Cosmological Constant}",
      journal = {\aj},
         year = 1998,
        month = sep,
       volume = {116},
       number = {3},
        pages = {1009-1038},
          doi = {10.1086/300499},
archivePrefix = {arXiv}
}

@ARTICLE{Phillips1993,
       author = {{Phillips}, M.~M.},
        title = "{The Absolute Magnitudes of Type Ia Supernovae}",
      journal = {\apjl},
         year = 1993,
        month = aug,
       volume = {413},
        pages = {L105},
          doi = {10.1086/186970},
       adsurl = {https://ui.adsabs.harvard.edu/abs/1993ApJ...413L.105P}
}

@INCOLLECTION{Taubenberger2017,
       author = {{Taubenberger}, Stefan},
        title = "{The Extremes of Thermonuclear Supernovae}",
    booktitle = {Handbook of Supernovae},
         year = 2017,
       editor = {{Alsabti}, Athem W. and {Murdin}, Paul},
        pages = {317},
          doi = {10.1007/978-3-319-21846-5_37},
       adsurl = {https://ui.adsabs.harvard.edu/abs/2017hsn..book..317T}
}

@ARTICLE{Filippenko1997,
       author = {{Filippenko}, Alexei V.},
        title = "{Optical Spectra of Supernovae}",
      journal = {\araa},
         year = 1997,
        month = jan,
       volume = {35},
        pages = {309-355},
          doi = {10.1146/annurev.astro.35.1.309},
       adsurl = {https://ui.adsabs.harvard.edu/abs/1997ARA&A..35..309F}
}

@ARTICLE{Chugai1987,
       author = {{Chugai}, N.~N.},
        title = "{Scattering of L-alpha photons in an infinite expanding medium when there is absorption in the continuum}",
      journal = {Astrofizika},
         year = 1987,
        month = jul,
       volume = {26},
        pages = {89-96},
       adsurl = {https://ui.adsabs.harvard.edu/abs/1987Afz....26...89C}
}

@ARTICLE{Hummer1985,
       author = {{Hummer}, D.~G. and {Rybicki}, G.~B.},
        title = "{The Sobolev approximation for line formation with continuous opacity}",
      journal = {\apj},
         year = 1985,
        month = jun,
       volume = {293},
        pages = {258-267},
          doi = {10.1086/163232},
       adsurl = {https://ui.adsabs.harvard.edu/abs/1985ApJ...293..258H}
}

@ARTICLE{Sobolev1957,
       author = {{Sobolev}, V.~V.},
        title = "{The Diffusion of L{\ensuremath{\alpha}} Radiation in Nebulae and Stellar Envelopes.}",
      journal = {\sovast},
         year = 1957,
        month = oct,
       volume = {1},
        pages = {678},
       adsurl = {https://ui.adsabs.harvard.edu/abs/1957SvA.....1..678S}
}

@ARTICLE{Castor1970,
       author = {{Castor}, J.~I.},
        title = "{Spectral line formation in Wolf-Rayet envelopes.}",
      journal = {\mnras},
         year = 1970,
        month = jan,
       volume = {149},
        pages = {111},
          doi = {10.1093/mnras/149.2.111},
       adsurl = {https://ui.adsabs.harvard.edu/abs/1970MNRAS.149..111C}
}

@ARTICLE{Lucy2002,
       author = {{Lucy}, L.~B.},
        title = "{Monte Carlo transition probabilities}",
      journal = {\aap},
         year = 2002,
        month = mar,
       volume = {384},
        pages = {725-735},
          doi = {10.1051/0004-6361:20011756},
archivePrefix = {arXiv},
       eprint = {astro-ph/0107377},
 primaryClass = {astro-ph},
       adsurl = {https://ui.adsabs.harvard.edu/abs/2002A&A...384..725L}
}

@ARTICLE{Kozyreva2020,
       author = {{Kozyreva}, Alexandra and {Shingles}, Luke and {Mironov}, Alexey and {Baklanov}, Petr and {Blinnikov}, Sergey},
        title = "{The influence of line opacity treatment in STELLA on supernova light curves}",
      journal = {\mnras},
         year = 2020,
        month = dec,
       volume = {499},
       number = {3},
        pages = {4312-4324},
          doi = {10.1093/mnras/staa2704},
archivePrefix = {arXiv},
       eprint = {2009.01566},
 primaryClass = {astro-ph.HE},
       adsurl = {https://ui.adsabs.harvard.edu/abs/2020MNRAS.499.4312K}
}

@ARTICLE{Mazzali2001,
       author = {{Mazzali}, Paolo A. and {Nomoto}, Ken'ichi and {Patat}, Ferdinando and {Maeda}, Keiichi},
        title = "{The Nebular Spectra of the Hypernova SN 1998bw and Evidence for Asymmetry}",
      journal = {\apj},
         year = 2001,
        month = oct,
       volume = {559},
       number = {2},
        pages = {1047-1053},
          doi = {10.1086/322420},
archivePrefix = {arXiv},
       eprint = {astro-ph/0106095},
 primaryClass = {astro-ph},
       adsurl = {https://ui.adsabs.harvard.edu/abs/2001ApJ...559.1047M}
}

@ARTICLE{Hoflich1996,
       author = {{Hoeflich}, P. and {Khokhlov}, A.},
        title = "{Explosion Models for Type Ia Supernovae: A Comparison with Observed Light Curves, Distances, H 0, and Q 0}",
      journal = {\apj},
         year = 1996,
        month = feb,
       volume = {457},
        pages = {500},
          doi = {10.1086/176748},
archivePrefix = {arXiv},
       eprint = {astro-ph/9602025},
 primaryClass = {astro-ph},
       adsurl = {https://ui.adsabs.harvard.edu/abs/1996ApJ...457..500H}
}

@ARTICLE{Pinto2000,
       author = {{Pinto}, Philip A. and {Eastman}, Ronald G.},
        title = "{The Physics of Type Ia Supernova Light Curves. I. Analytic Results and Time Dependence}",
      journal = {\apj},
         year = 2000,
        month = feb,
       volume = {530},
       number = {2},
        pages = {744-756},
          doi = {10.1086/308376},
       adsurl = {https://ui.adsabs.harvard.edu/abs/2000ApJ...530..744P}
}

@ARTICLE{Hauschildt1999,
       author = {{Hauschildt}, P.~H. and {Baron}, E.},
        title = "{Numerical solution of the expanding stellar atmosphere problem}",
      journal = {J. Comput. Appl. Math.},
         year = 1999,
        month = sep,
       volume = {109},
       number = {1},
        pages = {41-63},
          doi = {10.48550/arXiv.astro-ph/9808182},
archivePrefix = {arXiv},
       eprint = {astro-ph/9808182},
 primaryClass = {astro-ph},
       adsurl = {https://ui.adsabs.harvard.edu/abs/1999JCoAM.109...41H}
}

@ARTICLE{Bulla2019,
       author = {{Bulla}, M.},
        title = "{POSSIS: predicting spectra, light curves, and polarization for multidimensional models of supernovae and kilonovae}",
      journal = {\mnras},
         year = 2019,
        month = nov,
       volume = {489},
       number = {4},
        pages = {5037-5045},
          doi = {10.1093/mnras/stz2495},
archivePrefix = {arXiv},
       eprint = {1906.04205},
 primaryClass = {astro-ph.HE},
       adsurl = {https://ui.adsabs.harvard.edu/abs/2019MNRAS.489.5037B}
}

@ARTICLE{Baron1996,
       author = {{Baron}, E. and {Hauschildt}, P.~H. and {Nugent}, P. and {Branch}, D.},
        title = "{Non-local thermodynamic equilibrium effects in modelling of supernovae near maximum light.}",
      journal = {\mnras},
         year = 1996,
        month = nov,
       volume = {283},
       number = {1},
        pages = {297-315},
          doi = {10.1093/mnras/283.1.297},
       adsurl = {https://ui.adsabs.harvard.edu/abs/1996MNRAS.283..297B}
}

@ARTICLE{Wygoda2019,
       author = {{Wygoda}, Nahliel and {Elbaz}, Yonatan and {Katz}, Boaz},
        title = "{Type Ia supernovae have two physical width-luminosity relations and they favour sub-Chandrasekhar and direct collision models - I. Bolometric}",
      journal = {\mnras},
         year = 2019,
        month = apr,
       volume = {484},
       number = {3},
        pages = {3941-3950},
          doi = {10.1093/mnras/stz145},
archivePrefix = {arXiv},
       eprint = {1711.00969},
 primaryClass = {astro-ph.HE},
       adsurl = {https://ui.adsabs.harvard.edu/abs/2019MNRAS.484.3941W}
}

@ARTICLE{Gillanders2022,
       author = {{Gillanders}, J.~H. and {Smartt}, S.~J. and {Sim}, S.~A. and {Bauswein}, A. and {Goriely}, S.},
        title = "{Modelling the spectra of the kilonova AT2017gfo - I. The photospheric epochs}",
      journal = {\mnras},
         year = 2022,
        month = sep,
       volume = {515},
       number = {1},
        pages = {631-651},
          doi = {10.1093/mnras/stac1258},
archivePrefix = {arXiv},
       eprint = {2202.01786},
 primaryClass = {astro-ph.HE},
       adsurl = {https://ui.adsabs.harvard.edu/abs/2022MNRAS.515..631G}
}

@ARTICLE{Shen2021,
       author = {{Shen}, Ken J. and {Blondin}, St{\'e}phane and {Kasen}, Daniel and {Dessart}, Luc and {Townsley}, Dean M. and {Boos}, Samuel and {Hillier}, D. John},
        title = "{Non-local Thermodynamic Equilibrium Radiative Transfer Simulations of Sub-Chandrasekhar-mass White Dwarf Detonations}",
      journal = {\apjl},
         year = 2021,
        month = mar,
       volume = {909},
       number = {2},
          eid = {L18},
        pages = {L18},
          doi = {10.3847/2041-8213/abe69b},
archivePrefix = {arXiv},
       eprint = {2102.08238},
 primaryClass = {astro-ph.HE},
       adsurl = {https://ui.adsabs.harvard.edu/abs/2021ApJ...909L..18S}
}

@ARTICLE{Kasen2007,
       author = {{Kasen}, Daniel and {Plewa}, Tomasz},
        title = "{Detonating Failed Deflagration Model of Thermonuclear Supernovae. II. Comparison to Observations}",
      journal = {\apj},
         year = 2007,
        month = jun,
       volume = {662},
       number = {1},
        pages = {459-471},
          doi = {10.1086/516834},
archivePrefix = {arXiv},
       eprint = {astro-ph/0612198},
 primaryClass = {astro-ph},
       adsurl = {https://ui.adsabs.harvard.edu/abs/2007ApJ...662..459K}
}

@ARTICLE{Blondin2013,
       author = {{Blondin}, St{\'e}phane and {Dessart}, Luc and {Hillier}, D. John and {Khokhlov}, Alexei M.},
        title = "{One-dimensional delayed-detonation models of Type Ia supernovae: confrontation to observations at bolometric maximum}",
      journal = {\mnras},
         year = 2013,
        month = mar,
       volume = {429},
       number = {3},
        pages = {2127-2142},
          doi = {10.1093/mnras/sts484},
archivePrefix = {arXiv},
       eprint = {1211.5892},
 primaryClass = {astro-ph.SR},
       adsurl = {https://ui.adsabs.harvard.edu/abs/2013MNRAS.429.2127B}
}

@INPROCEEDINGS{McCray1996,
       author = {{McCray}, R.},
        title = "{Inferring Abundances from the Spectra of Supernovae}",
    booktitle = {Cosmic Abundances},
         year = 1996,
       editor = {{Holt}, Stephen S. and {Sonneborn}, George},
       series = {Astronomical Society of the Pacific Conference Series},
       volume = {99},
        month = jan,
        pages = {273},
       adsurl = {https://ui.adsabs.harvard.edu/abs/1996ASPC...99..273M}
}

@ARTICLE{Maurer2011,
       author = {{Maurer}, I. and {Jerkstrand}, A. and {Mazzali}, P.~A. and {Taubenberger}, S. and {Hachinger}, S. and {Kromer}, M. and {Sim}, S. and {Hillebrandt}, W.},
        title = "{NERO- a post-maximum supernova radiation transport code}",
      journal = {\mnras},
         year = 2011,
        month = dec,
       volume = {418},
       number = {3},
        pages = {1517-1525},
          doi = {10.1111/j.1365-2966.2011.19376.x},
archivePrefix = {arXiv},
       eprint = {1105.3049},
 primaryClass = {astro-ph.HE},
       adsurl = {https://ui.adsabs.harvard.edu/abs/2011MNRAS.418.1517M}
}

@BOOK{Mihalas1970,
       author = {{Mihalas}, Dimitri},
        title = "{Stellar atmospheres}",
      address = {San Francisco},
    publisher = {W.H. Freeman},
         year = 1970,
       adsurl = {https://ui.adsabs.harvard.edu/abs/1970stat.book.....M}
}

@ARTICLE{Lucy1970,
       author = {{Lucy}, L.~B. and {Solomon}, P.~M.},
        title = "{Mass Loss by Hot Stars}",
      journal = {\apj},
         year = 1970,
        month = mar,
       volume = {159},
        pages = {879},
          doi = {10.1086/150365},
       adsurl = {https://ui.adsabs.harvard.edu/abs/1970ApJ...159..879L}
}

@ARTICLE{Blondin2023,
       author = {{Blondin}, S. and {Dessart}, L. and {Hillier}, D.~J. and {Ramsbottom}, C.~A. and {Storey}, P.~J.},
        title = "{Nebular spectra from Type Ia supernova explosion models compared to JWST observations of SN 2021aefx}",
      journal = {\aap},
         year = 2023,
        month = oct,
       volume = {678},
          eid = {A170},
        pages = {A170},
          doi = {10.1051/0004-6361/202347147},
archivePrefix = {arXiv},
       eprint = {2306.07116},
 primaryClass = {astro-ph.HE},
       adsurl = {https://ui.adsabs.harvard.edu/abs/2023A&A...678A.170B}
}

@ARTICLE{Anderson1989,
       author = {{Anderson}, Lawrence S.},
        title = "{Line Blanketing without Local Thermodynamic Equilibrium. II. A Solar-Type Model in Radiative Equilibrium}",
      journal = {\apj},
         year = 1989,
        month = apr,
       volume = {339},
        pages = {558},
          doi = {10.1086/167317},
       adsurl = {https://ui.adsabs.harvard.edu/abs/1989ApJ...339..558A}
}

@ARTICLE{Botyanszki2018,
       author = {{Boty{\'a}nszki}, J{\'a}nos and {Kasen}, Daniel and {Plewa}, Tomasz},
        title = "{Multidimensional Models of Type Ia Supernova Nebular Spectra: Strong Emission Lines from Stripped Companion Gas Rule Out Classic Single-degenerate Systems}",
      journal = {\apjl},
         year = 2018,
        month = jan,
       volume = {852},
       number = {1},
          eid = {L6},
        pages = {L6},
          doi = {10.3847/2041-8213/aaa07b},
archivePrefix = {arXiv},
       eprint = {1712.03274},
 primaryClass = {astro-ph.SR},
       adsurl = {https://ui.adsabs.harvard.edu/abs/2018ApJ...852L...6B}
}

@ARTICLE{Botyanszki2017,
       author = {{Boty{\'a}nszki}, J{\'a}nos and {Kasen}, Daniel},
        title = "{How Do Type Ia Supernova Nebular Spectra Depend on Explosion Properties? Insights from Systematic Non-LTE Modeling}",
      journal = {\apj},
         year = 2017,
        month = aug,
       volume = {845},
       number = {2},
          eid = {176},
        pages = {176},
          doi = {10.3847/1538-4357/aa81d8},
archivePrefix = {arXiv},
       eprint = {1704.06275},
 primaryClass = {astro-ph.HE},
       adsurl = {https://ui.adsabs.harvard.edu/abs/2017ApJ...845..176B}
}

@ARTICLE{Hillier1989,
       author = {{Hillier}, D.~J.},
        title = "{WC Stars: Hot Stars with Cold Winds}",
      journal = {\apj},
         year = 1989,
        month = dec,
       volume = {347},
        pages = {392},
          doi = {10.1086/168127},
       adsurl = {https://ui.adsabs.harvard.edu/abs/1989ApJ...347..392H}
}

@ARTICLE{Hillier1983,
       author = {{Hillier}, D.~J. and {Jones}, T.~J. and {Hyland}, A.~R.},
        title = "{Infrared spectra of WN stars. I. HD 50896.}",
      journal = {\apj},
         year = 1983,
        month = aug,
       volume = {271},
        pages = {221-236},
          doi = {10.1086/161189},
       adsurl = {https://ui.adsabs.harvard.edu/abs/1983ApJ...271..221H}
}

@ARTICLE{Rho2021,
       author = {{Rho}, J. and {Evans}, A. and {Geballe}, T.~R. and {Banerjee}, D.~P.~K. and {Hoeflich}, P. and {Shahbandeh}, M. and {Valenti}, S. and {Yoon}, S. -C. and {Jin}, H. and {Williamson}, M. and {Modjaz}, M. and {Hiramatsu}, D. and {Howell}, D.~A. and {Pellegrino}, C. and {Vink{\'o}}, J. and {Cartier}, R. and {Burke}, J. and {McCully}, C. and {An}, H. and {Cha}, H. and {Pritchard}, T. and {Wang}, X. and {Andrews}, J. and {Galbany}, L. and {Van Dyk}, S. and {Graham}, M.~L. and {Blinnikov}, S. and {Joshi}, V. and {P{\'a}l}, A. and {Kriskovics}, L. and {Ordasi}, A. and {Szakats}, R. and {Vida}, K. and {Chen}, Z. and {Li}, X. and {Zhang}, J. and {Yan}, S.},
        title = "{Near-infrared and Optical Observations of Type Ic SN 2020oi and Broad-lined Type Ic SN 2020bvc: Carbon Monoxide, Dust, and High-velocity Supernova Ejecta}",
      journal = {\apj},
         year = 2021,
        month = feb,
       volume = {908},
       number = {2},
          eid = {232},
        pages = {232},
          doi = {10.3847/1538-4357/abd850},
archivePrefix = {arXiv},
       eprint = {2010.00662},
 primaryClass = {astro-ph.SR},
       adsurl = {https://ui.adsabs.harvard.edu/abs/2021ApJ...908..232R}
}

@ARTICLE{Liu1995,
       author = {{Liu}, Weihong and {Dalgarno}, A.},
        title = "{The Oxygen Temperature of SN 1987A}",
      journal = {\apj},
         year = 1995,
        month = nov,
       volume = {454},
        pages = {472},
          doi = {10.1086/176498},
       adsurl = {https://ui.adsabs.harvard.edu/abs/1995ApJ...454..472L}
}

@ARTICLE{Liu1992,
       author = {{Liu}, Weihong and {Dalgarno}, A. and {Lepp}, S.},
        title = "{Carbon Monoxide in SN 1987A}",
      journal = {\apj},
         year = 1992,
        month = sep,
       volume = {396},
        pages = {679},
          doi = {10.1086/171749},
       adsurl = {https://ui.adsabs.harvard.edu/abs/1992ApJ...396..679L}
}

@ARTICLE{Utrobin2005,
       author = {{Utrobin}, V.~P. and {Chugai}, N.~N.},
        title = "{Strong effects of time-dependent ionization in early SN 1987A}",
      journal = {\aap},
         year = 2005,
        month = oct,
       volume = {441},
       number = {1},
        pages = {271-281},
          doi = {10.1051/0004-6361:20042599},
archivePrefix = {arXiv},
       eprint = {astro-ph/0501036},
 primaryClass = {astro-ph},
       adsurl = {https://ui.adsabs.harvard.edu/abs/2005A&A...441..271U}
}

@ARTICLE{Dessart2008,
       author = {{Dessart}, Luc and {Hillier}, D. John},
        title = "{Time-dependent effects in photospheric-phase Type II supernova spectra}",
      journal = {\mnras},
         year = 2008,
        month = jan,
       volume = {383},
       number = {1},
        pages = {57-74},
          doi = {10.1111/j.1365-2966.2007.12538.x},
archivePrefix = {arXiv},
       eprint = {0710.0784},
 primaryClass = {astro-ph},
       adsurl = {https://ui.adsabs.harvard.edu/abs/2008MNRAS.383...57D}
}

@ARTICLE{Fransson1993,
       author = {{Fransson}, Claes and {Kozma}, Cecilia},
        title = "{The Freeze-out Phase of SN 1987A: Implications for the Light Curve}",
      journal = {\apjl},
         year = 1993,
        month = may,
       volume = {408},
        pages = {L25},
          doi = {10.1086/186822},
       adsurl = {https://ui.adsabs.harvard.edu/abs/1993ApJ...408L..25F}
}

@ARTICLE{Dalgarno1999,
       author = {{Dalgarno}, A. and {Yan}, Min and {Liu}, Weihong},
        title = "{Electron Energy Deposition in a Gas Mixture of Atomic and Molecular Hydrogen and Helium}",
      journal = {\apjs},
         year = 1999,
        month = nov,
       volume = {125},
       number = {1},
        pages = {237-256},
          doi = {10.1086/313267},
       adsurl = {https://ui.adsabs.harvard.edu/abs/1999ApJS..125..237D}
}

@ARTICLE{Waxman2019,
       author = {{Waxman}, Eli and {Ofek}, Eran O. and {Kushnir}, Doron},
        title = "{Late-time Kilonova Light Curves and Implications to GW170817}",
      journal = {\apj},
         year = 2019,
        month = jun,
       volume = {878},
       number = {2},
          eid = {93},
        pages = {93},
          doi = {10.3847/1538-4357/ab1f71},
archivePrefix = {arXiv},
       eprint = {1902.01197},
 primaryClass = {astro-ph.HE},
       adsurl = {https://ui.adsabs.harvard.edu/abs/2019ApJ...878...93W}
}

@ARTICLE{Hotokezaka2020,
       author = {{Hotokezaka}, Kenta and {Nakar}, Ehud},
        title = "{Radioactive Heating Rate of r-process Elements and Macronova Light Curve}",
      journal = {\apj},
         year = 2020,
        month = mar,
       volume = {891},
       number = {2},
          eid = {152},
        pages = {152},
          doi = {10.3847/1538-4357/ab6a98},
archivePrefix = {arXiv},
       eprint = {1909.02581},
 primaryClass = {astro-ph.HE},
       adsurl = {https://ui.adsabs.harvard.edu/abs/2020ApJ...891..152H}
}

@ARTICLE{Kasen2019,
       author = {{Kasen}, Daniel and {Barnes}, Jennifer},
        title = "{Radioactive Heating and Late Time Kilonova Light Curves}",
      journal = {\apj},
         year = 2019,
        month = may,
       volume = {876},
       number = {2},
          eid = {128},
        pages = {128},
          doi = {10.3847/1538-4357/ab06c2},
archivePrefix = {arXiv},
       eprint = {1807.03319},
 primaryClass = {astro-ph.HE},
       adsurl = {https://ui.adsabs.harvard.edu/abs/2019ApJ...876..128K}
}

@ARTICLE{Waxman2018,
       author = {{Waxman}, Eli and {Ofek}, Eran O. and {Kushnir}, Doron and {Gal-Yam}, Avishay},
        title = "{Constraints on the ejecta of the GW170817 neutron star merger from its electromagnetic emission}",
      journal = {\mnras},
         year = 2018,
        month = dec,
       volume = {481},
       number = {3},
        pages = {3423-3441},
          doi = {10.1093/mnras/sty2441},
archivePrefix = {arXiv},
       eprint = {1711.09638},
 primaryClass = {astro-ph.HE},
       adsurl = {https://ui.adsabs.harvard.edu/abs/2018MNRAS.481.3423W}
}

@ARTICLE{Seltzer1986,
       author = {{Seltzer}, S.~M. and {Berger}, M.~J.},
        title = "{Bremsstrahlung Energy Spectra from Electrons with Kinetic Energy 1 keV-10 GeV Incident on Screened Nuclei and Orbital Electrons of Neutral Atoms with Z = 1-100}",
      journal = {Atomic Data and Nuclear Data Tables},
         year = 1986,
        month = jan,
       volume = {35},
        pages = {345},
          doi = {10.1016/0092-640X(86)90014-8},
       adsurl = {https://ui.adsabs.harvard.edu/abs/1986ADNDT..35..345S}
}

@BOOK{Heath2005,
author = {{Heath}, M.},
title = "{Scientific computing: An introductory survey}",
year = 2002,
IGNOREnote = {http://books.google.de/books?id=gwBrMAEACAAJ},
publisher = {McGraw-Hill},
address = {New York}
}

@BOOK{Mott1949,
       author = {{Mott}, N.~F. and {Massey}, Harrie Stewart Wilson},
        title = "{The theory of atomic collisions}",
        address = {Oxford},
        publisher = {Clarendon Press},
         year = 1949,
       adsurl = {https://ui.adsabs.harvard.edu/abs/1949tac..book.....M}
}

@ARTICLE{vanRegemorter1962,
       author = {{van Regemorter}, Henri},
        title = "{Rate of Collisional Excitation in Stellar Atmospheres}",
      journal = {\apj},
         year = 1962,
        month = nov,
       volume = {136},
        pages = {906},
          doi = {10.1086/147445},
       adsurl = {https://ui.adsabs.harvard.edu/abs/1962ApJ...136..906V}
}

@ARTICLE{Lucy1991,
       author = {{Lucy}, L.~B.},
        title = "{Nonthermal Excitation of Helium in Type Ib Supernovae}",
      journal = {\apj},
         year = 1991,
        month = dec,
       volume = {383},
        pages = {308},
          doi = {10.1086/170787},
       adsurl = {https://ui.adsabs.harvard.edu/abs/1991ApJ...383..308L}
}

@ARTICLE{Manson1975,
       author = {{Manson}, Steven T. and {Toburen}, L.~H. and {Madison}, D.~H. and {Stolterfoht}, N.},
        title = "{Energy and angular distribution of electrons ejected from helium by fast protons and electrons: Theory and experiment}",
      journal = {\pra},
         year = 1975,
        month = jul,
       volume = {12},
       number = {1},
        pages = {60-79},
          doi = {10.1103/PhysRevA.12.60},
       adsurl = {https://ui.adsabs.harvard.edu/abs/1975PhRvA..12...60M}
}

@ARTICLE{Lotz1967,
       author = {{Lotz}, Wolfgang},
        title = "{Electron-Impact Ionization Cross-Sections and Ionization Rate Coefficients for Atoms and Ions}",
      journal = {\apjs},
         year = 1967,
        month = may,
       volume = {14},
        pages = {207},
          doi = {10.1086/190154},
       adsurl = {https://ui.adsabs.harvard.edu/abs/1967ApJS...14..207L}
}

@ARTICLE{Meyerott1978,
       author = {{Meyerott}, R.~E.},
        title = "{On the interpretation of the spectra of type I supernovae.}",
      journal = {\apj},
         year = 1978,
        month = may,
       volume = {221},
        pages = {975-989},
          doi = {10.1086/156103},
       adsurl = {https://ui.adsabs.harvard.edu/abs/1978ApJ...221..975M}
}

@ARTICLE{Spencer1954,
       author = {{Spencer}, L.~V. and {Fano}, U.},
        title = "{Energy Spectrum Resulting from Electron Slowing Down}",
      journal = {Physical Review},
         year = 1954,
        month = mar,
       volume = {93},
       number = {6},
        pages = {1172-1181},
          doi = {10.1103/PhysRev.93.1172},
       adsurl = {https://ui.adsabs.harvard.edu/abs/1954PhRv...93.1172S}
}

@ARTICLE{Fano1963,
       author = {{Fano}, U.},
        title = "{Penetration of Protons, Alpha Particles, and Mesons}",
      journal = {Annu. Rev. Nucl. Part. Sci.},
         year = 1963,
        month = jan,
       volume = {13},
        pages = {1-66},
          doi = {10.1146/annurev.ns.13.120163.000245},
       adsurl = {https://ui.adsabs.harvard.edu/abs/1963ARNPS..13....1F}
}

@BOOK{Spitzer1962,
       author = {{Spitzer}, L.},
        title = "{Physics of Fully Ionized Gases}",
        address = {New York},
        publisher = {Interscience},
         year = 1962,
       adsurl = {https://ui.adsabs.harvard.edu/abs/1962pfig.book.....S}
}

@ARTICLE{Spitzer1969,
       author = {{Spitzer}, Lyman, Jr. and {Scott}, Eugene H.},
        title = "{Heating of H i Regions by Energetic Particles. II. Interaction Between Secondaries and Thermal Electrons}",
      journal = {\apj},
         year = 1969,
        month = oct,
       volume = {158},
        pages = {161},
          doi = {10.1086/150180},
       adsurl = {https://ui.adsabs.harvard.edu/abs/1969ApJ...158..161S}
}

@ARTICLE{Inokuti1978,
       author = {{Inokuti}, Mitio and {Itikawa}, Yukikazu and {Turner}, James E.},
        title = "{Addenda: Inelastic collisions of fast charged particles with atoms and molecules{\textemdash}The Bethe theory revisited}",
      journal = {Reviews of Modern Physics},
         year = 1978,
        month = jan,
       volume = {50},
       number = {1},
        pages = {23-35},
          doi = {10.1103/RevModPhys.50.23},
       adsurl = {https://ui.adsabs.harvard.edu/abs/1978RvMP...50...23I}
}

@ARTICLE{Schunk1971,
       author = {{Schunk}, Robert W. and {Hays}, Paul B.},
        title = "{Photoelectron energy losses to thermal electrons}",
      journal = {\planss},
         year = 1971,
        month = jan,
       volume = {19},
       number = {1},
        pages = {113-117},
          doi = {10.1016/0032-0633(71)90071-7},
       adsurl = {https://ui.adsabs.harvard.edu/abs/1971P&SS...19..113S}
}

@ARTICLE{Barnes2016,
       author = {{Barnes}, Jennifer and {Kasen}, Daniel and {Wu}, Meng-Ru and {Mart{\'\i}nez-Pinedo}, Gabriel},
        title = "{Radioactivity and Thermalization in the Ejecta of Compact Object Mergers and Their Impact on Kilonova Light Curves}",
      journal = {\apj},
         year = 2016,
        month = oct,
       volume = {829},
       number = {2},
          eid = {110},
        pages = {110},
          doi = {10.3847/0004-637X/829/2/110},
archivePrefix = {arXiv},
       eprint = {1605.07218},
 primaryClass = {astro-ph.HE},
       adsurl = {https://ui.adsabs.harvard.edu/abs/2016ApJ...829..110B}
}

@BOOK{Press1992,
       author = {{Press}, William H. and {Teukolsky}, Saul A. and {Vetterling}, William T. and {Flannery}, Brian P.},
        title = "{Numerical recipes in FORTRAN. The art of scientific computing}",
        publisher = {Cambridge University Press},
         year = 1992,
       adsurl = {https://ui.adsabs.harvard.edu/abs/1992nrfa.book.....P}
}

@ARTICLE{Shingles2020,
       author = {{Shingles}, L.~J. and {Sim}, S.~A. and {Kromer}, M. and {Maguire}, K. and {Bulla}, M. and {Collins}, C. and {Ballance}, C.~P. and {Michel}, A.~S. and {Ramsbottom}, C.~A. and {R{\"o}pke}, F.~K. and {Seitenzahl}, I.~R. and {Tyndall}, N.~B.},
        title = "{Monte Carlo radiative transfer for the nebular phase of Type Ia supernovae}",
      journal = {\mnras},
         year = 2020,
        month = feb,
       volume = {492},
       number = {2},
        pages = {2029-2043},
          doi = {10.1093/mnras/stz3412},
archivePrefix = {arXiv},
       eprint = {1912.02214},
 primaryClass = {astro-ph.HE},
       adsurl = {https://ui.adsabs.harvard.edu/abs/2020MNRAS.492.2029S}
}

@ARTICLE{Hillier2012,
       author = {{Hillier}, D. John and {Dessart}, Luc},
        title = "{Time-dependent radiative transfer calculations for supernovae}",
      journal = {\mnras},
         year = 2012,
        month = jul,
       volume = {424},
       number = {1},
        pages = {252-271},
          doi = {10.1111/j.1365-2966.2012.21192.x},
archivePrefix = {arXiv},
       eprint = {1204.0527},
 primaryClass = {astro-ph.SR},
       adsurl = {https://ui.adsabs.harvard.edu/abs/2012MNRAS.424..252H}
}

@ARTICLE{Bersten2011,
       author = {{Bersten}, Melina C. and {Benvenuto}, Omar and {Hamuy}, Mario},
        title = "{Hydrodynamical Models of Type II Plateau Supernovae}",
      journal = {\apj},
         year = 2011,
        month = mar,
       volume = {729},
       number = {1},
          eid = {61},
        pages = {61},
          doi = {10.1088/0004-637X/729/1/61},
archivePrefix = {arXiv},
       eprint = {1101.0467},
 primaryClass = {astro-ph.SR},
       adsurl = {https://ui.adsabs.harvard.edu/abs/2011ApJ...729...61B}
}

@ARTICLE{Utrobin2004,
       author = {{Utrobin}, V.~P.},
        title = "{The Light Curve of Supernova 1987A: The Structure of the Presupernova and Radioactive Nickel Mixing}",
      journal = {Astron. Lett.},
         year = 2004,
        month = may,
       volume = {30},
        pages = {293-308},
          doi = {10.1134/1.1738152},
archivePrefix = {arXiv},
       eprint = {astro-ph/0406410},
 primaryClass = {astro-ph},
       adsurl = {https://ui.adsabs.harvard.edu/abs/2004AstL...30..293U}
}

@ARTICLE{Blinnikov1998,
       author = {{Blinnikov}, S.~I. and {Eastman}, R. and {Bartunov}, O.~S. and {Popolitov}, V.~A. and {Woosley}, S.~E.},
        title = "{A Comparative Modeling of Supernova 1993J}",
      journal = {\apj},
         year = 1998,
        month = mar,
       volume = {496},
       number = {1},
        pages = {454-472},
          doi = {10.1086/305375},
archivePrefix = {arXiv},
       eprint = {astro-ph/9711055},
 primaryClass = {astro-ph},
       adsurl = {https://ui.adsabs.harvard.edu/abs/1998ApJ...496..454B}
}

@ARTICLE{Morozova2015,
       author = {{Morozova}, Viktoriya and {Piro}, Anthony L. and {Renzo}, Mathieu and {Ott}, Christian D. and {Clausen}, Drew and {Couch}, Sean M. and {Ellis}, Justin and {Roberts}, Luke F.},
        title = "{Light Curves of Core-collapse Supernovae with Substantial Mass Loss Using the New Open-source SuperNova Explosion Code (SNEC)}",
      journal = {\apj},
         year = 2015,
        month = nov,
       volume = {814},
       number = {1},
          eid = {63},
        pages = {63},
          doi = {10.1088/0004-637X/814/1/63},
archivePrefix = {arXiv},
       eprint = {1505.06746},
 primaryClass = {astro-ph.HE},
       adsurl = {https://ui.adsabs.harvard.edu/abs/2015ApJ...814...63M}
}

@ARTICLE{Shingles2023,
       author = {{Shingles}, Luke J. and {Collins}, Christine E. and {Vijayan}, Vimal and {Fl{\"o}rs}, Andreas and {Just}, Oliver and {Leck}, Gerrit and {Xiong}, Zewei and {Bauswein}, Andreas and {Mart{\'\i}nez-Pinedo}, Gabriel and {Sim}, Stuart A.},
        title = "{Self-consistent 3D Radiative Transfer for Kilonovae: Directional Spectra from Merger Simulations}",
      journal = {\apjl},
         year = 2023,
        month = sep,
       volume = {954},
       number = {2},
          eid = {L41},
        pages = {L41},
          doi = {10.3847/2041-8213/acf29a},
archivePrefix = {arXiv},
       eprint = {2306.17612},
 primaryClass = {astro-ph.HE},
       adsurl = {https://ui.adsabs.harvard.edu/abs/2023ApJ...954L..41S}
}

@INCOLLECTION{Jerkstrand2017handbook,
       author = {{Jerkstrand}, Anders},
        title = "{Spectra of Supernovae in the Nebular Phase}",
    booktitle = {Handbook of Supernovae},
         year = 2017,
       editor = {{Alsabti}, Athem W. and {Murdin}, Paul},
        pages = {795},
          doi = {10.1007/978-3-319-21846-5_29},
       adsurl = {https://ui.adsabs.harvard.edu/abs/2017hsn..book..795J}
}

@ARTICLE{Shibata2019,
       author = {{Shibata}, Masaru and {Hotokezaka}, Kenta},
        title = "{Merger and Mass Ejection of Neutron Star Binaries}",
      journal = {Annu. Rev. Nucl. Part. Sci.},
         year = 2019,
        month = oct,
       volume = {69},
        pages = {41-64},
          doi = {10.1146/annurev-nucl-101918-023625},
archivePrefix = {arXiv},
       eprint = {1908.02350},
 primaryClass = {astro-ph.HE},
       adsurl = {https://ui.adsabs.harvard.edu/abs/2019ARNPS..69...41S}
}

@ARTICLE{Hotokezaka2023,
       author = {{Hotokezaka}, Kenta and {Tanaka}, Masaomi and {Kato}, Daiji and {Gaigalas}, Gediminas},
        title = "{Tellurium emission line in kilonova AT 2017gfo}",
      journal = {\mnras},
         year = 2023,
        month = nov,
       volume = {526},
       number = {1},
        pages = {L155-L159},
          doi = {10.1093/mnrasl/slad128},
archivePrefix = {arXiv},
       eprint = {2307.00988},
 primaryClass = {astro-ph.HE},
       adsurl = {https://ui.adsabs.harvard.edu/abs/2023MNRAS.526L.155H}
}

@ARTICLE{Pognan2023,
       author = {{Pognan}, Quentin and {Grumer}, Jon and {Jerkstrand}, Anders and {Wanajo}, Shinya},
        title = "{NLTE Spectra of Kilonovae}",
      journal = {\mnras},
         year = 2023,
        month = oct,
          doi = {10.1093/mnras/stad3106},
archivePrefix = {arXiv},
       eprint = {2309.01134},
 primaryClass = {astro-ph.HE},
       adsurl = {https://ui.adsabs.harvard.edu/abs/2023MNRAS.tmp.2989P}
}

@ARTICLE{Sneppen2023,
       author = {{Sneppen}, Albert and {Watson}, Darach},
        title = "{Discovery of a 760 nm P Cygni line in AT2017gfo: Identification of yttrium in the kilonova photosphere}",
      journal = {\aap},
         year = 2023,
        month = jul,
       volume = {675},
          eid = {A194},
        pages = {A194},
          doi = {10.1051/0004-6361/202346421},
archivePrefix = {arXiv},
       eprint = {2306.14942},
 primaryClass = {astro-ph.HE},
       adsurl = {https://ui.adsabs.harvard.edu/abs/2023A&A...675A.194S}
}

@ARTICLE{Mazzali2000,
       author = {{Mazzali}, P.~A.},
        title = "{Applications of an improved Monte Carlo code to the synthesis of early-time Supernova spectra}",
      journal = {\aap},
         year = 2000,
        month = nov,
       volume = {363},
        pages = {705-716},
       adsurl = {https://ui.adsabs.harvard.edu/abs/2000A&A...363..705M}
}

@ARTICLE{Maeda2002,
       author = {{Maeda}, Keiichi and {Nakamura}, Takayoshi and {Nomoto}, Ken'ichi and {Mazzali}, Paolo A. and {Patat}, Ferdinando and {Hachisu}, Izumi},
        title = "{Explosive Nucleosynthesis in Aspherical Hypernova Explosions and Late-Time Spectra of SN 1998bw}",
      journal = {\apj},
         year = 2002,
        month = jan,
       volume = {565},
       number = {1},
        pages = {405-412},
          doi = {10.1086/324487},
       adsurl = {https://ui.adsabs.harvard.edu/abs/2002ApJ...565..405M}
}

@ARTICLE{Maeda2006,
       author = {{Maeda}, K. and {Nomoto}, K. and {Mazzali}, P.~A. and {Deng}, J.},
        title = "{Nebular Spectra of SN 1998bw Revisited: Detailed Study by One- and Two-dimensional Models}",
      journal = {\apj},
         year = 2006,
        month = apr,
       volume = {640},
       number = {2},
        pages = {854-877},
          doi = {10.1086/500187},
archivePrefix = {arXiv},
       eprint = {astro-ph/0508373},
 primaryClass = {astro-ph},
       adsurl = {https://ui.adsabs.harvard.edu/abs/2006ApJ...640..854M}
}

@ARTICLE{Hoflich2002,
       author = {{H{\"o}flich}, Peter and {Gerardy}, Christopher L. and {Fesen}, Robert A. and {Sakai}, Shoko},
        title = "{Infrared Spectra of the Subluminous Type Ia Supernova SN 1999by}",
      journal = {\apj},
         year = 2002,
        month = apr,
       volume = {568},
       number = {2},
        pages = {791-806},
          doi = {10.1086/339063},
archivePrefix = {arXiv},
       eprint = {astro-ph/0112126},
 primaryClass = {astro-ph},
       adsurl = {https://ui.adsabs.harvard.edu/abs/2002ApJ...568..791H}
}

@ARTICLE{Hoflich1998,
       author = {{H{\"o}flich}, P. and {Wheeler}, J.~C. and {Thielemann}, F.~K.},
        title = "{Type Ia Supernovae: Influence of the Initial Composition on the Nucleosynthesis, Light Curves, and Spectra and Consequences for the Determination of {\ensuremath{\Omega}}$_{M}$ and {\ensuremath{\Lambda}}}",
      journal = {\apj},
         year = 1998,
        month = mar,
       volume = {495},
       number = {2},
        pages = {617-629},
          doi = {10.1086/305327},
archivePrefix = {arXiv},
       eprint = {astro-ph/9709233},
 primaryClass = {astro-ph},
       adsurl = {https://ui.adsabs.harvard.edu/abs/1998ApJ...495..617H}
}

@ARTICLE{Kasen2005,
       author = {{Kasen}, Daniel and {Plewa}, Tomasz},
        title = "{Spectral Signatures of Gravitationally Confined Thermonuclear Supernova Explosions}",
      journal = {\apjl},
         year = 2005,
        month = mar,
       volume = {622},
       number = {1},
        pages = {L41-L44},
          doi = {10.1086/429375},
archivePrefix = {arXiv},
       eprint = {astro-ph/0501453},
 primaryClass = {astro-ph},
       adsurl = {https://ui.adsabs.harvard.edu/abs/2005ApJ...622L..41K}
}

@ARTICLE{Kromer2013,
       author = {{Kromer}, M. and {Fink}, M. and {Stanishev}, V. and {Taubenberger}, S. and {Ciaraldi-Schoolman}, F. and {Pakmor}, R. and {R{\"o}pke}, F.~K. and {Ruiter}, A.~J. and {Seitenzahl}, I.~R. and {Sim}, S.~A. and {Blanc}, G. and {Elias-Rosa}, N. and {Hillebrandt}, W.},
        title = "{3D deflagration simulations leaving bound remnants: a model for 2002cx-like Type Ia supernovae}",
      journal = {\mnras},
         year = 2013,
        month = mar,
       volume = {429},
       number = {3},
        pages = {2287-2297},
          doi = {10.1093/mnras/sts498},
archivePrefix = {arXiv},
       eprint = {1210.5243},
 primaryClass = {astro-ph.HE},
       adsurl = {https://ui.adsabs.harvard.edu/abs/2013MNRAS.429.2287K}
}

@ARTICLE{Kromer2010,
       author = {{Kromer}, M. and {Sim}, S.~A. and {Fink}, M. and {R{\"o}pke}, F.~K. and {Seitenzahl}, I.~R. and {Hillebrandt}, W.},
        title = "{Double-detonation Sub-Chandrasekhar Supernovae: Synthetic Observables for Minimum Helium Shell Mass Models}",
      journal = {\apj},
         year = 2010,
        month = aug,
       volume = {719},
       number = {2},
        pages = {1067-1082},
          doi = {10.1088/0004-637X/719/2/1067},
archivePrefix = {arXiv},
       eprint = {1006.4489},
 primaryClass = {astro-ph.HE},
       adsurl = {https://ui.adsabs.harvard.edu/abs/2010ApJ...719.1067K}
}

@ARTICLE{Sim2013,
       author = {{Sim}, S.~A. and {Seitenzahl}, I.~R. and {Kromer}, M. and {Ciaraldi-Schoolmann}, F. and {R{\"o}pke}, F.~K. and {Fink}, M. and {Hillebrandt}, W. and {Pakmor}, R. and {Ruiter}, A.~J. and {Taubenberger}, S.},
        title = "{Synthetic light curves and spectra for three-dimensional delayed-detonation models of Type Ia supernovae}",
      journal = {\mnras},
         year = 2013,
        month = nov,
       volume = {436},
       number = {1},
        pages = {333-347},
          doi = {10.1093/mnras/stt1574},
archivePrefix = {arXiv},
       eprint = {1308.4833},
 primaryClass = {astro-ph.HE},
       adsurl = {https://ui.adsabs.harvard.edu/abs/2013MNRAS.436..333S}
}

@ARTICLE{Sim2012,
       author = {{Sim}, S.~A. and {Fink}, M. and {Kromer}, M. and {R{\"o}pke}, F.~K. and {Ruiter}, A.~J. and {Hillebrandt}, W.},
        title = "{2D simulations of the double-detonation model for thermonuclear transients from low-mass carbon-oxygen white dwarfs}",
      journal = {\mnras},
         year = 2012,
        month = mar,
       volume = {420},
       number = {4},
        pages = {3003-3016},
          doi = {10.1111/j.1365-2966.2011.20162.x},
archivePrefix = {arXiv},
       eprint = {1111.2117},
 primaryClass = {astro-ph.HE},
       adsurl = {https://ui.adsabs.harvard.edu/abs/2012MNRAS.420.3003S}
}

@ARTICLE{Sim2010,
       author = {{Sim}, S.~A. and {R{\"o}pke}, F.~K. and {Hillebrandt}, W. and {Kromer}, M. and {Pakmor}, R. and {Fink}, M. and {Ruiter}, A.~J. and {Seitenzahl}, I.~R.},
        title = "{Detonations in Sub-Chandrasekhar-mass C+O White Dwarfs}",
      journal = {\apjl},
         year = 2010,
        month = may,
       volume = {714},
       number = {1},
        pages = {L52-L57},
          doi = {10.1088/2041-8205/714/1/L52},
archivePrefix = {arXiv},
       eprint = {1003.2917},
 primaryClass = {astro-ph.HE},
       adsurl = {https://ui.adsabs.harvard.edu/abs/2010ApJ...714L..52S}
}

@ARTICLE{Jerkstrand2016,
       author = {{Jerkstrand}, A. and {Smartt}, S.~J. and {Heger}, A.},
        title = "{Nebular spectra of pair-instability supernovae}",
      journal = {\mnras},
         year = 2016,
        month = jan,
       volume = {455},
       number = {3},
        pages = {3207-3229},
          doi = {10.1093/mnras/stv2369},
archivePrefix = {arXiv},
       eprint = {1510.02698},
 primaryClass = {astro-ph.SR},
       adsurl = {https://ui.adsabs.harvard.edu/abs/2016MNRAS.455.3207J}
}

@ARTICLE{Dessart2016,
       author = {{Dessart}, Luc and {Hillier}, D. John and {Audit}, Edouard and {Livne}, Eli and {Waldman}, Roni},
        title = "{Models of interacting supernovae and their spectral diversity}",
      journal = {\mnras},
         year = 2016,
        month = may,
       volume = {458},
       number = {2},
        pages = {2094-2121},
          doi = {10.1093/mnras/stw336},
archivePrefix = {arXiv},
       eprint = {1602.02977},
 primaryClass = {astro-ph.SR},
       adsurl = {https://ui.adsabs.harvard.edu/abs/2016MNRAS.458.2094D}
}

@ARTICLE{Dessart2013,
       author = {{Dessart}, Luc and {Hillier}, D. John and {Waldman}, Roni and {Livne}, Eli},
        title = "{Type II-Plateau supernova radiation: dependences on progenitor and explosion properties}",
      journal = {\mnras},
         year = 2013,
        month = aug,
       volume = {433},
       number = {2},
        pages = {1745-1763},
          doi = {10.1093/mnras/stt861},
archivePrefix = {arXiv},
       eprint = {1305.3386},
 primaryClass = {astro-ph.SR},
       adsurl = {https://ui.adsabs.harvard.edu/abs/2013MNRAS.433.1745D}
}

@ARTICLE{Maguire2018,
       author = {{Maguire}, K. and {Sim}, S.~A. and {Shingles}, L. and {Spyromilio}, J. and {Jerkstrand}, A. and {Sullivan}, M. and {Chen}, T. -W. and {Cartier}, R. and {Dimitriadis}, G. and {Frohmaier}, C. and {Galbany}, L. and {Guti{\'e}rrez}, C.~P. and {Hosseinzadeh}, G. and {Howell}, D.~A. and {Inserra}, C. and {Rudy}, R. and {Sollerman}, J.},
        title = "{Using late-time optical and near-infrared spectra to constrain Type Ia supernova explosion properties}",
      journal = {\mnras},
         year = 2018,
        month = jul,
       volume = {477},
       number = {3},
        pages = {3567-3582},
          doi = {10.1093/mnras/sty820},
archivePrefix = {arXiv},
       eprint = {1803.10252},
 primaryClass = {astro-ph.HE},
       adsurl = {https://ui.adsabs.harvard.edu/abs/2018MNRAS.477.3567M}
}

@ARTICLE{Tominaga2007,
       author = {{Tominaga}, Nozomu and {Umeda}, Hideyuki and {Nomoto}, Ken'ichi},
        title = "{Supernova Nucleosynthesis in Population III 13-50 M$_{\odot}$ Stars and Abundance Patterns of Extremely Metal-poor Stars}",
      journal = {\apj},
         year = 2007,
        month = may,
       volume = {660},
       number = {1},
        pages = {516-540},
          doi = {10.1086/513063},
archivePrefix = {arXiv},
       eprint = {astro-ph/0701381},
 primaryClass = {astro-ph},
       adsurl = {https://ui.adsabs.harvard.edu/abs/2007ApJ...660..516T}
}

@ARTICLE{Rauscher2002,
       author = {{Rauscher}, T. and {Heger}, A. and {Hoffman}, R.~D. and {Woosley}, S.~E.},
        title = "{Nucleosynthesis in Massive Stars with Improved Nuclear and Stellar Physics}",
      journal = {\apj},
         year = 2002,
        month = sep,
       volume = {576},
       number = {1},
        pages = {323-348},
          doi = {10.1086/341728},
archivePrefix = {arXiv},
       eprint = {astro-ph/0112478},
 primaryClass = {astro-ph},
       adsurl = {https://ui.adsabs.harvard.edu/abs/2002ApJ...576..323R}
}

@ARTICLE{Chiappini1997,
       author = {{Chiappini}, C. and {Matteucci}, F. and {Gratton}, R.},
        title = "{The Chemical Evolution of the Galaxy: The Two-Infall Model}",
      journal = {\apj},
         year = 1997,
        month = mar,
       volume = {477},
       number = {2},
        pages = {765-780},
          doi = {10.1086/303726},
archivePrefix = {arXiv},
       eprint = {astro-ph/9609199},
 primaryClass = {astro-ph},
       adsurl = {https://ui.adsabs.harvard.edu/abs/1997ApJ...477..765C}
}

@ARTICLE{Ergon2018,
       author = {{Ergon}, M. and {Fransson}, C. and {Jerkstrand}, A. and {Kozma}, C. and {Kromer}, M. and {Spricer}, K.},
        title = "{Monte-Carlo methods for NLTE spectral synthesis of supernovae}",
      journal = {\aap},
         year = 2018,
        month = dec,
       volume = {620},
          eid = {A156},
        pages = {A156},
          doi = {10.1051/0004-6361/201833043},
archivePrefix = {arXiv},
       eprint = {1810.07165},
 primaryClass = {astro-ph.SR},
       adsurl = {https://ui.adsabs.harvard.edu/abs/2018A&A...620A.156E}
}

@ARTICLE{Dalgarno1972,
       author = {{Dalgarno}, A. and {McCray}, R.~A.},
        title = "{Heating and Ionization of HI Regions}",
      journal = {\araa},
         year = 1972,
        month = jan,
       volume = {10},
        pages = {375},
          doi = {10.1146/annurev.aa.10.090172.002111},
       adsurl = {https://ui.adsabs.harvard.edu/abs/1972ARA&A..10..375D}
}

@ARTICLE{Kasen2006,
       author = {{Kasen}, Daniel and {Thomas}, R.~C. and {Nugent}, P.},
        title = "{Time-dependent Monte Carlo Radiative Transfer Calculations for Three-dimensional Supernova Spectra, Light Curves, and Polarization}",
      journal = {\apj},
         year = 2006,
        month = nov,
       volume = {651},
       number = {1},
        pages = {366-380},
          doi = {10.1086/506190},
archivePrefix = {arXiv},
       eprint = {astro-ph/0606111},
 primaryClass = {astro-ph},
       adsurl = {https://ui.adsabs.harvard.edu/abs/2006ApJ...651..366K}
}

@ARTICLE{Metzger2010,
       author = {{Metzger}, B.~D. and {Mart{\'\i}nez-Pinedo}, G. and {Darbha}, S. and {Quataert}, E. and {Arcones}, A. and {Kasen}, D. and {Thomas}, R. and {Nugent}, P. and {Panov}, I.~V. and {Zinner}, N.~T.},
        title = "{Electromagnetic counterparts of compact object mergers powered by the radioactive decay of r-process nuclei}",
      journal = {\mnras},
         year = 2010,
        month = aug,
       volume = {406},
       number = {4},
        pages = {2650-2662},
          doi = {10.1111/j.1365-2966.2010.16864.x},
archivePrefix = {arXiv},
       eprint = {1001.5029},
 primaryClass = {astro-ph.HE},
       adsurl = {https://ui.adsabs.harvard.edu/abs/2010MNRAS.406.2650M}
}

@ARTICLE{Fransson1989,
       author = {{Fransson}, Claes and {Chevalier}, Roger A.},
        title = "{Late Emission from Supernovae: A Window on Stellar Nucleosynthesis}",
      journal = {\apj},
         year = 1989,
        month = aug,
       volume = {343},
        pages = {323},
          doi = {10.1086/167707},
       adsurl = {https://ui.adsabs.harvard.edu/abs/1989ApJ...343..323F}
}

@ARTICLE{Fransson1987,
       author = {{Fransson}, C. and {Chevalier}, R.~A.},
        title = "{Late Emission from SN 1987A}",
      journal = {\apjl},
         year = 1987,
        month = nov,
       volume = {322},
        pages = {L15},
          doi = {10.1086/185028},
       adsurl = {https://ui.adsabs.harvard.edu/abs/1987ApJ...322L..15F}
}

@ARTICLE{Mazzali1993,
       author = {{Mazzali}, P.~A. and {Lucy}, L.~B.},
        title = "{The application of Monte Carlo methods to the synthesis of early-time supernovae spectra.}",
      journal = {\aap},
         year = 1993,
        month = nov,
       volume = {279},
        pages = {447-456},
       adsurl = {https://ui.adsabs.harvard.edu/abs/1993A&A...279..447M}
}

@ARTICLE{RuizLapuente1992,
       author = {{Ruiz-Lapuente}, P. and {Cappellaro}, E. and {Turatto}, M. and {Gouiffes}, C. and {Danziger}, I.~J. and {della Valle}, M. and {Lucy}, L.~B.},
        title = "{Modeling the Iron-dominated Spectra of the Type Ia Supernova SN 1991T at Premaximum}",
      journal = {\apjl},
         year = 1992,
        month = mar,
       volume = {387},
        pages = {L33},
          doi = {10.1086/186299},
       adsurl = {https://ui.adsabs.harvard.edu/abs/1992ApJ...387L..33R}
}

@ARTICLE{Lucy1987,
       author = {{Lucy}, L.~B.},
        title = "{Computed ultraviolet spectra for SN 1987A.}",
      journal = {\aap},
         year = 1987,
        month = aug,
       volume = {182},
        pages = {L31-L33},
       adsurl = {https://ui.adsabs.harvard.edu/abs/1987A&A...182L..31L}
}

@ARTICLE{Branch1983,
       author = {{Branch}, D. and {Lacy}, C.~H. and {McCall}, M.~L. and {Sutherland}, P.~G. and {Uomoto}, A. and {Wheeler}, J.~C. and {Wills}, B.~J.},
        title = "{The type I supernova 1981b in NGC 4536 : the first 100 days.}",
      journal = {\apj},
         year = 1983,
        month = jul,
       volume = {270},
        pages = {123-139},
          doi = {10.1086/161103},
       adsurl = {https://ui.adsabs.harvard.edu/abs/1983ApJ...270..123B}
}

@ARTICLE{Branch1985,
       author = {{Branch}, D. and {Doggett}, J.~B. and {Nomoto}, K. and {Thielemann}, F. -K.},
        title = "{Accreting white dwarf models for type I supernovae. IV. The optical spectrum of a carbon-deflagration supernova.}",
      journal = {\apj},
         year = 1985,
        month = jul,
       volume = {294},
        pages = {619-625},
          doi = {10.1086/163329},
       adsurl = {https://ui.adsabs.harvard.edu/abs/1985ApJ...294..619B}
}

@INPROCEEDINGS{Branch1980,
       author = {{Branch}, David},
        title = "{Synthetic spectra of supernovae}",
    booktitle = {Supernovae Spectra},
         year = 1980,
       editor = {{Meyerott}, Roland and {Gillespie}, H. George},
       series = {American Institute of Physics Conference Series},
       volume = {63},
        month = nov,
        pages = {39-48},
          doi = {10.1063/1.32213},
       adsurl = {https://ui.adsabs.harvard.edu/abs/1980AIPC...63...39B}
}

@ARTICLE{Gaskell1986,
       author = {{Gaskell}, C.~M. and {Cappellaro}, E. and {Dinerstein}, H.~L. and {Garnett}, D.~R. and {Harkness}, R.~P. and {Wheeler}, J.~C.},
        title = "{Type Ib Supernovae 1983n and 1985f: Oxygen-rich Late Time Spectra}",
      journal = {\apjl},
         year = 1986,
        month = jul,
       volume = {306},
        pages = {L77},
          doi = {10.1086/184709},
       adsurl = {https://ui.adsabs.harvard.edu/abs/1986ApJ...306L..77G}
}

@ARTICLE{Filippenko1986,
       author = {{Filippenko}, A.~V. and {Sargent}, W.~L.~W.},
        title = "{The unique supernova (1985f) in NGC 4618.}",
      journal = {\aj},
         year = 1986,
        month = apr,
       volume = {91},
        pages = {691-696},
          doi = {10.1086/114051},
       adsurl = {https://ui.adsabs.harvard.edu/abs/1986AJ.....91..691F}
}

@ARTICLE{Wheeler1985,
       author = {{Wheeler}, J.~C. and {Levreault}, R.},
        title = "{The peculiar type I supernova in NGC 991.}",
      journal = {\apjl},
         year = 1985,
        month = jul,
       volume = {294},
        pages = {L17-L20},
          doi = {10.1086/184500},
       adsurl = {https://ui.adsabs.harvard.edu/abs/1985ApJ...294L..17W}
}

@ARTICLE{Li2011,
       author = {{Li}, Weidong and {Leaman}, Jesse and {Chornock}, Ryan and {Filippenko}, Alexei V. and {Poznanski}, Dovi and {Ganeshalingam}, Mohan and {Wang}, Xiaofeng and {Modjaz}, Maryam and {Jha}, Saurabh and {Foley}, Ryan J. and {Smith}, Nathan},
        title = "{Nearby supernova rates from the Lick Observatory Supernova Search - II. The observed luminosity functions and fractions of supernovae in a complete sample}",
      journal = {\mnras},
         year = 2011,
        month = apr,
       volume = {412},
       number = {3},
        pages = {1441-1472},
          doi = {10.1111/j.1365-2966.2011.18160.x},
archivePrefix = {arXiv},
       eprint = {1006.4612},
 primaryClass = {astro-ph.SR},
       adsurl = {https://ui.adsabs.harvard.edu/abs/2011MNRAS.412.1441L}
}

@ARTICLE{Baade1934,
       author = {{Baade}, W. and {Zwicky}, F.},
        title = "{On Super-novae}",
      journal = {PNAS},
         year = 1934,
        month = may,
       volume = {20},
       number = {5},
        pages = {254-259},
          doi = {10.1073/pnas.20.5.254},
       adsurl = {https://ui.adsabs.harvard.edu/abs/1934PNAS...20..254B}
}

@ARTICLE{Eastman1993,
       author = {{Eastman}, Ronald G. and {Pinto}, Philip A.},
        title = "{Spectrum Formation in Supernovae: Numerical Techniques}",
      journal = {\apj},
         year = 1993,
        month = aug,
       volume = {412},
        pages = {731},
          doi = {10.1086/172957},
       adsurl = {https://ui.adsabs.harvard.edu/abs/1993ApJ...412..731E}
}

@ARTICLE{Weaver1978,
       author = {{Weaver}, T.~A. and {Zimmerman}, G.~B. and {Woosley}, S.~E.},
        title = "{Presupernova evolution of massive stars}",
      journal = {\apj},
         year = 1978,
        month = nov,
       volume = {225},
        pages = {1021-1029},
          doi = {10.1086/156569},
       adsurl = {https://ui.adsabs.harvard.edu/abs/1978ApJ...225.1021W}
}

@ARTICLE{Hoflich1993,
       author = {{Hoeflich}, P. and {Mueller}, E. and {Khokhlov}, A.},
        title = "{Light curve models for type Ia supernovae: physical assumptions, their influence and validity}",
      journal = {\aap},
         year = 1993,
        month = feb,
       volume = {268},
       number = {2},
        pages = {570-590},
       adsurl = {https://ui.adsabs.harvard.edu/abs/1993A&A...268..570H}
}

@ARTICLE{Stone1992,
       author = {{Stone}, James M. and {Mihalas}, Dimitri and {Norman}, Michael L.},
        title = "{ZEUS-2D: A Radiation Magnetohydrodynamics Code for Astrophysical Flows in Two Space Dimensions. III. The Radiation Hydrodynamic Algorithms and Tests}",
      journal = {\apjs},
         year = 1992,
        month = jun,
       volume = {80},
        pages = {819},
          doi = {10.1086/191682},
       adsurl = {https://ui.adsabs.harvard.edu/abs/1992ApJS...80..819S}
}

@ARTICLE{Bulla2023,
       author = {{Bulla}, Mattia},
        title = "{The critical role of nuclear heating rates, thermalization efficiencies, and opacities for kilonova modelling and parameter inference}",
      journal = {\mnras},
         year = 2023,
        month = apr,
       volume = {520},
       number = {2},
        pages = {2558-2570},
          doi = {10.1093/mnras/stad232},
archivePrefix = {arXiv},
       eprint = {2211.14348},
 primaryClass = {astro-ph.HE},
       adsurl = {https://ui.adsabs.harvard.edu/abs/2023MNRAS.520.2558B}
}

@ARTICLE{Tanaka2013,
       author = {{Tanaka}, Masaomi and {Hotokezaka}, Kenta},
        title = "{Radiative Transfer Simulations of Neutron Star Merger Ejecta}",
      journal = {\apj},
         year = 2013,
        month = oct,
       volume = {775},
       number = {2},
          eid = {113},
        pages = {113},
          doi = {10.1088/0004-637X/775/2/113},
archivePrefix = {arXiv},
       eprint = {1306.3742},
 primaryClass = {astro-ph.HE},
       adsurl = {https://ui.adsabs.harvard.edu/abs/2013ApJ...775..113T}
}

@ARTICLE{Kerzendorf2014,
       author = {{Kerzendorf}, Wolfgang E. and {Sim}, Stuart A.},
        title = "{A spectral synthesis code for rapid modelling of supernovae}",
      journal = {\mnras},
         year = 2014,
        month = may,
       volume = {440},
       number = {1},
        pages = {387-404},
          doi = {10.1093/mnras/stu055},
archivePrefix = {arXiv},
       eprint = {1401.5469},
 primaryClass = {astro-ph.SR},
       adsurl = {https://ui.adsabs.harvard.edu/abs/2014MNRAS.440..387K}
}

@BOOK{Mihalas1984,
       author = {{Mihalas}, D. and {Mihalas}, B.~W.},
        title = "{Foundations of radiation hydrodynamics}",
      address = {New York},
    publisher = {Oxford University Press},
         year = 1984,
       adsurl = {https://ui.adsabs.harvard.edu/abs/1984oup..book.....M}
}

@BOOK{Hubeny2014,
       author = {{Hubeny}, Ivan and {Mihalas}, Dimitri},
        title = "{Theory of Stellar Atmospheres}",
         year = 2014,
      address = {Princeton, NJ},
    publisher = {Princeton University Press},
       adsurl = {https://ui.adsabs.harvard.edu/abs/2014tsa..book.....H}
}

@ARTICLE{Kozma1998-I,
       author = {{Kozma}, Cecilia and {Fransson}, Claes},
        title = "{Late Spectral Evolution of SN 1987A. I. Temperature and Ionization}",
      journal = {\apj},
         year = 1998,
        month = mar,
       volume = {496},
       number = {2},
        pages = {946-966},
          doi = {10.1086/305409},
archivePrefix = {arXiv},
       eprint = {astro-ph/9712223},
 primaryClass = {astro-ph},
       adsurl = {https://ui.adsabs.harvard.edu/abs/1998ApJ...496..946K}
}

@INPROCEEDINGS{Hillier2003,
       author = {{Hillier}, D.~J.},
        title = "{On the Solution of the Statistical Equilibrium Equations}",
    booktitle = {Stellar Atmosphere Modeling},
         year = 2003,
       editor = {{Hubeny}, Ivan and {Mihalas}, Dimitri and {Werner}, Klaus},
    publisher = {Astronomical Society of the Pacific},
    series = {ASP Conference Series},
       volume = {288},
        month = jan,
        pages = {199},
       adsurl = {https://ui.adsabs.harvard.edu/abs/2003ASPC..288..199H}
}

@INCOLLECTION{Sim2017,
       author = {{Sim}, Stuart A.},
        title = "{Spectra of Supernovae During the Photospheric Phase}",
    booktitle = {Handbook of Supernovae},
         year = 2017,
       editor = {{Alsabti}, Athem W. and {Murdin}, Paul},
        pages = {769},
          doi = {10.1007/978-3-319-21846-5_28},
       adsurl = {https://ui.adsabs.harvard.edu/abs/2017hsn..book..769S}
}

@ARTICLE{Noebauer2019,
       author = {{Noebauer}, Ulrich M. and {Sim}, Stuart A.},
        title = "{Monte Carlo radiative transfer}",
      journal = {Living Rev. Comput. Astrophys.},
         year = 2019,
        month = jun,
       volume = {5},
       number = {1},
          eid = {1},
        pages = {1},
          doi = {10.1007/s41115-019-0004-9},
archivePrefix = {arXiv},
       eprint = {1907.09840},
 primaryClass = {astro-ph.IM},
       adsurl = {https://ui.adsabs.harvard.edu/abs/2019LRCA....5....1N}
}

@ARTICLE{Lucy1999a,
       author = {{Lucy}, L.~B.},
        title = "{Computing radiative equilibria with Monte Carlo techniques}",
      journal = {\aap},
         year = 1999,
        month = apr,
       volume = {344},
        pages = {282-288},
       adsurl = {https://ui.adsabs.harvard.edu/abs/1999A&A...344..282L}
}

@ARTICLE{Lucy2005,
       author = {{Lucy}, L.~B.},
        title = "{Monte Carlo techniques for time-dependent radiative transfer  in 3-D supernovae}",
      journal = {\aap},
         year = 2005,
        month = jan,
       volume = {429},
        pages = {19-30},
          doi = {10.1051/0004-6361:20041656},
archivePrefix = {arXiv},
       eprint = {astro-ph/0409249},
 primaryClass = {astro-ph},
       adsurl = {https://ui.adsabs.harvard.edu/abs/2005A&A...429...19L}
}

@ARTICLE{Lucy2003,
       author = {{Lucy}, L.~B.},
        title = "{Monte Carlo transition probabilities. II.}",
      journal = {\aap},
         year = 2003,
        month = may,
       volume = {403},
        pages = {261-275},
          doi = {10.1051/0004-6361:20030357},
archivePrefix = {arXiv},
       eprint = {astro-ph/0303202},
 primaryClass = {astro-ph},
       adsurl = {https://ui.adsabs.harvard.edu/abs/2003A&A...403..261L}
}

@ARTICLE{Wollaeger2013,
       author = {{Wollaeger}, Ryan T. and {van Rossum}, Daniel R. and {Graziani}, Carlo and {Couch}, Sean M. and {Jordan}, George C., IV and {Lamb}, Donald Q. and {Moses}, Gregory A.},
        title = "{Radiation Transport for Explosive Outflows: A Multigroup Hybrid Monte Carlo Method}",
      journal = {\apjs},
         year = 2013,
        month = dec,
       volume = {209},
       number = {2},
          eid = {36},
        pages = {36},
          doi = {10.1088/0067-0049/209/2/36},
archivePrefix = {arXiv},
       eprint = {1306.5700},
 primaryClass = {astro-ph.HE},
       adsurl = {https://ui.adsabs.harvard.edu/abs/2013ApJS..209...36W}
}

@ARTICLE{Kromer2009,
       author = {{Kromer}, M. and {Sim}, S.~A.},
        title = "{Time-dependent three-dimensional spectrum synthesis for Type Ia supernovae}",
      journal = {\mnras},
         year = 2009,
        month = oct,
       volume = {398},
       number = {4},
        pages = {1809-1826},
          doi = {10.1111/j.1365-2966.2009.15256.x},
archivePrefix = {arXiv},
       eprint = {0906.3152},
 primaryClass = {astro-ph.HE},
       adsurl = {https://ui.adsabs.harvard.edu/abs/2009MNRAS.398.1809K}
}

@ARTICLE{Opal1971,
       author = {{Opal}, C.~B. and {Peterson}, W.~K. and {Beaty}, E.~C.},
        title = "{Measurements of Secondary-Electron Spectra Produced by Electron Impact Ionization of a Number of Simple Gases}",
      journal = {\jcp},
         year = 1971,
        month = oct,
       volume = {55},
       number = {8},
        pages = {4100-4106},
          doi = {10.1063/1.1676707},
       adsurl = {https://ui.adsabs.harvard.edu/abs/1971JChPh..55.4100O}
}

@BOOK{Longair2011,
       author = {{Longair}, Malcolm S.},
        title = "{High Energy Astrophysics}",
    publisher = {Cambridge University Press},
         year = 2011,
          doi = {10.1017/CBO9780511778346},
       adsurl = {https://ui.adsabs.harvard.edu/abs/2011hea..book.....L}
}

@ARTICLE{Xu1991,
       author = {{Xu}, Yueming and {McCray}, Richard},
        title = "{Energy Degradation of Fast Electrons in Hydrogen Gas}",
      journal = {\apj},
         year = 1991,
        month = jul,
       volume = {375},
        pages = {190},
          doi = {10.1086/170180},
       adsurl = {https://ui.adsabs.harvard.edu/abs/1991ApJ...375..190X}
}

@ARTICLE{Kozma1992,
       author = {{Kozma}, Cecilia and {Fransson}, Claes},
        title = "{Gamma-Ray Deposition and Nonthermal Excitation in Supernovae}",
      journal = {\apj},
         year = 1992,
        month = may,
       volume = {390},
        pages = {602},
          doi = {10.1086/171311},
       adsurl = {https://ui.adsabs.harvard.edu/abs/1992ApJ...390..602K}
}

@PHDTHESIS{Axelrod1980,
       author = {{Axelrod}, T.~S.},
        title = "{Late Time Optical Spectra from the NICKEL(56) Model for Type I Supernovae}",
       school = {University of California, Santa Cruz},
         year = 1980,
        month = jan,
       adsurl = {https://ui.adsabs.harvard.edu/abs/1980PhDT.........1A}
}

@ARTICLE{Hillier1990,
       author = {{Hillier}, D.~J.},
        title = "{An iterative method for the solution of the statistical and radiativeequilibrium equations in expanding atmospheres}",
      journal = {\aap},
         year = 1990,
        month = may,
       volume = {231},
        pages = {116-124},
       adsurl = {https://ui.adsabs.harvard.edu/abs/1990A&A...231..116H}
}

@ARTICLE{Blondin2022,
     author = {{Blondin}, St{\'e}phane and {Blinnikov}, Sergei and {Callan}, Fionntan P. and {Collins}, Christine E. and {Dessart}, Luc and {Even}, Wesley and {Fl{\"o}rs}, Andreas and {Fullard}, Andrew G. and {Hillier}, D. John and {Jerkstrand}, Anders and {Kasen}, Daniel and {Katz}, Boaz and {Kerzendorf}, Wolfgang and {Kozyreva}, Alexandra and {O'Brien}, Jack and {P{\'a}ssaro}, Ezequiel A. and {Roth}, Nathaniel and {Shen}, Ken J. and {Shingles}, Luke and {Sim}, Stuart A. and {Singhal}, Jaladh and {Smith}, Isaac G. and {Sorokina}, Elena and {Utrobin}, Victor P. and {Vogl}, Christian and {Williamson}, Marc and {Wollaeger}, Ryan and {Woosley}, Stan E. and {Wygoda}, Nahliel},
        title = "{StaNdaRT: a repository of standardised test models and outputs for supernova radiative transfer}",
      journal = {\aap},
         year = 2022,
        month = dec,
       volume = {668},
          eid = {A163},
        pages = {A163},
          doi = {10.1051/0004-6361/202244134},
archivePrefix = {arXiv},
       eprint = {2209.11671},
 primaryClass = {astro-ph.SR},
       adsurl = {https://ui.adsabs.harvard.edu/abs/2022A&A...668A.163B}
}

@ARTICLE{Fransson2015,
       author = {{Fransson}, Claes and {Jerkstrand}, Anders},
        title = "{Reconciling the Infrared Catastrophe and Observations of SN 2011fe}",
      journal = {\apjl},
         year = 2015,
        month = nov,
       volume = {814},
       number = {1},
          eid = {L2},
        pages = {L2},
          doi = {10.1088/2041-8205/814/1/L2},
archivePrefix = {arXiv},
       eprint = {1511.00245},
 primaryClass = {astro-ph.SR},
       adsurl = {https://ui.adsabs.harvard.edu/abs/2015ApJ...814L...2F}
}

@ARTICLE{Liljegren2023,
       author = {{Liljegren}, S. and {Jerkstrand}, A. and {Barklem}, P.~S. and {Nyman}, G. and {Brady}, R. and {Yurchenko}, S.~N.},
        title = "{The molecular chemistry of Type Ibc supernovae and diagnostic potential with the James Webb Space Telescope}",
      journal = {\aap},
         year = 2023,
        month = jun,
       volume = {674},
          eid = {A184},
        pages = {A184},
          doi = {10.1051/0004-6361/202243491},
archivePrefix = {arXiv},
       eprint = {2203.07021},
 primaryClass = {astro-ph.SR},
       adsurl = {https://ui.adsabs.harvard.edu/abs/2023A&A...674A.184L}
}

@ARTICLE{Liljegren2020,
       author = {{Liljegren}, S. and {Jerkstrand}, A. and {Grumer}, J.},
        title = "{Carbon monoxide formation and cooling in supernovae}",
      journal = {\aap},
         year = 2020,
        month = oct,
       volume = {642},
          eid = {A135},
        pages = {A135},
          doi = {10.1051/0004-6361/202038116},
archivePrefix = {arXiv},
       eprint = {2008.03160},
 primaryClass = {astro-ph.SR},
       adsurl = {https://ui.adsabs.harvard.edu/abs/2020A&A...642A.135L}
}

@ARTICLE{Jerkstrand2018,
       author = {{Jerkstrand}, A. and {Ertl}, T. and {Janka}, H. -T. and {M{\"u}ller}, E. and {Sukhbold}, T. and {Woosley}, S.~E.},
        title = "{Emission line models for the lowest mass core-collapse supernovae - I. Case study of a 9 M$_{{\ensuremath{\odot}}}$ one-dimensional neutrino-driven explosion}",
      journal = {\mnras},
         year = 2018,
        month = mar,
       volume = {475},
       number = {1},
        pages = {277-305},
          doi = {10.1093/mnras/stx2877},
archivePrefix = {arXiv},
       eprint = {1710.04508},
 primaryClass = {astro-ph.SR},
       adsurl = {https://ui.adsabs.harvard.edu/abs/2018MNRAS.475..277J}
}

@ARTICLE{Jerkstrand2011,
       author = {{Jerkstrand}, A. and {Fransson}, C. and {Kozma}, C.},
        title = "{The $^{44}$Ti-powered spectrum of SN 1987A}",
      journal = {\aap},
         year = 2011,
        month = jun,
       volume = {530},
          eid = {A45},
        pages = {A45},
          doi = {10.1051/0004-6361/201015937},
archivePrefix = {arXiv},
       eprint = {1103.3653},
 primaryClass = {astro-ph.HE},
       adsurl = {https://ui.adsabs.harvard.edu/abs/2011A&A...530A..45J}
}

@PHDTHESIS{Jerkstrand2011PhD,
       author = {{Jerkstrand}, Anders},
        title = "{Spectral modeling of nebular-phase supernovae}",
       school = {Stockholm University},
         year = 2011,
        month = dec,
       adsurl = {https://ui.adsabs.harvard.edu/abs/2011PhDT........90J}
}

@ARTICLE{Jerkstrand2017,
       author = {{Jerkstrand}, A. and {Smartt}, S.~J. and {Inserra}, C. and {Nicholl}, M. and {Chen}, T. -W. and {Kr{\"u}hler}, T. and {Sollerman}, J. and {Taubenberger}, S. and {Gal-Yam}, A. and {Kankare}, E. and {Maguire}, K. and {Fraser}, M. and {Valenti}, S. and {Sullivan}, M. and {Cartier}, R. and {Young}, D.~R.},
        title = "{Long-duration Superluminous Supernovae at Late Times}",
      journal = {\apj},
         year = 2017,
        month = jan,
       volume = {835},
       number = {1},
          eid = {13},
        pages = {13},
          doi = {10.3847/1538-4357/835/1/13},
archivePrefix = {arXiv},
       eprint = {1608.02994},
 primaryClass = {astro-ph.HE},
       adsurl = {https://ui.adsabs.harvard.edu/abs/2017ApJ...835...13J},
      IGNOREnote = {This paper derives two in particularly important results on superluminous Type Ic supernovae: 1) They have a strong similarity with GRB SNe, revealed once galaxy contamination is recognized and removed. 2) Robustly inferred large (largely model-independent) masses of oxygen in the ejecta, which implicates very massive Wolf-Rayet star progenitors. Combined with LIGO results that black holes in the mass range $15-40~M_\odot$ exist, and almost certainly imply failed SNe, it now seems compelling that WR stars sometimes make black holes but sometimes explode as SLSNe or GRB SNe.}
}

@ARTICLE{Jerkstrand2015a,
       author = {{Jerkstrand}, A. and {Ergon}, M. and {Smartt}, S.~J. and {Fransson}, C. and {Sollerman}, J. and {Taubenberger}, S. and {Bersten}, M. and {Spyromilio}, J.},
        title = "{Late-time spectral line formation in Type IIb supernovae, with application to SN 1993J, SN 2008ax, and SN 2011dh}",
      journal = {\aap},
         year = 2015,
        month = jan,
       volume = {573},
          eid = {A12},
        pages = {A12},
          doi = {10.1051/0004-6361/201423983},
archivePrefix = {arXiv},
       eprint = {1408.0732},
 primaryClass = {astro-ph.HE},
       adsurl = {https://ui.adsabs.harvard.edu/abs/2015A\&A...573A..12J},
      IGNOREnote = {Modelling of Type IIb SNe, with particular focus on SN 2011dh. The [O I] 6300, 6364 lines constrain the progenitors of SN 1993J, SN 2008ax, and SN 2011dh to the 12-16 $M_\odot$ range. The paper also studies magnesium line formation, and derives semi-analytical methods to use the Mg I 1.50 $\mu$m luminosity combined with the oxygen recombination line luminosities to determine the magnesium mass (0.02 - 0.14 $M_\odot$ for SN 2011dh). The paper also studies radiative transfer effects; the line blocking of the metal core comes out as a very plausible explanation for the line blue shifts seen on most stripped envelope SNe.}
}

@ARTICLE{Jerkstrand2012,
       author = {{Jerkstrand}, A. and {Fransson}, C. and {Maguire}, K. and {Smartt}, S. and {Ergon}, M. and {Spyromilio}, J.},
        title = "{The progenitor mass of the Type IIP supernova SN 2004et from late-time spectral modeling}",
      journal = {\aap},
         year = 2012,
        month = oct,
       volume = {546},
          eid = {A28},
        pages = {A28},
          doi = {10.1051/0004-6361/201219528},
archivePrefix = {arXiv},
       eprint = {1208.2183},
 primaryClass = {astro-ph.HE},
       adsurl = {https://ui.adsabs.harvard.edu/abs/2012A\&A...546A..28J},
      IGNOREnote = {The first paper in a series aimed at analysing the oxygen yields in Type IIP SNe, and from that progenitor masses. Here we study SN 2004et, favouring a progenitor around 15 $M_\odot$. We also made detailed modelling of the unique set of MIR spectra available for SN 2004et. We could use the fact that many iron-group MIR lines are in LTE and optically thick to determine the fraction of the core occupied by the $^{56}$Ni bubble. This is the only SN apart from SN 1987A for which this has been done. This paper also contains the code developments that lead to the complete coupling of NLTE, radiation transport, and non-thermal effects.}
}

@ARTICLE{Jerkstrand2015b,
       author = {{Jerkstrand}, A. and {Smartt}, S.~J. and {Sollerman}, J. and {Inserra}, C. and {Fraser}, M. and {Spyromilio}, J. and {Fransson}, C. and {Chen}, T. -W. and {Barbarino}, C. and {Dall'Ora}, M. and {Botticella}, M.~T. and {Della Valle}, M. and {Gal-Yam}, A. and {Valenti}, S. and {Maguire}, K. and {Mazzali}, P. and {Tomasella}, L.},
        title = "{Supersolar Ni/Fe production in the Type IIP SN 2012ec}",
      journal = {\mnras},
         year = 2015,
        month = apr,
       volume = {448},
       number = {3},
        pages = {2482-2494},
          doi = {10.1093/mnras/stv087},
archivePrefix = {arXiv},
       eprint = {1410.8394},
 primaryClass = {astro-ph.SR},
       adsurl = {https://ui.adsabs.harvard.edu/abs/2015MNRAS.448.2482J},
      IGNOREnote = {Third paper in the Type IIP series, studying SN 2012ec. While the original aim was to study the oxygen lines, the paper took another turn as SN 2012ec turned out to have highly unusual nickel lines ([Ni II] 7378 and [Ni II] 1.939 mum). These lines may be used to show that a high production of stable nickel has occurred (mass 6E-3 Msun). The paper also presents a simple analytical technique one may use to determine the Ni/Fe ratio from luminosities in [Ni II] 7378 and [Fe II] 7378. SN 2012ec has a Ni/Fe ratio of over 3 times the solar value.}
}

@ARTICLE{Perlmutter1999,
       author = {{Perlmutter}, S. and {Aldering}, G. and {Goldhaber}, G. and {Knop}, R.~A. and {Nugent}, P. and {Castro}, P.~G. and {Deustua}, S. and {Fabbro}, S. and {Goobar}, A. and {Groom}, D.~E. and {Hook}, I.~M. and {Kim}, A.~G. and {Kim}, M.~Y. and {Lee}, J.~C. and {Nunes}, N.~J. and {Pain}, R. and {Pennypacker}, C.~R. and {Quimby}, R. and {Lidman}, C. and {Ellis}, R.~S. and {Irwin}, M. and {McMahon}, R.~G. and {Ruiz-Lapuente}, P. and {Walton}, N. and {Schaefer}, B. and {Boyle}, B.~J. and {Filippenko}, A.~V. and {Matheson}, T. and {Fruchter}, A.~S. and {Panagia}, N. and {Newberg}, H.~J.~M. and {Couch}, W.~J. and {Project}, The Supernova Cosmology},
        title = "{Measurements of {\ensuremath{\Omega}} and {\ensuremath{\Lambda}} from 42 High-Redshift Supernovae}",
      journal = {\apj},
         year = 1999,
        month = jun,
       volume = {517},
       number = {2},
        pages = {565-586},
          doi = {10.1086/307221},
archivePrefix = {arXiv},
       eprint = {astro-ph/9812133},
 primaryClass = {astro-ph},
       adsurl = {https://ui.adsabs.harvard.edu/abs/1999ApJ...517..565P}
}

@misc{kenta_hotokezaka_2020_3601589,
  author       = {{Hotokezaka}, Kenta and {Nakar}, Ehud},
  title        = "{Radioactive Heating Rate and Macronova (Kilonova) Light Curve}",
  year         = 2020,
  publisher    = {Zenodo},
  version      = {v1.0.0},
  doi          = {10.5281/zenodo.3601589},
  note         = {source code}
}

@ARTICLE{Hotokezaka2021,
       author = {{Hotokezaka}, Kenta and {Tanaka}, Masaomi and {Kato}, Daiji and {Gaigalas}, Gediminas},
        title = "{Nebular emission from lanthanide-rich ejecta of neutron star merger}",
      journal = {\mnras},
         year = 2021,
        month = oct,
       volume = {506},
       number = {4},
        pages = {5863-5877},
          doi = {10.1093/mnras/stab1975},
archivePrefix = {arXiv},
       eprint = {2102.07879},
 primaryClass = {astro-ph.HE},
       adsurl = {https://ui.adsabs.harvard.edu/abs/2021MNRAS.506.5863H}
}

@ARTICLE{Pognan2022b,
       author = {{Pognan}, Quentin and {Jerkstrand}, Anders and {Grumer}, Jon},
        title = "{NLTE effects on kilonova expansion opacities}",
      journal = {\mnras},
         year = 2022,
        month = jul,
       volume = {513},
       number = {4},
        pages = {5174-5197},
          doi = {10.1093/mnras/stac1253},
archivePrefix = {arXiv},
       eprint = {2202.09245},
 primaryClass = {astro-ph.HE},
       adsurl = {https://ui.adsabs.harvard.edu/abs/2022MNRAS.513.5174P}
}

@ARTICLE{Pognan2022a,
       author = {{Pognan}, Quentin and {Jerkstrand}, Anders and {Grumer}, Jon},
        title = "{On the validity of steady-state for nebular phase kilonovae}",
      journal = {\mnras},
         year = 2022,
        month = mar,
       volume = {510},
       number = {3},
        pages = {3806-3837},
          doi = {10.1093/mnras/stab3674},
archivePrefix = {arXiv},
       eprint = {2112.07484},
 primaryClass = {astro-ph.HE},
       adsurl = {https://ui.adsabs.harvard.edu/abs/2022MNRAS.510.3806P}
}

@ARTICLE{Watson2019,
       author = {{Watson}, Darach and {Hansen}, Camilla J. and {Selsing}, Jonatan and {Koch}, Andreas and {Malesani}, Daniele B. and {Andersen}, Anja C. and {Fynbo}, Johan P.~U. and {Arcones}, Almudena and {Bauswein}, Andreas and {Covino}, Stefano and {Grado}, Aniello and {Heintz}, Kasper E. and {Hunt}, Leslie and {Kouveliotou}, Chryssa and {Leloudas}, Giorgos and {Levan}, Andrew J. and {Mazzali}, Paolo and {Pian}, Elena},
        title = "{Identification of strontium in the merger of two neutron stars}",
      journal = {\nat},
         year = 2019,
        month = oct,
       volume = {574},
       number = {7779},
        pages = {497-500},
          doi = {10.1038/s41586-019-1676-3},
archivePrefix = {arXiv},
       eprint = {1910.10510},
 primaryClass = {astro-ph.HE},
       adsurl = {https://ui.adsabs.harvard.edu/abs/2019Natur.574..497W}
}

@ARTICLE{Domoto2021,
       author = {{Domoto}, Nanae and {Tanaka}, Masaomi and {Wanajo}, Shinya and {Kawaguchi}, Kyohei},
        title = "{Signatures of r-process Elements in Kilonova Spectra}",
      journal = {\apj},
         year = 2021,
        month = may,
       volume = {913},
       number = {1},
          eid = {26},
        pages = {26},
          doi = {10.3847/1538-4357/abf358},
archivePrefix = {arXiv},
       eprint = {2103.15284},
 primaryClass = {astro-ph.HE},
       adsurl = {https://ui.adsabs.harvard.edu/abs/2021ApJ...913...26D}
}

@ARTICLE{Domoto2022,
       author = {{Domoto}, Nanae and {Tanaka}, Masaomi and {Kato}, Daiji and {Kawaguchi}, Kyohei and {Hotokezaka}, Kenta and {Wanajo}, Shinya},
        title = "{Lanthanide Features in Near-infrared Spectra of Kilonovae}",
      journal = {\apj},
         year = 2022,
        month = nov,
       volume = {939},
       number = {1},
          eid = {8},
        pages = {8},
          doi = {10.3847/1538-4357/ac8c36},
archivePrefix = {arXiv},
       eprint = {2206.04232},
 primaryClass = {astro-ph.HE},
       adsurl = {https://ui.adsabs.harvard.edu/abs/2022ApJ...939....8D}
}

@ARTICLE{Hotokezaka2022,
       author = {{Hotokezaka}, Kenta and {Tanaka}, Masaomi and {Kato}, Daiji and {Gaigalas}, Gediminas},
        title = "{Tungsten versus Selenium as a potential source of kilonova nebular emission observed by Spitzer}",
      journal = {\mnras},
         year = 2022,
        month = sep,
       volume = {515},
       number = {1},
        pages = {L89-L93},
          doi = {10.1093/mnrasl/slac071},
archivePrefix = {arXiv},
       eprint = {2204.00737},
 primaryClass = {astro-ph.HE},
       adsurl = {https://ui.adsabs.harvard.edu/abs/2022MNRAS.515L..89H}
}

@ARTICLE{vanBaal2023,
       author = {{van Baal}, Bart F.~A. and {Jerkstrand}, Anders and {Wongwathanarat}, Annop and {Janka}, Hans-Thomas},
        title = "{Modelling supernova nebular lines in 3D with EXTRASS}",
      journal = {\mnras},
         year = 2023,
        month = jul,
       volume = {523},
       number = {1},
        pages = {954-973},
          doi = {10.1093/mnras/stad1488},
archivePrefix = {arXiv},
       eprint = {2305.08933},
 primaryClass = {astro-ph.HE},
       adsurl = {https://ui.adsabs.harvard.edu/abs/2023MNRAS.523..954V}
}

\end{document}